\documentclass[12pt,preprint]{aastex}

\newcommand{\MBH}{{\rm M}_\bullet}
\newcommand{\ML}{\Upsilon}
\newcommand{\MSun}{{\rm M}_\odot}
\newcommand{\kms}{km s$^{-1}$}
\newcommand{\Vcor}{$V_{\rm corr}$}
\newcommand{\NIRcor}{NIR$_{\rm corr}$}
\newcommand {\pc}{\mbox{pc}}

\shorttitle{NGC 4258 Black Hole Mass}
\shortauthors{Siopis et al.}

\begin{document}

\title{A Stellar Dynamical Measurement of the Black Hole Mass \\ 
     in the Maser Galaxy NGC 4258}

\author{Christos Siopis\altaffilmark{1,2},
Karl Gebhardt\altaffilmark{3},
Tod R.\ Lauer\altaffilmark{4},
John Kormendy\altaffilmark{3},
Jason Pinkney\altaffilmark{5},
Douglas Richstone\altaffilmark{1},
S.\ M.\ Faber\altaffilmark{6},
Scott Tremaine\altaffilmark{7},
M.\ C.\ Aller\altaffilmark{1},
Ralf Bender\altaffilmark{8},
Gary Bower\altaffilmark{9},
Alan Dressler\altaffilmark{10},
Alexei V.\ Filippenko\altaffilmark{11},
Richard Green\altaffilmark{12},
Luis C.\ Ho\altaffilmark{10}, and
John Magorrian\altaffilmark{13}}

\altaffiltext{1}{Department of Astronomy, University of Michigan, Ann
  Arbor, MI 48109-1042; siopis@umich.edu, maller@umich.edu,
  dor@umich.edu}

\altaffiltext{2}{Institut d'Astronomie et d'Astrophysique, 
   Universit\'e Libre de Bruxelles, CP 226, Boulevard du Triomphe, 
   1050 Bruxelles, Belgium; christos.siopis@ulb.ac.be}

\altaffiltext{3}{Department of Astronomy, University of Texas, 
RLM 15.308, Austin, TX 78712; gebhardt@astro.as.utexas.edu,
kormendy@astro.as.utexas.edu}

\altaffiltext{4}{National Optical Astronomy Observatories, P. O. Box
26732, Tucson, AZ 85726; lauer@noao.edu}

\altaffiltext{5}{Department of Physics and Astronomy, Ohio Northern
  University, 525 S Main St, Ada, OH 45810; j-pinkney@onu.edu}

\altaffiltext{6}{UCO/Lick Observatory, University of California,
Santa Cruz, CA 95064; faber@ucolick.org}

\altaffiltext{7}{School of Natural Sciences, Institute for Advanced 
    Study, Einstein Drive, Princeton, NJ 08540; tremaine@ias.edu}

\altaffiltext{8}{Universit\"ats-Sternwarte, Scheinerstra\ss e 1,
M\"unchen 81679, Germany; bender@usm.uni-muenchen.de}

\altaffiltext{9}{Computer Sciences Corporation, Space Telescope Science
Institute, 3700 San Martin Drive, Baltimore, MD 21218; bower@stsci.edu}

\altaffiltext{10}{The Observatories of the Carnegie Institution of
Washington, 813 Santa Barbara St., Pasadena, CA 91101;
dressler@ociw.edu, lho@ociw.edu}

\altaffiltext{11}{Department of Astronomy, University of California,
Berkeley, CA 94720-3411; alex@astro.berkeley.edu}

\altaffiltext{12}{Large Binocular Telescope Observatory, 933
  N. Cherry, Tucson, AZ 85721-0065; rgreen@as.arizona.edu}

\altaffiltext{13}{Rudolf Peierls Centre for Theoretical Physics, 
   University of Oxford, Keble Road,
  Oxford, OX1 3RH, UK; j.magorrian1@physics.ox.ac.uk}

%%%%%%%%%%%%%%%%%%%%%%%%%%%%%%%%%%%%%%%%%%%%%%%%%%%%%%%%%%%%%%%%%%%%%%%%%%%
%%%%%%%%%%%%%%%%%%%%%%%%%%%%%%%%%%%%%%%%%%%%%%%%%%%%%%%%%%%%%%%%%%%%%%%%%%%

\begin{abstract}

  We determine the mass of the black hole at the center of the spiral
  galaxy NGC 4258 by constructing axisymmetric dynamical models of the
  galaxy. These models are constrained by high spatial resolution
  imaging and long-slit spectroscopy of the nuclear region obtained
  with the {\em Hubble Space Telescope}, complemented by ground-based
  observations extending to larger radii. Our best mass estimate is
  $\MBH = (3.3 \pm 0.2) \times 10^7 \MSun $ for a distance of 7.28 Mpc
  (statistical errors only).  This is within 15\% of $ (3.82\pm 0.01)
  \times 10^7 \MSun$, the mass determined from the kinematics of water
  masers (rescaled to the same distance) assuming they are in
  Keplerian rotation in a warped disk.  The construction of accurate
  dynamical models of NGC 4258 is somewhat compromised by an
  unresolved active nucleus and color gradients, the latter caused by
  variations in the stellar population and/or obscuring
  dust. Depending on how these effects are treated, as well as on
  assumptions about the ellipticity and inclination of the galaxy, we
  obtain black hole masses ranging from $2.4 \times 10^7 \MSun$ to
  $3.6\times10^7 \MSun$. This spread is mainly due to uncertainties in
  the stellar mass profile inside the central 2\arcsec\ ($\sim$70 pc).
  Obscuration of high-velocity stars by circumnuclear dust (possibly
  associated with the masing disk) could lead to an underestimate of
  the black hole mass which is hard to correct. These problems are not
  present in the $\sim 30$ other black hole mass determinations from
  stellar dynamics that have been published by us and other groups;
  thus, the relatively close agreement between the stellar dynamical
  mass and the maser mass in NGC 4258 enhances our confidence in the
  black hole masses determined in other galaxies from stellar dynamics
  using similar methods and data of comparable quality.

\end{abstract}


%%%%%%%%%%%%%%%%%%%%%%%%%%%%%%%%%%%%%%%%%%%%%%%%%%%%%%%%%%%%%%%%%%%%%%%%%%%
%%%%%%%%%%%%%%%%%%%%%%%%%%%%%%%%%%%%%%%%%%%%%%%%%%%%%%%%%%%%%%%%%%%%%%%%%%%

\section{INTRODUCTION}

The characterization of massive dark objects (hereafter ``black
holes'') at the centers of nearby galaxies has advanced considerably
in the past decade.  Nonetheless the dynamical modeling of both
stellar and gas kinematics remains subject to uncertainties. These can
arise from, among other things, systematic uncertainties and noise
amplification in the deprojection of the observables from two to three
dimensions, the presence of dust that can affect both the
mass-to-light ratio ($M/L$) and the observed kinematics, variations in
$M/L$ due to stellar population gradients, and the presence of
nonstellar point and/or continuum sources in the central galactic
regions which may swamp the light from the surrounding stars or
gas. Gas kinematics can also be affected by nongravitational forces,
random velocities and pressure support, which are difficult to model
accurately, while the complex phase-space structure of stellar orbits
can make it difficult to properly sample initial conditions and the
phase-space density distribution of the orbits in (stellar) galactic
equilibria. Reverberation mapping presents a third possibility for
determining black hole masses but comes with its own set of
limitations \citep[e.g.,][]{BlaMcK82,Pet93,NetPet97,GebEtal00b}.

It is therefore desirable to test the reliability of these
methods. Unfortunately, comparing stellar dynamical estimates to 
gas dynamical estimates of black hole masses is not usually
possible, as few galaxies have properties that permit more than one
method to be applied with confidence.  Even in those cases
where it is possible, a detailed comparison is not always
illuminating due to uncertainties in the measurements.  One case where
gas dynamical and stellar dynamical mass measurements are available
for the same galaxy is not reassuring: the gas dynamical mass for the
black hole in IC 1459 is a factor of $\sim 5$ smaller than the
stellar dynamical mass \citep{VerEtal00,CapEtal02}. Better
agreement, within a factor of $\sim 2$, was found between the mass
estimate of the black hole in Centaurus A by \cite{SilEtal05} from
stellar dynamics, and by \cite{MarEtal06} from gas dynamics. Even in
the case of the supermassive black hole in the nucleus of our own 
Milky Way Galaxy, the mass estimates from individual
stellar orbits \citep{GheEtal05} and from a 
statistical analysis of radial velocities and proper motions
\citep{ChaSah01} differ by about a factor of two.  

NGC 4258 presents a unique opportunity to test the reliability of
black hole mass estimators. \citet{MiyEtal95} and \citet{HerEtal99}
used the Very Long Baseline Array (VLBA) to observe water maser
emission in a thin, warped, and nearly edge-on gas annulus located at
0.16--0.28 pc from the center of this galaxy, and determined that a
central black hole with a mass $\MBH = (3.9\pm0.1) \times 10^7 \MSun$
is needed to account for the observed velocity profile, proper
motions, and accelerations. This black hole mass is thought to be
quite reliable for several reasons. (i) The maser emission lines are
both strong and narrow, lending themselves to VLBA observations of
much higher resolution than possible with the {\slshape Hubble Space
  Telescope (HST)} or ground-based telescopes. (ii) The masers seem to
lie in a disk that is very thin (height/radius $\approx 0.02$), so
the corrections to the rotation curve due to pressure or random
velocities are negligible. (iii) Most importantly, the maser velocities
scale very nearly as $v(r) \propto r^{-1/2}$ with distance $r$ from
the center, so that they exhibit an almost perfect Keplerian motion, 
which makes the deduction of the black hole mass easy and almost free
of assumptions. (iv) Finally, the measurements of the mass and
distance from the radial velocities, proper motions, and accelerations
of the masers are all consistent. More recently, \citet{HerEtal05},
using a maximum likelihood analysis of the maser positions and
velocities, quantified the deviation from Keplerian motion.
A number of different models that can explain
this discrepancy yield central masses in the range (3.59--3.88)
$\times 10^7 \MSun$ rescaled to a distance of 7.28 Mpc. The 
warped disk model, which the
authors consider the most probable, implies $\MBH = (3.82 \pm 0.01)
\times 10^7 \MSun$ at this distance.  

In this paper, we estimate the mass of the central black hole of NGC
4258 via modeling of the stellar dynamics of the galaxy. This is
accomplished using the orbit superposition tools developed by our
collaboration (the ``Nuker'' team)
\citep{GebEtal00a,BowEtal01,GebEtal03,ThoEtal04}. Alternative orbit
superposition tools have been developed by \cite{MarEtal98},
\cite{CreEtal99}, and \cite{ValEtal04}. Because the maser black hole
mass estimate seemed very accurate, we intended this to be a thorough,
end-to-end test of the accuracy of our method, at least for this
galaxy. Unfortunately, we encountered a number of problems that
hindered our ability to determine unambiguously the mass density
profile in the central region of the galaxy. In particular, (i) it was
not possible to accurately subtract the light of the unresolved active
galactic nucleus (AGN) at the center of the galaxy to recover the
underlying stellar luminosity profile; (ii) it was not possible to
reliably determine the stellar $M/L$ profile within $\sim$2\arcsec\ of
the nucleus due to a color gradient, presumably caused by diffuse dust
and/or a spatial variation in the stellar population; and (iii) it was
not possible to estimate the extinction caused by a compact dust patch
near the nucleus, which may be obscuring light from fast-moving stars
very close to the black hole, thus causing a systematic underestimate
of the black hole mass.

Despite these difficulties, our ``most trusted'' estimate, $\MBH =
(3.3 \pm 0.2) \times 10^7 \MSun$, is in reasonable agreement with the
maser mass estimate. The quoted error margin corresponds to the
``1$\sigma$'' confidence band, and reflects statistical errors
only. Uncertainties in the mass profile discussed in the previous
paragraph widen the error margin to $2.4 \times 10^7 \MSun < \MBH <
3.6 \times 10^7 \MSun$. This range still does not include the
potential effect of obscuring circumnuclear dust, which cannot be
easily estimated, or any other systematic effects such as deviations
from axisymmetry that cannot be properly treated by our dynamical
modeling method. These factors are discussed in detail in the
following sections.

\citet{past07} estimate the mass of the black hole in NGC 4258 using
gas kinematics, but their 95\% confidence intervals span a factor of
ten in mass so this is not a strong test of the reliability of
black-hole mass estimates.

For this work, we use 7.28 Mpc for the distance to NGC 4258
\citep{TonEtal01}, which is in excellent agreement with the
maser-derived distance of 7.2 Mpc. At that distance, $1\arcsec \simeq
35$ pc.

In \S\ref{obs} we present the photometric and kinematic datasets,
which consist of {\em HST} WFPC2 and NICMOS imaging and STIS
spectroscopy, complemented by ground-based imaging and long-slit
spectroscopy that extend data coverage to larger radii. We discuss the
large-scale morphology of the galaxy in \S\ref{ObsMorph}, mainly as it
pertains to the determination of the contribution of the stars to the
gravitational potential. In \S\ref{VarML} we consider the consequences
of the color gradient in the central 2\arcsec. \S\ref{DynMod} presents
the dynamical models.  We conclude in \S\ref{discussion}.

%%%%%%%%%%%%%%%%%%%%%%%%%%%%%%%%%%%%%%%%%%%%%%%%%%%%%%%%%%%%%%%%%%%%%%%%%%%
%%%%%%%%%%%%%%%%%%%%%%%%%%%%%%%%%%%%%%%%%%%%%%%%%%%%%%%%%%%%%%%%%%%%%%%%%%%

\section{OBSERVATIONS AND REDUCTION} \label{obs}

\subsection{Photometry} \label{phot}

Here we describe the acquisition and reduction of the
photometric data, which are needed mainly to determine the stellar
mass distribution and the kinematic tracer profile, in the way
described in \S\ref{DynMethod} and \S\ref{AmbigP}.

A journal of the imaging observations, from which the surface
photometry was derived, is shown in Table \ref{PhotoDataTable}. A wide
variety of data are available in different colors, both from {\slshape
  HST} and from the ground. For reasons explained in
\S\ref{VarMLga01}, we produced three luminosity profiles. (i) The
near-infrared (NIR) profile was created by combining ground-based and
{\slshape HST} $J$, $H$, and $K$-band images, as described in
\S\ref{RedNIRData}. (ii) The $V$-band profile was generated from
images in the $V$ band ({\slshape HST}) and in the $R$ band
(ground-based), as described in \S\ref{RedVIData}. (iii) The $I$-band
profile was created from the {\slshape HST} $I$-band image, as
described in \S\ref{RedVIData}. Although this image was saturated in
the nucleus, it was useful to consult for the luminosity weighting
(normalization) of the line-of-sight velocity distributions (LOSVDs),
as it identifies the kinematic tracer population (\S\ref{AmbigP}).

\subsubsection{Reduction of the $J$, $H$, and $K$-Band
  Data} \label{RedNIRData} The 2MASS $J$, $H$, and $K$-band images
(Figure \ref{GroundImages-NIR}, top row) were kindly provided to us by
Tom Jarrett \citep{JarEtal03}, and extend out to $R \approx 8\farcm6$
from the center\footnote{Throughout the paper, $R$ refers to distance
from the center of the galaxy, while $r$ refers to distance along the
isophotal semimajor axis.}. For the morphological study of the
large-scale properties of the galaxy (\S\ref{ObsMorph}), the
signal-to-noise ratio (S/N) was boosted by constructing a composite
$J$+$H$+$K$ image, which was then smoothed with a Gaussian
point-spread function (PSF) having full width at half maximum (FWHM)
of 7\arcsec\ for $r \leq 60\arcsec$ and FWHM = 11\arcsec\ for $r >
60\arcsec$. The 2MASS $K$-band image was reduced in a manner similar
to that of the $J$+$H$+$K$ image. It proved to have sufficiently high
S/N to give a reasonably adequate measurement of the outer disk. A
comparison of the $K$-band and $J$+$H$+$K$-band photometry provides an
estimate of the uncertainties; these are dominated by the effects of
patchy star formation and other irregularities, not by photon
statistics. Therefore, we decided to discard the $J$+$H$+$K$-band
profile as a tracer of stellar mass in the dynamical modeling, and to
use the $K$-band profile instead, as being less affected by these
irregularities.

The KPNO 2.1-m telescope $K$-band image extends out to $R \approx
3\arcmin$ from the center (Figure \ref{GroundImages-NIR}, bottom
row). Comparison of the 2MASS $K$-band profile with the KPNO $K$-band
profile, which has a higher spatial resolution, showed that the 2MASS
profile is affected by smoothing only at $r < 32\arcsec$, so the 2MASS
profile was truncated inside this radius. Thus, the KPNO profile
was used for $2\arcsec\ < r < 32\arcsec$ and the 2MASS profile
at $r \ge 32\arcsec$.

\citet{ChaEtal00} obtained {\slshape HST} NICMOS images of NGC 4258 in
bandpasses that approximate $J$, $H$, and $K$. Eric Becklin and Ranga
Chary kindly made the reduced images available to us (Figure
\ref{HSTImages-NIR}), along with the corresponding PSF images. Details
of the reductions can be found in \citet{ChaEtal00}. The {\slshape
  HST} WFPC2 $V$-band image (\S\ref{RedVIData}) shows extensive dust
at $r \gtrsim 2\arcsec$ on the SW side of the center and a few thinner
dust patches on the NE side. There is also a scatter of faint sources
in the bulge, apparently associated with some ongoing star
formation. In the NICMOS $K$-band image, these faint sources are not
visible, and the dust patches on the NE side are essentially invisible
as well. However, the dust on the SW side, although less conspicuous
in $K$ than in $V$, remains a substantial problem that we had to
correct. After considering a number of alternatives, we chose the
strategy of creating a symmetrized image of the galaxy, whereby the SW
side was replaced with the NE side flipped (not rotated) across the
major axis.  As a check, the profiles from
both the symmetrized and the unsymmetrized $K$-band image were
compared and found to agree extremely well at $r \la
2\arcsec$. Therefore, this symmetrization is not a factor in the $V-K$
color gradient discussed in \S\ref{VarMLga01}. We applied the same
procedure to the $J$- and $H$-band images, as well.

The nucleus of NGC 4258 emits broad permitted lines \citep{HoEtal97},
and is classified as a Type 1.9 Seyfert. The AGN at the center is very
red \citep{ChaEtal00} and, in the $K$ band, it appears as a bright,
unresolved point source that clobbers the central brightness
distribution of the stars and needs to be subtracted. Unfortunately,
the PSF provided by \citet{ChaEtal00} was determined using the TinyTim
routine and not from actual images taken at the same time as the
galaxy observations. It proves to be a mediocre match to the AGN at $r
< 0\farcs2$, in the sense that there is a residual diffraction pattern
no matter how the PSF is scaled. We applied the following procedure to
improve the quality of the PSF subtraction (cf.\ Figure
\ref{PSF}). Using {\tt VISTA} \citep{Lau83}, we scaled the TinyTim PSF
image by various factors and measured the resulting surface brightness
profiles in the residual image using the unsymmetrized galaxy
image. We also examined the residual images to see which ones had the
small-scale structure, illustrated in the top left image of Figure
\ref{PSF}, optimally subtracted. Most of the weight in the choice of
the final PSF scale factor came from how well the first diffraction
ring of the AGN was subtracted. The scale factor that we finally
adopted in this way yields ellipticity and position angle profiles
that are both smooth and well behaved, which is reassuring. This is
not the case with scale factors that differ by more than $\pm10\%$
from the adopted value. However, ultimately no value of the scale
factor allows both a convincing subtraction of the first diffraction
ring and a smooth profile at $r < 0\farcs2$ that follows the power law
seen at larger radii. This is probably a result of the imperfect match
of the TinyTim-generated PSF to the $K$-band data. It is unlikely that
a real galaxy feature is involved, considering that no such feature is
detectable in the higher-resolution $V$-band image in the radial range
$0\farcs05<r<0\farcs2$ (cf.\ \S\ref{RedVIData}).

We also applied forty iterations of Lucy deconvolution
\citep{Luc74,Ric72} to the {\slshape HST} NICMOS $J$, $H$, and
$K$-band images using the PSFs from \citet{ChaEtal00}. Because the
PSFs are not ideal, we applied the deconvolution to images with the
PSF of the AGN subtracted as explained above. Still, the deconvolved
$K$-band image showed bad ``ringing,'' so we truncated it inside $r =
0\farcs46$. The $J$-band profile appeared better behaved, at least
partly since the AGN is much dimmer at this bluer wavelength,
although there was also some sign of a residual AGN at $r <
0\farcs1$. The $J$, $H$, and $K$-band deconvolved profiles agree very
well at $r \ga 0\farcs5$, where the AGN is not a problem, and the
ellipticity profiles are all well behaved.

Based on the preceding analysis, we decided (i) to use the original
(not deconvolved) $K$-band photometry for $r \geq 2\farcs8$ (KPNO data
at $2\farcs8 \le r < 32\arcsec$, and 2MASS data at $32\arcsec \le r
\le 560\arcsec$), (ii) to average the deconvolved NICMOS $K$-band and
$J$-band photometry for $0\farcs2 \leq r < 2\farcs8$, and (iii) to use
deconvolved $J$-band data for $0\farcs038 \leq r < 0\farcs2$ in order
to avoid the AGN as much as possible. The resulting profile is shown
in Figure \ref{CumPhotom}. However, even in $J$, the points at $r <
0\farcs11$ are probably still affected by the AGN. We 
elaborate on this point in \S\ref{VarMLla01}.

\subsubsection{Reduction of the $V$ and $I$-Band
  Data} \label{RedVIData} We used the MDM 1.3-m telescope to obtain an
image of the galaxy in the $R$ band (Figure \ref{GroundImages-R}). The
image was reduced and dust-clipped in the usual way. The luminosity
profile was then extracted and matched to the $V$-band {\slshape HST}
profile, discussed below, in the range $5\arcsec \la r \la 7\arcsec$,
using a $V$-band zeropoint. There were no significant color,
ellipticity, or position angle (PA) gradients between the $V$-band and
$R$-band profiles over the radial range in which they were
matched. Henceforth, we will refer to this composite profile, made by
stitching together the {\slshape HST} $V$-band profile with the
$R$-band profile shifted by a constant offset, as the $V$-band
profile. It extends out to $r \approx 150\arcsec$.

We obtained {\slshape HST} WFPC2 images of the NGC 4258 nucleus
through the F547M filter under program GO--8591. The F547M filter was
selected to exclude any bright emission lines associated with the 
AGN; it is roughly equivalent to a narrow $V$-band filter, and is
referred to as such in the remainder of this paper. The galaxy was
centered in the PC1 chip and dithered in a $2\times2$ square raster of
$\sim$0.5 pixel steps over four separate 400~s exposures to
maximize the spatial resolution of the complete data set. The four
dithered images were combined into a single Nyquist-sampled ``super
image" using the algorithm of \cite{Lau99a}, which provides for the
optimal combination of undersampled images without any associated
degradation of the resolution. This final image has a pixel scale of
$0\farcs0228,$ twice as fine as that of the original images. However, 
the Nyquist image remains modulated by the {\slshape HST} PSF;
thus, it was deconvolved prior to further analysis using 80 iterations
of the \cite{Luc74} and \cite{Ric72} algorithm. The PSF was estimated
using the TinyTim package, but incorporates the WFPC2 pixel-response
function recovered by \cite{Lau99b}. The final deconvolved $V$-band
image of NGC 4258 is shown in Figure \ref{HSTImages-V}.

The $V$-band surface brightness profile of NGC 4258, shown in Figure
\ref{CumPhotom}, was estimated from the F547M Nyquist image using a
combination of the high spatial resolution algorithm of \cite{Lau85}
and the least-squares estimator of \cite{Lau86}. The patchy dust seen
throughout the bulge of NGC 4258 complicates measurement of the
brightness profile; as is visible in Figure \ref{HSTImages-V}, there
is a compact dust feature just slightly offset from the nuclear source
that was especially problematic. For all dust features other than the
central patch, pixels affected by dust could simply be excluded from
the least-squares profile estimator. In practice, this was done
iteratively by comparing the image to a model reconstructed from the
profile to isolate faint dust absorption that may have been missed
during the initial inspection of the image. Unfortunately, pixels
could not be masked out of the high-resolution profile algorithm,
which is used in preference to the least-squares estimator for
$r<0\farcs5,$ so the nuclear dust patch was first filled in with an
estimated correction provided by the lower-resolution least-squares
algorithm. This initial correction was then refined by replacing it
with an improved estimate provided by a model reconstructed by the
high-resolution profile. 

Figure \ref{HSTImages-V} shows the central portion of the $V$-band
image divided by a model reconstructed from the final brightness
profile. Apart from the nuclear dust patch, the residuals are flat,
showing that the brightness profile extracted is faithful to the
central structure of NGC 4258. The nuclear dust patch, itself, is
clearly compact, affecting only a single quadrant of the nucleus,
which again allowed it to be isolated from the profile estimation. As
a ``sanity check'' on the brightness profile, we compared the profile
to an intensity trace taken along the bulge semimajor axis on
the side of the nucleus opposite the nuclear dust patch. The two
measures agreed extremely well, showing that the final profile is not
strongly affected by any residual dust features or the pixels excluded
from the analysis.

In addition to the {\slshape HST} F547M data, we also analyzed two
archival PC1 F791W ($I$ band) images obtained by \cite{CecEtal00}
under program GO--6563. Unfortunately, both images were saturated at
$r<0\farcs09$ from the nucleus, and thus provide information only at
somewhat larger radii. The images were combined and deconvolved;
however, without a full accounting of the light contributed by the
nucleus, the $I$-band profile should be regarded with caution for
$r<0\farcs5$.

Using the F547M and the F791W images, we created a F547M/F791W color
map of the nuclear region of the galaxy, approximating $V-I$ (Figure
\ref{ColorMap}). The color map shows more clearly the distribution of
dust in the center, and will be further discussed in
\S\ref{VarMLga01}. A larger-scale view of the central region of the
galaxy is shown in Figure \ref{HSTImages-color}, which is a composite
color image created by combining the F547M and the F791W images along
with an archival F300W ($U$-band) image, also obtained by
\cite{CecEtal00}.

%%%%%%%%%%%%%%%%%%%%%%%%%%%%%%%%%%%%%%%%%%%%%%%%%%%%%%%%%%%%%%%%%%%%%%%%%%%

\subsection{Spectroscopy} \label{Spectr}

We used the \ion{Ca}{2} triplet absorption line (8498~\AA, 8542~\AA,
8662~\AA; hereafter CaT) to determine the stellar LOSVDs at the center
of NGC 4258, and at various positions along the major axis (out to $R
= 18\farcs18$) as well as along the minor axis (out to $R =
11\farcs69$). The spectra were obtained from the ground as well as
with the {\slshape HST} Space Telescope Imaging Spectrograph
(STIS). Spectrograph configurations are shown in Table
\ref{SpectrConfigsTable}.

In particular, we made 24 cosmic-ray-split exposures of NGC 4258 with
STIS to obtain high-quality nucleus-centered CaT spectra along the
major axis of the galaxy, out to $R=0\farcs80$. The STIS slit was
placed at PA = 140\arcdeg\ instead of directly on the major axis (PA =
150\arcdeg) because of guide-star constraints. Our reduction procedure
closely followed that of \citet{PinEtal03}. The procedure bypasses the
pipeline (``Calstis'') reduction so that we can employ ``self-darks,''
i.e., dark frames that are constructed by combining our dithered
datasets with the galaxy light rejected. Our method works best if all
of the exposures are taken within a short period of time because of
the rapidly changing dark pattern on the STIS CCD. This was not quite
the case here, since the exposures were taken on March 12 and 21,
2001. Nevertheless, a single self-dark worked well.

There is one small difference between our STIS reduction procedure and
that described by \citet{PinEtal03}: the initial dark subtraction was
done with an archival, ``weekly'' dark rather than by a sum of the
galaxy frames with a region around the galaxy masked out. This created
a smoother dark on which to begin further iterations using the same
``self-dark'' strategy.

We also used the Modspec spectrograph on the MDM 2.4-m telescope to
obtain ground-based CaT spectra along both the major and minor
axes at larger radii. The configuration is again shown in Table
\ref{SpectrConfigsTable}. The seeing varied between 1\arcsec\ and
1\farcs5 (FWHM). Kinematic parameters were extracted out to $R =
18\farcs18$ and $R = 11\farcs69$ along the major and minor axis,
respectively. Again, the reduction procedure followed closely that
described by \citet{PinEtal03}.

Figures \ref{SpectraHST} and \ref{SpectraGround} show spectra
extracted at several radial positions from the STIS and the
ground-based data, respectively. The emission-line feature visible
near 8620~\AA\ (rest frame) in the nuclear STIS spectrum in Figure
\ref{SpectraHST} is most probably the \ion{Fe}{2} 8618~\AA\ line, and
had to be removed before the LOSVD fitting could be done. Consequently,
there was a concern that the central (or central few) LOSVDs would be
less reliable, but it proved easy to interpolate under the line (see
Figure \ref{SpectraHST}), so in the end this was not a problem.

We decided to discard the central ground-based LOSVD along the major
axis, because of concerns about template mismatch and AGN
contamination.  In retrospect, the central ground-based LOSVD on both
the major and minor axes should have been dropped since the STIS
data had comparable statistical quality and much better understood
PSFs.  In fact, we used the minor axis central LOSVD, except in one
experiment, where its exclusion had no effect (see \S\ref{AmbigK}).

In a dust-free axisymmetric galaxy that also exhibits symmetry about
the equatorial plane, such as assumed by our modeling method,
kinematic quantities also manifest symmetries about the center. This
entitles us to ``symmetrize,'' with respect to the center of the
galaxy, the LOSVDs along the major and minor axes, for both STIS and
Modspec. This process helps alleviate potential biases in the
extraction of the velocity profile. \citet{GebEtal03} provide more
details on how this was done. Unfortunately, NGC 4258 contains
substantial quantities of dust, especially near the dynamically
important central region (Figures \ref{HSTImages-V}, \ref{ColorMap}),
which calls into question the validity of this assumption. Potential
implications are discussed in \S\ref{discussion}.

We used the maximum penalized likelihood method (MPL), as described by
\citet{PinEtal03}, to derive LOSVDs from all the CaT spectra. A
nonparametric form of the LOSVDs, rather than Gauss-Hermite moments,
was used for the dynamical modeling. Each LOSVD was represented by 13
equally spaced velocity bins, with the uncertainty in each bin
determined from Monte Carlo simulations.

Figure \ref{VSigmaH3H4} illustrates  $V, \sigma, H3$, and $H4$, the
fitting parameters from the truncated Gauss-Hermite series expansions
of the LOSVDs \citep{Ger93,MarFra93}, along the major and minor axes
of the galaxy, both from STIS and from the ground. The same parameters
in tabular form can be found in Table \ref{SpectrTable}. We stress
again that a nonparametric form of the LOSVDs was actually used for
the dynamical modeling.  We anticipate our best fitting model
discussed in \S 5.1 by illustrating its projected Gauss-Hermite
moments here.   

%%%%%%%%%%%%%%%%%%%%%%%%%%%%%%%%%%%%%%%%%%%%%%%%%%%%%%%%%%%%%%%%%%%%%%%%%%%
%%%%%%%%%%%%%%%%%%%%%%%%%%%%%%%%%%%%%%%%%%%%%%%%%%%%%%%%%%%%%%%%%%%%%%%%%%%

\section{THE LARGE-SCALE MORPHOLOGY} \label{ObsMorph}

NGC 4258 is a typical oval-disk galaxy \citep{Kor82}, classified as
SAB(s)bc. Morphologically, the disk consists of two nested regions of
slowly varying surface brightness: the ``inner oval'' component (PA
$\simeq$ 156\arcdeg) is much brighter than the ``outer oval''
component (PA $\simeq$ 150\arcdeg), and both have relatively sharp
outer edges, at $R \simeq 200\arcsec$ and $R \simeq 560\arcsec$,
respectively (Figure \ref{GroundImages-NIR}). Embedded in the inner
oval component is a bulge (PA $\simeq$ 146\arcdeg) which remains
important out to $r \simeq 40\arcsec$, as well as a weak bar (PA
$\simeq$ 11\arcdeg) extending out to $R \simeq 150\arcsec$. These
components can also be identified using the profiles in Figure
\ref{CumPhotom}.

We are interested in the large-scale morphological properties of NGC
4258 primarily to establish that there are no major deviations from
the assumption of axial symmetry, which is built into our modeling
method (\S\ref{DynMethod}). Deviations from axisymmetry can be
important for two reasons. First, orbits in a non-axisymmetric
potential do not conserve any component of the angular
momentum. Therefore, the orbital structure of a non-axisymmetric
system can be considerably different from that of an axisymmetric
one. Second, non-axisymmetric deviations can induce kinematic
signatures in the radial velocity profile similar to those created by
a central mass concentration. Either effect can bias a black hole mass
estimate made under the assumption of axisymmetry.

The ellipticity profiles in $V$ and in the NIR are
quite different in the innermost 0\farcs2, but they agree rather well
further out. We consider the ellipticity profile in $V$ to be overall
more reliable than that in the NIR due to the higher resolution in
$V$. It becomes progressively flatter from the center out to $r \simeq
5\arcsec$, and then it remains approximately constant between $r
\simeq 5\arcsec$ and 12\arcsec. The ellipticity starts dropping again
out to $r \simeq 40\arcsec$, but this is probably caused by the weak
bar, which adds extra light in the minor-axis direction while probably
having no effect on the major-axis profile. The onset of the dominance
of the disk is signified by the increasing ellipticity outward of $r
\simeq 40\arcsec$.

If we wanted to rigorously test these morphological characterizations,
we would have to perform a full bulge-bar-disk photometric
decomposition. However, such a decomposition would be very uncertain
because of parameter coupling: the bar is very weak, and trading light
between the components while correspondingly changing their isophote
shapes would provide considerable freedom to tinker with the bulge
ellipticity.

In the absence of extinction, the projected isophotes of a galaxy that
has emissivity constant on coaxial and similar spheroids
(as assumed in our dynamical modeling) are concentric, coaxial, and
similar ellipses \citep{Con56,Fis61}. The absence of any strong
isophote twists out to $r \simeq 40\arcsec$, the beginning of the weak
bar, indicates that, to a good approximation, NGC 4258 is axisymmetric
out to that radius. Furthermore, the fact that the PA of the bulge and
of the outer disk agree quite well is evidence against a triaxial
bulge. However, there \emph{does} exist a strong isophote twist
between 40\arcsec\ and 150\arcsec, caused by the bar, while Figure
\ref{VSigmaH3H4} shows some minor-axis rotation in the bulge, outside
$R \simeq 10\arcsec$, which could be indicative of triaxiality at
intermediate radii.

Even though our current algorithm cannot model triaxial components,
the dynamical effect of the bar on the measured value of the black
hole mass is probably small, for at least three reasons:  (i)
$M_\bullet$ is mostly determined from the dynamics near the center,
where triaxial deviations are least significant;  (ii) bars generally
have relatively little effect on the kinematics --- for example, even
the much stronger bar of NGC 936 has little effect on the
velocity dispersion of that galaxy \citep{Kor83,Kor84}; 
(iii) and finally 
the velocities of bar stars that cross the slit, while probably
somewhat bigger than the local circular velocity (a $\sim$5--10\%
effect), are seen moving nearly perpendicular to the line-of-sight 
because the bar is strongly inclined
to the major axis, and hence they probably do not signicantly affect
the observed kinematics.

In summary, NGC 4258 contains non-axisymmetric structures which
probably {\it do} affect the kinematics, and hence the black hole mass
estimate, but probably only in a minor way.

%%%%%%%%%%%%%%%%%%%%%%%%%%%%%%%%%%%%%%%%%%%%%%%%%%%%%%%%%%%%%%%%%%%%%%%%%%%
%%%%%%%%%%%%%%%%%%%%%%%%%%%%%%%%%%%%%%%%%%%%%%%%%%%%%%%%%%%%%%%%%%%%%%%%%%%

\section{VARIATIONS IN ${M/L}$ RATIO} \label{VarML}

Stellar population gradients, dust extinction, and contamination by an
AGN can all induce variations in the $M/L$ ratio, which have to be
taken into account when deriving the stellar mass density profile from
the luminosity density profile. We first discuss variations in $M/L$
far from the AGN-dominated nuclear region, and then we describe the
complications due to the AGN.

%%%%%%%%%%%%%%%%%%%%%%%%%%%%%%%%%%%%%%%%%%%%%%%%%%%%%%%%%%%%%%%%%%%%%%%%%%%

\subsection{The Stellar ${M/L}$ Ratio Far from the AGN Point Source} \label{VarMLga01}
Color gradients in the outer disk of NGC 4258 are small, even between
the $R$ and $K$ bands (Figure \ref{CumPhotom}), implying that the
$M/L$ ratio in the disk is not seriously affected by star formation or
dust. However, inside $\sim$2\arcsec\ from the nucleus, the $V$-band
profile ``peels off'' from the NIR profile and becomes flatter. This
color gradient can be seen in Figure \ref{CumPhotom}, but becomes more
apparent in Figure \ref{CumPhotomCenter} which shows the inner
10\arcsec\ of the surface brightness profile. Also, the color map
shown in Figure \ref{ColorMap} clearly identifies the color gradient
as a ``red,'' diffuse area surrounding the nucleus out to $r \approx
2$\arcsec\ ($R \approx 1\arcsec - 2\arcsec$). This is an issue for the
dynamical modeling because it takes place in the dynamically important
central region, which affects the entire model: we must determine
which, if either, of the two profiles ($V$-band or NIR) traces stellar
mass.

Could the color gradient be caused by variations in the stellar
population? The strongest evidence {\em against} variations in the
stellar population is the absence of any marked color gradient between
the $J$ and $K$ profiles, which should be present if the central NIR
brightening were caused by either a metallicity gradient or red
supergiant stars. Furthermore, although some traces of scattered star
formation can be seen on the $V$-band image (Figure \ref{HSTImages-V})
as well as on the color map (Figure \ref{ColorMap}), they are absent
in the NIR images (Figure \ref{HSTImages-NIR}), even after accounting
for the different FWHM of the PSFs.

More information can be obtained from the equivalent width (EW)
profiles of the three CaT line components, which we computed from the
STIS spectra extending out to $1\farcs5$ from the center (Figure
\ref{EW}). A decreasing EW profile would be evidence of a young
stellar population or of a brightening $I$-band nonthermal
continuum. A drop is seen in the central $0\farcs3$, and is discussed
in the next subsection. Here we are concerned with possible trends
beyond that radius. For $r > 0\farcs3$, there is a hint of a slow
increase in the EW away from the center, but the data are also
consistent with a flat profile. However, the datapoints at $r >
0\farcs8$ suffer from the low S/N of the originating spectra, and we
have no data as far out as $\sim$2\arcsec, where the $V$-band profile
begins to peels off.

The other possible explanation is  the presence of diffuse dust, which
would produce  a small gradient  in $J-K$ but  a larger one  in $V-J$,
just as observed. The observed  smoothness in the $V-J$ color gradient
would then imply a similarly  smooth increase in the projected density
of diffuse dust toward the center.

In the end, we were unable to determine unequivocally whether the
origin of the color gradient is a variation in the stellar population
or the presence of a diffuse dust component. We decided to construct
models for both of these possibilities, and investigate the
uncertainties induced on the black hole mass (\S\ref{AmbigP}).

%%%%%%%%%%%%%%%%%%%%%%%%%%%%%%%%%%%%%%%%%%%%%%%%%%%%%%%%%%%%%%%%%%%%%%%%%%%

\subsection{The Stellar ${M/L}$ Ratio Near the AGN Point
  Source} \label{VarMLla01}

Dynamical modeling only strongly constrains the {\em total} mass,
$M_{\rm tot}(r_0) = \MBH + M_*(r_0)$, inside a small radius $r_0$
around the center, where $M_*(r_0)$ is the enclosed stellar mass, and
$r_0$ is comparable to the spatial resolution of the nuclear spectrum
($\sim$0\farcs1). Therefore, there is a degeneracy between $\MBH$ and
$M_*(r_0)$, which can be resolved only if we know $M_*(r_0)$
separately, i.e., only if we know the luminosity profile and the
stellar $M/L$ ratio inside $r_0$. The latter is assumed to be constant
throughout the galaxy and is predominantly constrained by the
dynamical model at large radii (\S\ref{DynMethod}). In essence, $\MBH
= M_{\rm tot}(r_0) - M_*(r_0)$ corresponds to the ``excess'' dynamical
mass inside $r_0$ after subtraction of the mass due to
stars. Therefore, we must again determine which, if either, of the two
luminosity profiles ($V$ band or NIR) traces stellar mass, $M_*$.

The $V$-band profile, which is well determined down to $r \simeq
0\farcs011$, manifests a progressively more shallow slope for $r \la
0\farcs15$, interrupted by a peak within 0\farcs04 of the
nucleus. Although the cause of the peak is not entirely clear, it is
unlikely that it is due to light from an old stellar population. The
CaT EW profile (Figure \ref{EW}) shows a $\sim$30\% decline within
0\farcs2 of the nucleus, which strongly suggests contamination by a
continuum source, possibly OB stars or a nonthermal AGN.

Notwithstanding the uncertain origin of the peak, it seems unlikely
that it traces stellar mass. Therefore, we replaced the innermost
three $V$-band datapoints in a way that extrapolates inward the
profile at $0\farcs04 \la r \la 0\farcs15$ (Figure
\ref{CumPhotomCenter}; \S\ref{AmbigP}). However, we also created
models {\em with} the central peak, to investigate its effect on
$\MBH$. Not surprisingly, the effect was negligible because the light
in the peak corresponds to a very small mass (\S\ref{AmbigP}).

In \S\ref{RedNIRData} we saw that, at $r \la 0\farcs11$, the $J$-band
profile begins to show signs of contamination by AGN light, and the
PSF of the AGN could not be subtracted reliably. We also saw earlier
that the central depression in the CaT EW profile begins at $r \simeq
0\farcs2$, and implies a brightening of the CaT continuum, which lies
in the $I$ band, inside that radius. The AGN is red, so it seems
likely that the $J$-band profile is even more affected by the AGN
continuum radiation than the $I$-band profile.  While luminous AGNs normally
have a blue featureless continuum, low-luminosity AGNs such as NGC
4258 tend to have emission spectra that peak in the mid-infrared 
\citep{Ho99}, probably as a consequence of their low accretion rates
\citep{Ho08}.   Taking all these
considerations into account, we decided that the $J$-band profile
inward of $r \simeq 0\farcs2$ cannot be trusted to trace mass (i.e.,
the old stellar population). Accordingly, inside $r \simeq 0\farcs2$
we replaced it with a profile that follows the $V$-band profile {\it
  without} the central peak. As explained above, this profile is
presumably not affected by the AGN. The related dynamical models are
discussed in \S\ref{AmbigP}.

%%%%%%%%%%%%%%%%%%%%%%%%%%%%%%%%%%%%%%%%%%%%%%%%%%%%%%%%%%%%%%%%%%%%%%%%%%%
%%%%%%%%%%%%%%%%%%%%%%%%%%%%%%%%%%%%%%%%%%%%%%%%%%%%%%%%%%%%%%%%%%%%%%%%%%%

\section{DYNAMICAL MODELING} \label{DynMod}

\subsection{The Method} \label{DynMethod}

A detailed account of our dynamical modeling method is provided in the
Appendix. Here we present a short description with emphasis on NGC
4258.

The first step involves a deprojection of the surface photometry to
obtain the three-dimensional luminosity density. This is done under
the assumption that emissivity is constant on similar, coaxial
spheroids.  The deprojection is
performed via a nonparametric Abel inversion \citep[cf.][]{GebEtal96}
assuming a value for the inclination angle ($i = 72 \arcdeg$ in the
base model).
 The gravitational
potential can then be recovered by specifying a central point mass
(``black hole'') $\MBH$ and the stellar $M/L$ ratio, $\ML$, assumed
constant throughout the galaxy. The dynamical models for NGC 4258
extend out to $r=100\arcsec$.

The dynamical models are constructed for a grid of $(\MBH,\ML)$
values. Each model consists of a superposition of representative
orbits, appropriately weighted to match the observed kinematics,
subject to constraints that force the stellar density distribution to
match the luminous density distribution of the galaxy.  
Each model is computed using $N_o \approx 7000$
orbits. Orbits are integrated for about 100 crossing times.  

Self-consistency is enforced by requiring that the model satisfy
exactly the photometric constraints on a polar grid of 15 radial
shells as described in the appendix, each subdivided into 4 bins in
polar angle. The grid used to solve Poisson's equation is 4X finer
than the grid used to bin the kinematic results.  For most orbits
energy is conserved to much better than 1\%.

For each model, the orbit weights that best fit the observed
kinematics are determined by minimizing
\begin{equation}
  \chi^2 = \sum_k
  \left[ l_{k,o} - 
   \sum m_{ik} w_i  
   \right]/   
     {\sigma_k^2} \,\, ,
\label{chi2} 
\end{equation}
where $l_{k,o}$ is the light at a specific
position and velocity (the index $k$ runs over both position and
velocity) with uncertainty $\sigma_k$, and where $m_{ik}$ is the mass
deposited by the $i^{th}$ orbit weighted by $w_i$ in the $k^{th}$
projected position and velocity  bin; cf.\ \S\ref{Spectr}). We use a 
maximum-entropy technique as described in the Appendix to minimize $\chi^2$
using non-negative orbit occupation numbers.  

For each dataset, we determine $\chi^2(\MBH,\ML)$.  The minimum 
determines the best (most probable) 
combination of $\MBH$ and $\ML$ for the galaxy. The $\Delta\chi^2=1$
contour on the $(\MBH,\ML)$ plane, where $\Delta\chi^2 \equiv
\chi^2-\chi^2_{\min}$, determines the nominal ``1$\sigma$''
uncertainty ranges for $\MBH$ and $\ML$.

The process of converting observational quantities to model input
entails assumptions and systematic uncertainties that have to be taken
into account when determining the uncertainty of the ``best'' $\MBH$,
which may be larger than the nominal ``1$\sigma$'' confidence band on
the $(\MBH,\ML)$ plane. Therefore, the previous steps must be repeated
for a number of plausible parameter choices (other than $\MBH$ and
$\ML$) in order to determine the overall uncertainty. We cannot afford
to make an exhaustive exploration of the parameter space because of
the time required to create the models. Instead, we opt to construct a
base model corresponding to our ``best guess'' for the parameter
values and investigate the effect on $\MBH$ of varying each parameter
independently.

The results from the dynamical modeling are summarized in Table
\ref{ModelTable}, and the corresponding $\chi^2$ contour maps are
shown in Figure \ref{ModelPlots}. The black hole mass for the base
model, which uses the $V$-band profile {\it without} the central peak
for both the mass and the kinematic tracer profiles, is $\MBH =
3.31^{+0.22}_{-0.17} \times 10^7 \MSun$ with a total ({\em not} per
degree of freedom) $\chi^2_{\min} \approx 290$. 

The total number of parameters is 27 (LOSVDs) times 13 (velocity
bins), which is 351. Typically, the reduced $\chi^2$ is 0.3 to 0.5 in
our models \citep{GebEtal03}, which is less than the expected value of
unity because of covariance in the LOSVD components. For NGC 4258, the
reduced $\chi^2$ is near unity, but most of the contribution to
$\chi^2$ comes from the minor axis. The velocity centroid on the minor
axis varies much more than the stated uncertainties (see Figure 11),
whereas an axisymmetric model would have zero velocity along the minor
axis.  Regardless of whether this contribution is due to
underestimated uncertainties or real non-axisymmetric motion, it is the main
cause for the inflated $\chi^2$. Along the major axis, the reduced
$\chi^2$ is about 0.4, which is typical for the orbit-based models.

%%%%%%%%%%%%%%%%%%%%%%%%%%%%%%%%%%%%%%%%%%%%%%%%%%%%%%%%%%%%%%%%%%%%%%%%%%%

\subsection{Ambiguities in the Deprojection Parameters} \label{AmbigD}

\subsubsection{Isophotal Ellipticity $(\epsilon)$} \label{DynModEll}
The (projected) ellipticity profile of NGC 4258 is discussed in
\S\ref{ObsMorph}. The profile varies with radius, but the origin of
the variation (disk, weak bar, genuine variations in the axis ratios
of the bulge isodensity spheroids) is not clear. Therefore, although
it is possible to incorporate an arbitrary ellipticity profile
$\epsilon(r)$ in our modeling algorithm, we preferred instead to make
models with a constant ellipticity profile, and examine the extent to
which $\MBH$ is affected by the adopted value.

For the base model we used $\epsilon=0.35$, corresponding to the mean
ellipticity in the radial range $2\arcsec \la r \la 65\arcsec$. The
effect of varying $\epsilon$ was examined in model D1, for which we
used $\epsilon=0.45$. We obtain $\MBH = 3.48^{+0.42}_{-0.38} \times
10^7 \MSun$, a slightly higher mass but within the same ``1$\sigma$''
confidence as the base model.

Nonetheless, the assumption of constant ellipticity is clearly
problematic because the ellipticity profile varies between $\sim$0.1
and $\sim$0.6.

\subsubsection{Disk Inclination Angle $(i)$} \label{DynModIncl}
We need to know the inclination of the disk to deduce the intrinsic
ellipticity of the isophotes from their projected ellipticity, under
the assumption that the equatorial plane of the bulge of the galaxy is
coplanar with the galactic disk.

The outer disk (beyond $r \approx 200\arcsec$) has average and maximum
ellipticities of $\epsilon=0.62$ and $0.64$, respectively. If this
disk is infinitesimally thin, the corresponding inclination angles are
$i=69\arcdeg$ and $70\arcdeg$, respectively, where edge-on corresponds
to $i=90\arcdeg$. If this disk were thick with an intrinsic axis ratio
of 0.25 \citep{SanFreSto70}, then the corresponding inclinations would
be $i=73\arcdeg$ and $75\arcdeg$, respectively. At large radii, the
optical PA is 150\arcdeg. \Citet{Alb80}, using H I kinematic
observations, derives the same PA and $i=72\arcdeg$, which is
consistent with the photometrically derived inclination given the
uncertainty in the intrinsic thickness of the disk.

We therefore adopted $i=72\arcdeg$ for the base model, but we also
examined the sensitivity of $\MBH$ to variations in $i$. Model D2 is
identical to the base model except $i=62\arcdeg$, a value close to the
lowest inclination angles that we found in the literature. We obtained
a $\sim$10\% higher mass, $\MBH = 3.62^{+0.38}_{-0.49} \times 10^7
\MSun$ which is consistent with a $\MBH \propto (\sin i)^{-1}$
dependence.  This would be expected in a rotationally supported
system.  It appears from our models that the inner part of the galaxy
is rotating quite rapidly.  For this model, $\chi^2_{\min} = 303$,
which is greater than 280, the $\chi^2_{\min}$ for the base
model. This could be interpreted as evidence that the base model is
more likely to be true, at least within the framework of assumptions
of the modeling method.

We did not attempt to obtain a better estimate for the errors due to
deprojection (by running models for more combinations or $i$ and
$\epsilon$) because the uncertainties in the luminosity profiles
(\S\ref{AmbigP}) dominate the errors.

%%%%%%%%%%%%%%%%%%%%%%%%%%%%%%%%%%%%%%%%%%%%%%%%%%%%%%%%%%%%%%%%%%%%%%%%%%%

\subsection{Ambiguities in the Kinematic Data} \label{AmbigK}

In \S\ref{Spectr} we mentioned that the presence of an emission line
near the CaT in the nuclear STIS spectrum may have affected the
reliability of the LOSVD extraction. In fact, the next two STIS
spectra (at $R = 0\farcs05$ and 0\farcs10) show (weaker) signatures of
the same emission line, as well. We also mentioned that we have some
concerns about the quality of the central LOSVD along the minor axis,
in the ground-based data. In this section we remove the suspect LOSVDs
from the $\chi^2$ fit, and examine the effect on $\MBH$.

In models K1 and K2 we isolate the uncertainty introduced to $\MBH$ by
the potentially unreliable LOSVD extraction. Model K1 is identical to
the base model, except the nuclear STIS LOSVD is not included as a
constraint. In model K2, the three innermost STIS LOSVDs are not
included. In model K1, the effect on $\MBH$ is very small. This is
reassuring, since it is the nuclear spectrum that would have been
affected the most. Model K2 yields a significantly lower $\MBH =
2.20^{+0.54}_{-0.31} \times 10^7 \MSun$ but with a much higher
uncertainty, which overlaps with the base model at the
$\sim$1.5$\sigma$ level.

An interesting question is whether {\slshape HST} spectroscopy is
required to find evidence for a black hole. Using ground-based
kinematic data alone should, of course, yield a black hole mass
consistent with the mass obtained by also including {\slshape HST}
data, albeit with larger uncertainties, depending primarily on the
angular size of the radius of influence of the black hole on the
sky. We address this question for NGC 4258 with model K3, which is
identical to the base model, except only ground-based kinematic data
are used. It turns out that the stellar $M/L$ ratio can be constrained
quite well, but the black hole mass, $\MBH = 1.03^{+1.00}_{-0.28}
\times 10^7 \MSun$, is very uncertain. This is not surprising,
considering that the radius of influence for the black hole in the
base model is $G\MBH/\sigma_e^2 \approx 0\farcs32$, where
$\sigma_e=105$ \kms\ is the velocity dispersion in the main body of
the bulge. This is substantially smaller than the spatial resolution
of the ground-based kinematic data, for which FWHM =
1\arcsec--1\farcs5. Therefore, {\slshape HST} kinematic data are
critical for a well-constrained estimate of the black hole mass in
this galaxy. {\slshape HST} photometric data are also critical for
galaxies where the luminosity profile near the center cannot be
reliably extrapolated from the profile farther out, as is also the
case with this galaxy (unfortunately, it is still not possible to
determine unambiguously the inner mass profile, as discussed in
\S\ref{AmbigP}).

We also constructed model K4, which uses all kinematic data
(STIS+Modspec) except the central ground-based LOSVD along the minor
axis (the central ground-based LOSVD along the major axis was not
included in {\em any} model). We find that $\MBH$
is essentially identical to that of the base model.

%%%%%%%%%%%%%%%%%%%%%%%%%%%%%%%%%%%%%%%%%%%%%%%%%%%%%%%%%%%%%%%%%%%%%%%%%%%

\subsection{Ambiguities in the Mass and Kinematic Tracer
  Profiles} \label{AmbigP}

In \S\ref{VarML} we saw that stellar population gradients and/or dust
extinction in the innermost $\sim$2\arcsec\ of the galaxy, as well as
light from the unresolved AGN, can induce variations in the measured
stellar $M/L$ ratio near the center of the galaxy. These have to be
taken into account to determine which, if any, of the available
luminosity profiles traces the mass density profile, and which, if
any, corresponds to the luminosity profile of the kinematic tracer
population. The latter profile is needed for the luminosity weighting
(normalization) of the LOSVDs. Unfortunately, we were unable to make
these determinations convincingly, for the reasons explained in
\S\ref{VarML}.

Consequently, we decided to identify three possible luminosity
profiles, and to make models that use them as either mass or kinematic
tracer profiles. The expectation is that these three profiles
correspond to limiting cases, and that the true profiles lie somewhere
in between. Then, the spread in $\MBH$ that we obtain from these
limiting cases would be a good indication of the $\MBH$ uncertainty
caused by our inability to determine the true profiles.

The profiles, which are depicted in Figure \ref{CumPhotomCenter}, are
the following:

(1) {\slshape The $V$ profile:} This is identical to the $V$-band
    profile as measured.

(2) {\slshape The \Vcor\ profile:} This is identical to the $V$
    profile, with the three innermost points replaced by a shallower
    core, as explained in \S\ref{VarMLla01}.

(3) {\slshape The \NIRcor\ profile:} For $r > 0\farcs2$, this is
    identical to the spliced $J$, $J$+$K$, and $K$ profiles in the top
    panel of Figure \ref{CumPhotom}. Inward of 0\farcs2, that profile
    is replaced by one that parallels (shows no color gradient with
    respect to) the \Vcor\ profile, for the reasons mentioned in
    \S\ref{VarMLla01}.

If the observed color gradient is mostly or entirely due to the
presence of diffuse dust or metallicity, then the \NIRcor\ profile is
a better tracer of mass than the $V$ or the \Vcor\ profiles. However,
the $V$-band profile has higher intrinsic resolution and a dimmer AGN,
and the WFPC2 is better understood photometrically than NICMOS. For
these reasons, we try all three versions as stellar mass profiles.

The kinematic tracer population was observed in the CaT wavelength,
which lies in the $I$ band. It would thus be natural to use the
$I$-band profile as the kinematic tracer profile. However, we saw in
\S\ref{RedVIData} that the $I$-band image is saturated in the central
0\farcs09, and thus cannot be deconvolved. The NIR profile shares the
same slope with the $I$-band profile at larger radii (where $I$ is not
saturated), and could arguably be used instead. Unfortunately, it is
also ill-determined near the center due to the AGN and reduced
resolution (\S\ref{VarMLga01}). Consequently, we decided to again test
both \NIRcor\ and \Vcor\ as the kinematic tracer profile, and examine
the induced uncertainty in $\MBH$. For the base model we adopted
\Vcor\ because of its higher resolution and smaller AGN
contamination. This is somewhat arbitrary, but the choice of the
kinematic tracer profile turns out to have a very small effect on
$\MBH$, as we will see below.

These scenarios were tested in models P1 through P4. Models P1--P3
substitute \NIRcor\ for \Vcor\ in the mass and/or kinematic tracer
profiles, and show that the choice of the stellar mass profile matters
much more than the choice of the tracer profile. When \Vcor\ is used
for the mass profile (base model and P2), $\MBH = 3.31^{+0.22}_{-0.17}
\times 10^7 \MSun$ or $\MBH = 3.33^{+0.18}_{-0.17} \times 10^7 \MSun$,
depending on whether \Vcor\ or \NIRcor\ is used, respectively, for the
tracer profile. These numbers become $2.49^{+0.09}_{-0.10} \times 10^7
\MSun$ and $2.43^{+0.21}_{-0.30} \times 10^7 \MSun$, respectively,
when \NIRcor\ is used for the mass profile (models P1 and P3). The
relative insensitivity to the choice of the tracer profile is not
surprising, considering that, e.g., for a spherical system, the tracer
profile $\nu(r)$ affects mass $M(r)$ within radius $r$ only via
$d\ln\nu/d\ln r$, which by construction is nearly identical for both
\Vcor\ and \NIRcor\ near the center of the galaxy.

Model P4 substitutes $V$ for \Vcor\ in the mass profile. It yields
$\MBH = 3.25^{+0.22}_{-0.14} \times 10^7 \MSun$, very similar to the
base model, reflecting the fact that the only difference between P4
and the base model is the small luminosity excess within 0\farcs04 of
the nucleus, well below the spatial resolution of the kinematic data
($\sim$0\farcs1). The stellar $M/L$ ratio is constrained at larger
radii, so that the only difference in the stellar mass profile,
$M_*(r)$, between P4 and the base model is due to this small
luminosity excess. From the $\MBH - M_*(r_0)$ degeneracy, discussed in
\S\ref{VarMLla01}, the only option for the modeling algorithm is to
reduce $\MBH$ by a (small) amount equal to the difference in stellar
mass inside $r_0 \approx 0\farcs1$ between P4 and the base model.

This point is illustrated in Figure \ref{MassProfile}, which shows the
stellar mass profile, $M_*(r)$, and the total mass profile, $\MBH +
M_*(r)$, for the base model and for models P1 and P4. The mass
profiles of these models correspond to the three profile choices
(\Vcor, \NIRcor, and $V$, respectively) discussed above. It is clear
that dynamic modeling assigns black hole masses to P4 and to the base
model such that they both contain the same total mass inside $r
\simeq 0\farcs04$ (since both models are constrained by the same
kinematic data and the same $V$-band mass profile beyond $r \simeq
0\farcs04$). This is not the case with model P1 which, although again
constrained by the same kinematic dataset, is characterized by a
different mass profile (\NIRcor) outside of $\sim0\farcs1$, the
spatial resolution of the kinematic data.

%%%%%%%%%%%%%%%%%%%%%%%%%%%%%%%%%%%%%%%%%%%%%%%%%%%%%%%%%%%%%%%%%%%%%%%%%%%

\section{DISCUSSION} \label{discussion}

The determination of the central black hole mass in NGC 4258 from
stellar dynamics proved harder than anticipated. The main source of
difficulty (and uncertainty in $\MBH$) comes from the presence of a
$V-J$ color gradient in the central $\sim$2\arcsec\ of the galaxy,
which prevents us from determining the stellar mass profile near the
center with confidence. The model that we trust most (``base model''
in Table \ref{ModelTable}) yields $\MBH = 3.31^{+0.22}_{-0.17} \times
10^7 \MSun$ and assumes that the stellar mass is traced by the
``corrected'' $V$-band profile, or \Vcor\ (Figure
\ref{CumPhotomCenter}). This is identical to the $V$-band profile
except for the removal of a small luminous ``bump'' in the central
$\sim$0\farcs04, which we attribute to the AGN. Using the uncorrected
$V$-band profile yields a very similar $\MBH = 3.25^{+0.22}_{-0.14}
\times 10^7 \MSun$ (model P4 in Table \ref{ModelTable}). This small
black hole mass ``deficit,'' compared to the base model, corresponds
approximately to the mass of stars in the luminosity ``bump'' and is a
consequence of the degeneracy between stellar mass and black hole mass
very near the center (\S\ref{VarMLla01}; Figure \ref{MassProfile}).

Using the steeper $J$-band (NIR) profile as the mass tracer (model P1),
we obtain $\MBH = 2.49^{+0.09}_{-0.10}\times10^7 \MSun$, a factor of
25\% lower than in the base case. Although $V$-band light is more
likely to be compromised as an indicator of the stellar mass
distribution than $J$-band light, due to extinction and metallicity
gradients, we have greater confidence in the base model because (i)
the $V$-band profile is better determined near the center owing to
the higher resolution of the $V$ images, (ii) the AGN is considerably
fainter in $V$ than in $J$, making it easier to subtract, and (iii) the
minimum $\chi^2$ for the base model is 280, significantly less than
the minimum $\chi^2$ for the P1 model using $J$-band data (which was
303).

The spread in $\MBH$ introduced by uncertainties in the deprojection
parameters (models D1 and D2) is only of order 10\%, so it is the
uncertainties in the stellar mass profile that dominate the errors.

After all these sources of error are taken into account, our black
hole mass determination is $\sim$15\% lower than the ``preferred''
maser determination $[(3.82\pm0.01) \times 10^7 \MSun$] and only 7\%
lower than the lowest maser determination $[(3.59\pm0.01) \times 10^7
\MSun]$ (rescaled to our distance for the galaxy). We view this level
of agreement (2$\sigma$) as evidence that stellar dynamical mass
determinations using similar methods to those in this paper, and with
data of comparable quality, are accurate at the 2$\sigma$ level or
better.  Our previous work has sought to avoid galaxies with dusty
centers such as NGC 4258.  In this case obscuration of high-velocity
stars by the compact nuclear dust patch (\S\ref{RedVIData}), possibly
associated with the masing disk, could be responsible for an
underestimate of the black hole mass in our work. To these problems
should be added possible systematic effects due to deviations from
axisymmetry, which cannot be properly treated by our axisymmetric
modeling code. We also note that the discrepancy between the mass
determination from stellar dynamics and the maser mass determination
in NGC 4258 is smaller than the discrepancy between the stellar
dynamical mass determination and the mass determination from
individual orbits for our own Galaxy, which amounts to about a factor
of two \citep{ChaSah01,GheEtal05}.  Most concerns about black hole
mass estimates from stellar dynamics have stressed the likelihood that
the masses are overestimated, whereas here the stellar dynamical mass
is smaller than the maser mass.

Finally, although we regard the maser mass in NGC 4258 as the gold
standard in black hole masses, it is conceivable, however unlikely,
that some unrecognized effect has led to an overestimate of the mass
determined from maser kinematics.  

%%%%%%%%%%%%%%%%%%%%%%%%%%%%%%%%%%%%%%%%%%%%%%%%%%%%%%%%%%%%%%%%%%%%%%%%%%%

\acknowledgments

We are grateful to Eric Becklin and Ranga Chary for providing the
{\slshape HST} NICMOS images, and to Tom Jarrett for providing the
2MASS images before publication. We thank the referee, Tim de Zeeuw,
for comments that improved the manuscript. C.\ S.\ is grateful to
Seppo Laine and Ranga Chary for useful discussions. Support for
proposal GO--8591 was provided by the National Aeronautics and Space
Administration (NASA) through a grant from the Space Telescope Science
Institute, which is operated by the Association of Universities for
Research in Astronomy, Inc., under NASA contract NAS 5-26555. Support
for this research was also provided by NASA grant NAG 5-8238 to the
University of Michigan. A.V.F. is grateful for the support of NSF
grant AST--0607485. This research has made use of the NASA/IPAC
Extragalactic Database (NED) which is operated by the Jet Propulsion
Laboratory, California Institute of Technology, under contract with
NASA.

\appendix

\section{Description of the Orbit-Superposition Method}

For over two decades, orbit-superposition methods have been used to
study the dynamics of galaxies. Originally invented to construct a
model of a triaxial galaxy \citep{sch}, the methods have become the
tool of choice for interpreting data in terms of equilibrium galaxy
models.  The most important feature of orbit-superposition models is
that they provide a constructive proof that a given set of photometric
and kinematic data can (or cannot) be reproduced by an equilibrium
stellar system in a specified gravitational field. The computer code
described in this appendix has been used to estimate black hole masses
in galaxies \citep{GebEtal03} and to determine the relative
contributions of dark and luminous matter in elliptical galaxies
\citep{ThoEtal04,ThoEtal05}. It is a descendant of the spherical
program described in \cite{richs88}, but improved in two major
respects: it treats the galaxy as axisymmetric rather than spherical,
and it matches the full LOSVD of the galaxy at specified positions on
the sky, rather than the second moment only. Similar programs have
been developed by \cite{MarEtal98}, \cite{CreEtal99}, and
\cite{ValEtal04}.

A number of preliminary steps must be taken to convert the reduced
data into inputs for the orbit-superposition method.  These include
the deprojection of the observed surface brightness to construct a light
distribution in three-dimensional space and the reduction of spectra
at different locations on the sky to projected velocity
distributions. Both of these activities require additional assumptions
or choices. We normally deproject under the assumption that level
surfaces in stellar density are coaxial spheroids \citep{GebEtal96}
and we normally use a penalized maximum-likelihood estimator to
construct the LOSVDs \citep{GebEtal00a,PinEtal03}. Next we compute the
gravitational field from the three-dimensional light distribution
under the assumption that the mass consists of a black hole of mass
$\MBH$ and stars having a $M/L$ ratio independent of
position.  Dark matter can be incorporated into such a model, but we
have not done so here.

Once the preliminaries are complete, the calculation of a dynamical
model by this method consists of two steps. First, a library of orbits
is constructed using initial conditions that cover all possible
locations in phase space. Each orbit's contribution to the surface
brightness and projected velocities is logged. Second, the orbits are
combined to match the light distribution and LOSVDs of the galaxy as
well as possible (using $\chi^2$ parameter to assess the goodness of
fit). This procedure is repeated for a set of black hole masses, and
limits on the latter are set from $\chi^2$ as a function of
$\MBH$. Note that although different orbit libraries are required for
each different ratio of $M/L$ to $\MBH$, a single library can be used
for various values of $\MBH$ so long as $M/L$ scales with $\MBH$.

\subsection{The Orbit Library}

We construct the model on a grid in coordinates $r, \eta$ where $r$ is
the radius and $\eta$ is the latitude.  We restrict ourselves to
potentials (and mass distributions) symmetric about the equator.  The
grid is evenly spaced in $\nu = \sin(\eta)$ from $\nu = 0$ to $1$, and
in $r$ in even intervals in $k$ defined by
\begin{equation}
k = {1 \over a} \log \left( 1 + {a \over b } r \right).  
\end{equation}
Bins in $(r,\eta)$ are bounded by the grid points.  Below we also use
standard cylindrical coordinates ($\varpi, z$).  It is essential to
explore the phase space of orbits with a resolution at least as fine
as the resolution used to construct the models.

Orbits in axisymmetric potentials always have two isolating integrals,
the energy $E$, and the $z$-axis angular momentum $J_z$. 
Regardless of
the possible presence of a third integral, conservation of $E$ and
$J_z$ confine the orbit within a region of $(\varpi, z)$ space
defined by a boundary where the orbits' velocities are zero:
\begin{equation}
    \Phi(\varpi,z) + {J_z^2 \over 2 \ \varpi^2}  < E.
  \end{equation}
\noindent
  The allowed volume intersects the equator only over a limited range
  in radius with an upper and lower limit. We choose $(E, J_z)$ pairs
  so that there is an orbit with upper and lower limits in the middle of
  each radial bin.  We associate an area with each grid point $(\Delta
  E \times \Delta J_z)$ by halving the distance to adjacent grid
  points.  For each specified $(E, J_z)$, the section of allowed phase
  space (we ignore $\phi$ because of the conservation of $J_z$) that
  lies on the equator is a two-dimensional surface with coordinates
  $\varpi$ and $v_\varpi$ (a surface of section). A regular orbit is
  defined by a curve on this surface. We systematically tile this
  surface in the manner invented by \cite{ThoEtal04}, launching an
  orbit from the center of each bin and assigning a phase space volume
  to each orbit by
\begin{equation}
\Delta  \Omega = 
        \Delta E \, \Delta J_z \int T (\varpi, v_\varpi) 
         d\varpi \times d v_\varpi,
\label{AppdOmega}
\end{equation}
based on the work of \cite{Bin85}, where $T$ is the time between
successive crossings of the equatorial plane.

The procedure above gives a starting position for each orbit in the
equatorial plane of the model with a well-defined $\varpi$,
$v_\varpi$, $E$, and $J_z$. We use $E$ to determine $v_z$, and we
integrate the motion of the orbit in the $(\varpi, z)$ plane. Assuming
that all orbits cross the equator, it is a complete survey of orbits.  We
follow the orbits of stars in spherical coordinates, to take advantage
of the near sphericity of the mass distribution in the region where a
black hole may dominate the gravitational field. Using $r$ as the
radial coordinate, $\eta$ as the equatorial angle, and $\phi$ as the
axial angle, the equations of motion are
\begin{eqnarray}
\frac{dv_r}{dt}   &  = &
                {v_\eta^2 + v_\phi^2 \over r} - 
                {\partial \Phi \over \partial r}, \\
\frac{dv_\eta}{dt}    &    = &
                -{ v_r v_\eta \over r} 
                - \tan\eta \, {v_\phi^2 \over r} 
                - {1 \over r} {\partial \Phi \over \partial \eta}, ~{\rm and} \\
v_\phi        &  = &
        {l_z \over r \cos\eta}.
\end{eqnarray}

The first two equations are integrated using a standard fourth-order
Runge-Kutta method with variable timestep, and $v_\phi$ is recovered
at each timestep from equation (A6). We obtained good energy
conservation by letting $dt = \epsilon r/v_{circ}(r)$, where $v_{circ}$
is the velocity of a circular orbit at that radius, with $\epsilon$
between 0.01 and 0.1.  Orbits are followed for about $100$ crossing
times.  We terminate the calculation of an orbit if its energy error exceeds
1\%. Typical energy errors are 0.1\%. 
The potential is set equal to zero at $\infty$.  

The accelerations for the orbit integrations are determined by
decomposing the assumed galaxy density distribution in spherical harmonics.
Specializing immediately to axisymmetry gives
\begin{equation}
\Phi({\bf x}) = 
        - 4 \pi G \sum_{l=0}^{l_{max}}
                \left( A_l r^{-(l+1)} 
             + B_l r^l  \right)
                        P_l (\nu),
\label{AppPhi}
\end{equation}
with
\begin{eqnarray}
A_l     &= &
        \int_0^r \left[ \int_0^1 \rho(s,\nu) P_l(\nu) d \nu \right]   
            s^l s^2 ds, \\
B_l &= &\int_r^\infty \left[ \int_0^1 \rho(s,\nu) P_l(\nu) d \nu \right]   
        s^{-(l+1)} s^2 ds, 
\end{eqnarray}
and where $\nu = \sin\eta$. The gravitational acceleration on the
orbit is determined by differentiating equation \ref{AppPhi} with
respect to $r$ and $\eta$. 

We tested our
expansion in a Kuzmin-Kutusov model \citep{DejDez88} with $a/c=0.5$, 
which corresponds to a very flat galaxy (about E5). This model
has analytic density, potential, and accelerations. Truncating the
expansion at $P_6$ accurately reproduces the exact accelerations to
1\% or better everywhere in the galaxy, so we set $l_{max} = 6$ in
equation (A7).  

The additional acceleration from a mass point is straightforward to
add to the radial component of the accelerations above. We store and
model the galaxy in the angular range $0^\circ \le \eta \le
90^\circ$. We store the three lowest-order internal moments in each
bin for each orbit (mass $m$, $m {\mathbf v}$, $m{{\mathbf v}{\mathbf
    v}}$) for later use.

Simultaneously, the projected line-of-sight velocities are binned in a
three-dimensional grid in projected radius $R$, angle $\beta$
(measured from the major axis of the galaxy on the plane perpendicular
to the line of sight), and velocity $v$. The
spacing in $R$ and $\beta$ is defined as above. Since the orbits are
being followed in $r$ and $\eta$, $\phi$ is not defined. At each
timestep, in order to project the orbit onto the line of sight, we
choose 100 values of $\phi$, evenly spaced from 0 to $\pi$.

\subsection{Matching the Light Distribution and Observed Dynamics }

The next step is to find the superposition of orbits consistent with
the stellar mass distribution that best matches the observed LOSVDs.
\cite{richs88} provide an efficient method to match the mass
distribution that can be augmented to minimize $\chi^2$. 
Their method maximizes an objective  function 
\begin{equation}
F = - \sum_i C_i(x_i)
\end{equation}
while satisfying a set of constraints 
\begin{equation}
Y_j = \sum_i m_{ij} x_i.
\end{equation}

We choose to maximize the objective function 
\begin{equation}
 S - \alpha \, \chi^2 =
            - \sum_{i=1}^{N_{orb}} 
          w_i \log \left({w_i \over \Delta \Omega_i} \right)
         - \alpha \sum_{k=1}^{N_d} {(l_k - l_{k,o})^2 \over \sigma_k^2},
\end{equation}
where $w_i$ is the weight of the $i^{th}$ orbit and  $l_k$ is the light in
the $k^{th}$ bin in the {\em observational} phase space composed of
the position on the sky and the line of sight velocity. The more
familiar LOSVD is the distribution of light in line of sight velocity
at a single position on the sky.  Thus the index $k$ identifies a
small range in both projected velocity and projected position.  The
variable $\alpha$ is an adjustable parameter discussed further below.
There are $N_{orb}$ orbits in the orbit library, and $k$ varies from
$1$ to $N_d$, where $N_d$ is the number of positions at which the
LOSVD has been measured times the number of velocity intervals at each
measurement. The first sum in equation (A12) is an entropy,
and the second is a $\chi^2$ that measures the goodness of fit of the
model LOSVDs to the observed LOSVDs.  The sum of the orbits must also
match the deprojected (3D) light distribution in the $N_{bin}$ bins:
\begin{equation}
  M_j 
  = \sum_{i=1}^{N_{orb}} m_{ij} w_i \ {\rm for} \  j = 1, N_{bin}, 
\label{AppL}
\end{equation}
where $m_{ij}$ is the mass contribution of the $i^{th}$ orbit to the
$j^{th}$ bin.  

In order to put equation (A12) in the form of equations (A10) and
(A11), we make the following substitutions and identifications.
First, we define $x_i = w_i$ for $i = 1, N_{orb}$, and add an
additional set of variables that measure the error in the light
projected into each LOSVD bin $x_{N_{orb} + k} = l_k - l_{k,o}$ for $k
= 1, N_d$.  Since the light in each bin of the model LOSVDs is
given by the contribution of each orbit to that 
{\em projected} velocity and position 
there is an additional set of constraint equations
of the form the form
\begin{equation}
  l_{k,o} = \sum_{i = 1}^{N_{orb}} m_{ij} w_i - x_j
  \ {\rm for} \  j = N_{bin} + 1,  N_{bin} +  N_d, 
\end{equation}
where $k = j - N_{orb}$ and 
$m_{ij}$ is the contribution of the $i^{th}$ orbit to the
$k^{th}$ component of the set of LOSVDs.  

Then we have 
\begin{eqnarray}
C_i = &  
          x_i \log \left({x_i / \Delta \Omega_i} \right)
          & {\rm for ~} i \le N_{orb}  \nonumber \\
C_i = &  \alpha x_i^2 / \sigma^2_{i - N_{orb}} 
          & {\rm for ~}  N_{orb}  < i \le N_{orb} + N_d 
\end{eqnarray}
corresponding to equation (A10) and equations (A13) and (A15)
correspond to equation (A11).  Note that for $i > N_{orb}$, $-C_i$ is
maximized at $x_i = 0$.  The first set of $C_i$ (for $i \le N_{orb}$)
effectively keep the orbit weights positive. The second set (for $ i >
N_{orb}$) minimizes the components of $\chi^2$.  We then use the
method described by \cite{richs88} to iteratively solve the equations.

 An important feature of this problem is the choice of the parameter
 $\alpha$. Although we have not devised a method to specify
 appropriate values of this parameter in advance, if the quantity
 $\chi^2$ reflects a satisfactory accounting of the estimates of
 errors in the recovery of the LOSVDs of the target galaxy, then the
 choice of $\alpha$ and interpretation of the results is reasonably
 straightforward.  A small $\alpha$ will generally produce a solution.
 Increasing $\alpha$ slowly decreases $\chi^2$ but not beyond some
 limit. In practice, we start with $\alpha = 0$ and increase it slowly
 until the change in $\chi^2$ is less than 0.02 in a single iteration.

\subsection{A Test with a Black Hole}

In this case we specify a dynamical model with a
gravitational potential and a distribution function (DF) 
for the stars (the starlight need not, and does not, 
determine the mass in this model).   The gravitational potential is 
\begin{equation}
  \Phi(r) = v_{\rm circ}^2 \log(r) - {G \MBH \over r}.  
\end{equation} 
This potential includes a point mass and a flat 
circular-velocity curve (with velocity $v_{\rm circ}$) at large $r$.  
We investigated Michie-like DFs of the 
form 
\begin{equation}
  f = A \exp \left\{ - \left[ {E + J_z^2/(2 \varpi_a^2) 
         \over  \sigma^2} 
       \right] \right\}
     J_z^{2N}
      \sqcap (E_1, E, E_2),  
\end{equation} 
where $A$ is normalization, $E$ and $J_z$ are energy and angular
momentum, $w
\varpi_a$ (an anisotropy distance), $N$, and $\sigma$ (a velocity
dispersion-like quantity) are model parameters, and $E_1$ and $E_2$ are
upper and lower limits on the energy.  The function $\sqcap$
is defined by
\begin{equation}
     \sqcap (E_1,E, E_2) = 
         \left\{ 
              \begin{array}{ll} 
                1 & \mbox{if ~~}  E_1 \le E \le E_2 \\
                0 &  \mbox{otherwise} . 
               \end{array}
         \right. 
\end{equation} 
The DF can be rewritten as 
\begin{equation}
  f = \pi A \varpi^{2N} \exp (- \Phi/\sigma^2) 
      \sqcap(v_1, v_\phi, v_2) 
      v_\phi^{2N}
      \exp(-v_\phi^2/(2 \tilde \sigma^2))
      \exp(-t^2/(2  \sigma^2))
       d t^2 dv_\phi,   
\end{equation}
where $t^2 = v_\varpi^2 + v_z^2$, $\tilde \sigma^2 = 
\sigma^2 (\varpi_a^2/(\varpi_a^2 + \varpi^2))$ and 
$v_i = \sqrt{2(E_i - \Phi)}$.  This function has a number of 
interesting properties:  $N > 0$ tends to create $v_\phi$ enhanced
anisotropy and a flattened density distribution;  $\tilde \varpi_a <
\infty$ tends to depress $v_\phi$ and create a prolate density
distribution.  The cutoffs in energy avert a  divergence in the
luminous matter distribution near the black hole.

The density is 
\begin{equation}
  \rho = 2 \pi \sigma^2 A 
        \varpi^{2N} \exp (- \Phi/\sigma^2) 
        \int _{v_1}^{v_2}      v_\phi^{2N}
      \exp(-v_\phi^2/(2 \tilde \sigma^2))
      [\exp(-t_1^2/(2  \sigma^2)) - \exp(-t_2^2/(2  \sigma^2))]
        dv_\phi,   
\end{equation}
where $t_2 = v_2^2 - v_\phi^2$ and $t_1 = \max(v_1^2 - v_\phi^2,0)$.

Note that the distribution functions with different choices of
parameters can be added to each other (in the same potential) and
similarly the densities and density-weighted moments are additive.
A Monte Carlo realization of the model is constructed by sampling each
density distribution and choosing the velocities
randomly from the velocity distribution.  Note that we have the
additional freedom of choosing an arbitrary fraction of the $\phi$
velocities in the positive sense, and hence setting the spin of any
model between a maximum amount and zero.  The total number of random
variates chosen sets the normalization.

For the test here, we adopted a potential with $v_{\rm circ} = 220
\,\mbox{\kms}$ and $\MBH = 1.126 \times 10^8 \MSun$.  We added two
DFs.  The first had $\sigma = 160 \,\mbox{\kms}$, $\varpi_a = 600
\,\pc$, $N = 0$, $E_1 = \Phi(10 \,\pc)$, $E_2 = \Phi(1000 \,\pc)$,
with equal numbers of the $\phi$ velocities chosen in the positive and
negative sense.  The second DF had $\sigma = 120 \,\mbox{\kms}$, $w_a
= 200 \,\pc$, $N = 2$, $E_1 = 10\, \pc$, $E_2 = 1000\, \pc$ and $3/4$ of
the $\phi$ velocities were chosen positive.  Each DF was assigned $
1/2 \times 10^9$ random points, so they had equal mass.  The second
DF, featuring a moderately spinning disky structure, dominates the
model near the center of the galaxy, but declines rapidly.  The first
DF, a mildly prolate structure with no net spin dominates further out.
No stars are found beyond a radius of $1000\, \pc$ from the center of
the model.  The $10^9$ random points each are converted to projected
positions and velocities (we assumed the galaxy was edge on), and
binned into ``observations'' mimicing our combined HST and ground-based data,
and also producing data mimicing data that would be obtained with an
integral field unit.  For the example below we used the IFU data with
a resolution of $10 \pc$ at the center, and lower resolution data
extending to a distance of $800\, \pc$ from the center.  At each
location the projected velocity variates were binned at a resolution
of $20 \,\mbox{\kms}$.  Gaussian noise was added to the data, in a
series of 20 realizations.  These datasets were then fed into the
modeling program, which produced (in each case) a $\chi^2$ map in
$v_{\rm circ}$ and $\MBH$.  

Marginalizing over each variable in turn produced the determinations
of $\MBH$ and $v_{\rm circ}$ illustrated in Figure 17.  Each dataset had
LOSVDs at 32 positions with 15 velocity bins, a product of 480.  For
each realization of this set, the best-fit model had a $\chi^2$ near
$500$.  Since the velocity bins are uncorrelated in these datasets
this is about the expected $\chi^2$.  For the presentation in Figure
17 the plots of $\chi^2$ against black hole mass and $v_{\rm circ}$ were
arbitrarily shifted to a minimum of $500$.  The error bars judged from
the individual $\chi^2$ profiles are about $20 \%$ smaller than the
error (above) determined from the spread in minima.  The mean values
of $\MBH$ and $v_{\rm circ}$ from these experiments were $(1.15 \pm 0.03)
\times 10^8 \MSun$ and $222 \pm 5 \,\mbox{\kms}$, in excellent
agreement with the target values.

An earlier version of the code suffered from a units conversion error
that led to an underestimate of black hole masses (and $M/L$).  All
black hole masses obtained with this program published in or prior
to 2003 should be multiplied by $1.09$.

%%%%%%%%%%%%%%%%%%%%%%%%%%%%%%%%%%%%%%%%%%%%%%%%%%%%%%%%%%%%%%%%%%%%%%%%%%%
% BIBLIOGRAPHY
%%%%%%%%%%%%%%%%%%%%%%%%%%%%%%%%%%%%%%%%%%%%%%%%%%%%%%%%%%%%%%%%%%%%%%%%%%%

\clearpage

%%%%%%%%%%%%%%%%%%%%%%%%%%%%%%%%%%%%%%%%%%%%%%%%%%%%%%%%%%%%%%%%%%%%%%%%%%%
% FIGURES
%%%%%%%%%%%%%%%%%%%%%%%%%%%%%%%%%%%%%%%%%%%%%%%%%%%%%%%%%%%%%%%%%%%%%%%%%%%
% Fig GroundImages-NIR

\begin{figure}

\caption{Ground-based NIR images of NGC 4258 identifying
  large-scale morphological characteristics of the galaxy: The bulge
  (PA $\simeq$ 146\arcdeg), which is embedded in the inner oval disk
  (PA $\simeq$ 156\arcdeg); the weak bar (PA $\simeq$ 11\arcdeg); and
  the outer oval disk (PA $\simeq$ 150\arcdeg). 
 {\slshape Top
    left:} 2MASS composite $J$+$H$+$K$ image. Intensity is
  proportional to the square root of surface brightness to illustrate
  the bulge and the small bar. The scale is 1\arcsec\ pixel$^{-1}$ and
  the image size is $17\farcm1 \times 17\farcm1$. {\slshape Top
    right:} As before, but intensity is proportional to surface
  brightness; the inner oval disk is now saturated and the outer oval
  disk becomes more prominent. {\slshape Bottom left:} $K$-band image
  using SQUID on the KPNO 2.1-m telescope. Intensity is proportional
  to the square root of surface brightness to illustrate the
  bulge. The scale is 0\farcs69~pixel$^{-1}$ and the image size is
  $5\farcm9 \times 5\farcm9$. {\slshape Bottom right:} As before, but
  intensity is proportional to surface brightness to illustrate the
  weak bar. \label{GroundImages-NIR}}

\end{figure}

%%%%%%%%%%%%%%%%%%%%%%%%%%%%%%%%%%%%%%%%%%%%%%%%%%%%%%%%%%%%%%%%%%%%%%%%%%%
% Fig HSTImages-NIR

\begin{figure}

\begin{center}
\plottwo{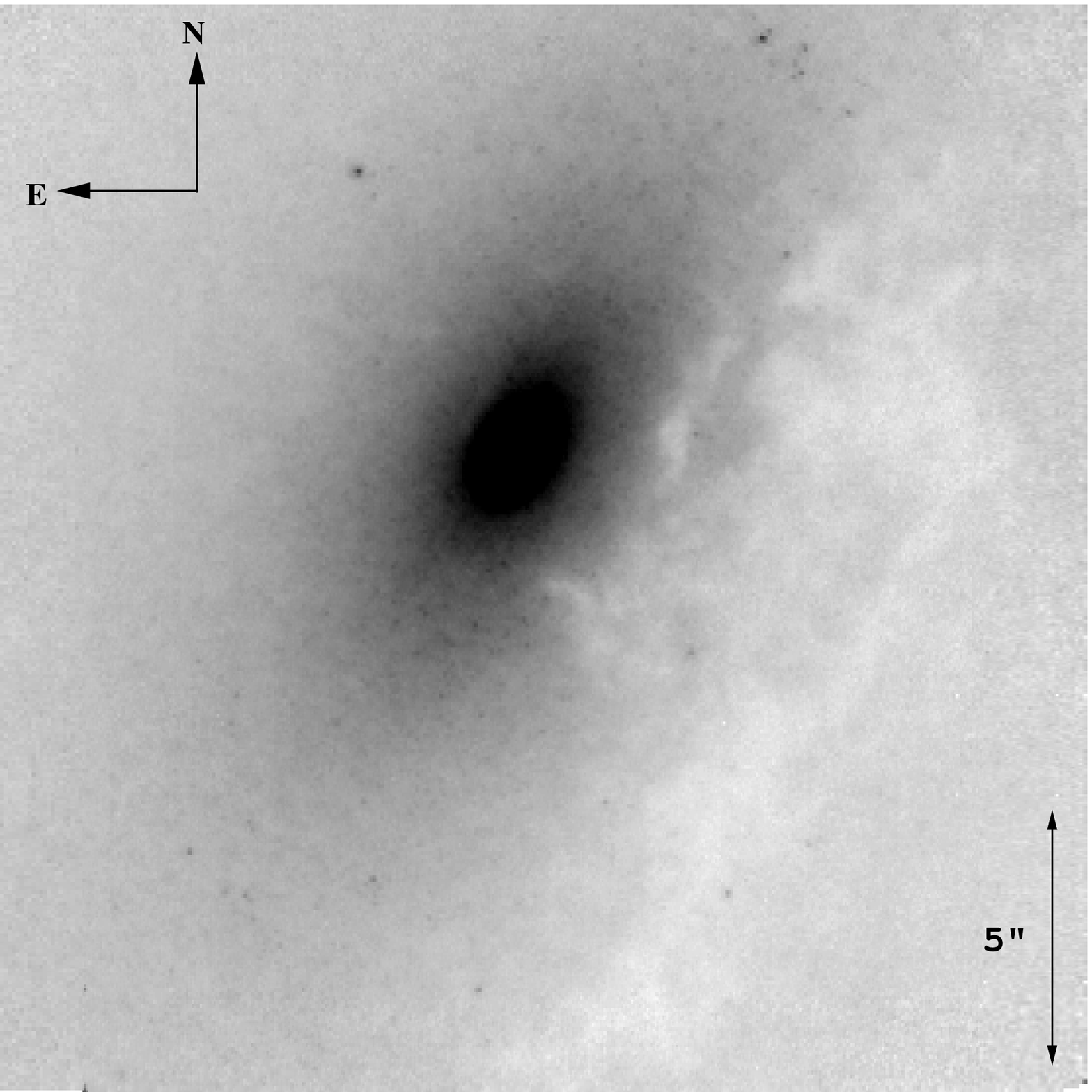}{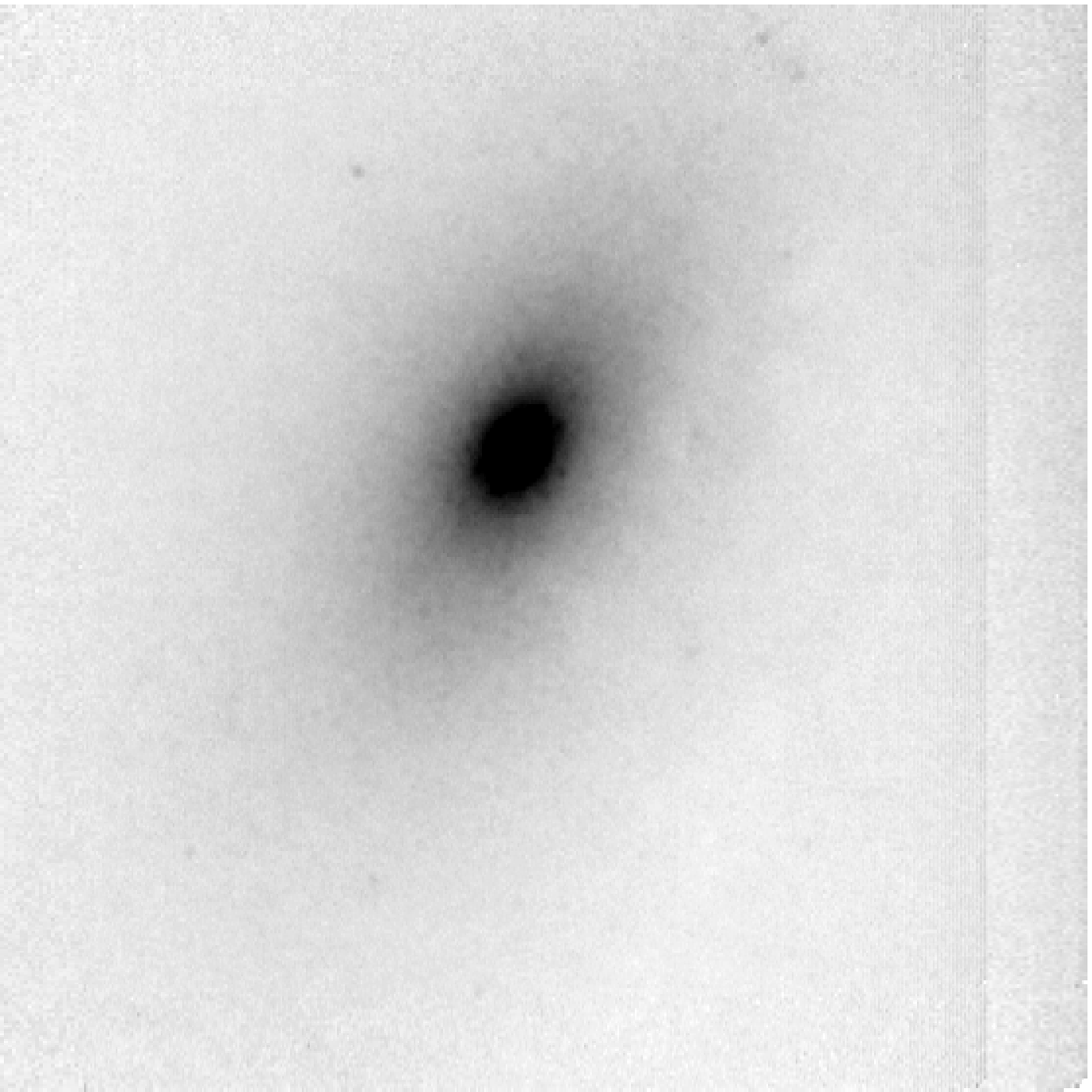}
\end{center}

\caption{{\slshape HST} NICMOS images of NGC 4258 in the $J$ band
(F110W) (left) and in the $K$ band (F222M) (right), from
\citet{ChaEtal00}. In both images, the scale and size are
$\sim$0\farcs038 pixel$^{-1}$ and $\sim21\arcsec \times
21\arcsec$, respectively. The images are not deconvolved. Intensity is
proportional to the square root of surface brightness.  A band of data
is missing on the right side of the $K$-band image. For the extraction
of stellar mass profiles from these images, the dust on the SW side
was corrected by replacing the SW side of the galaxy with the NE side
flipped across the major axis. \label{HSTImages-NIR}}

\end{figure}

%%%%%%%%%%%%%%%%%%%%%%%%%%%%%%%%%%%%%%%%%%%%%%%%%%%%%%%%%%%%%%%%%%%%%%%%%%%
% Fig PSF

\begin{figure}

\begin{center}
\includegraphics[scale=0.2]{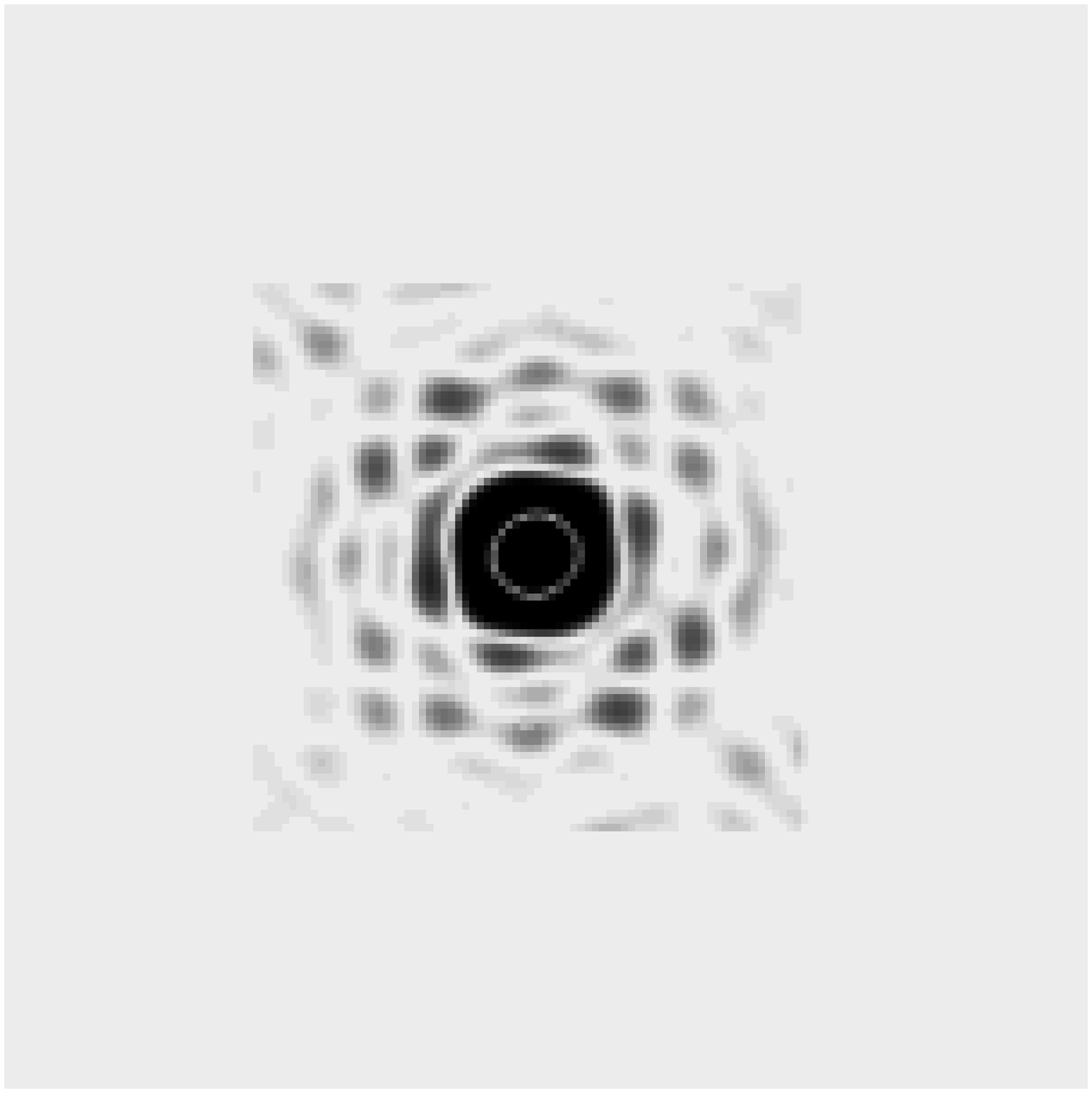}
\includegraphics[scale=0.2]{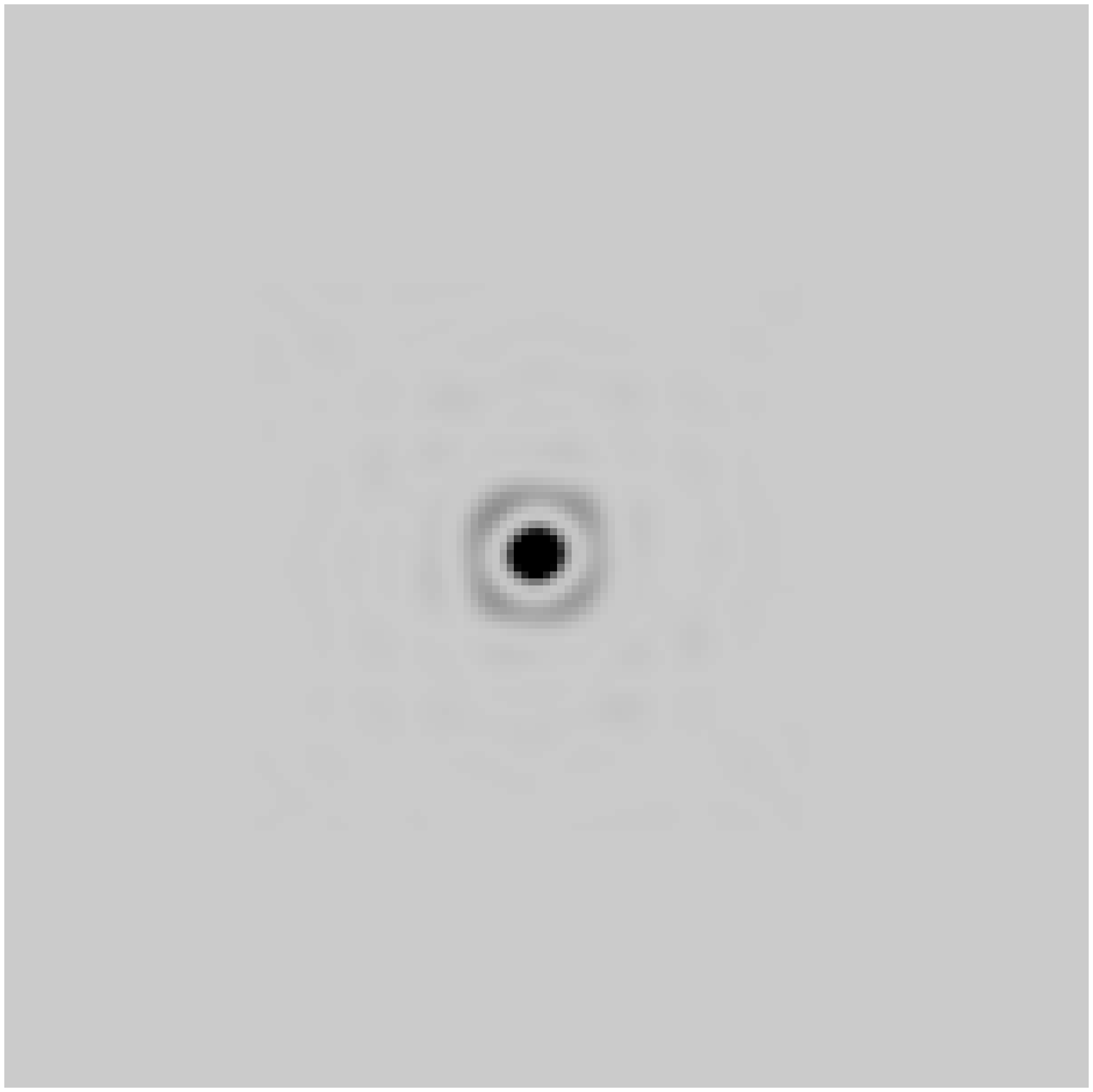}

\includegraphics[scale=0.2]{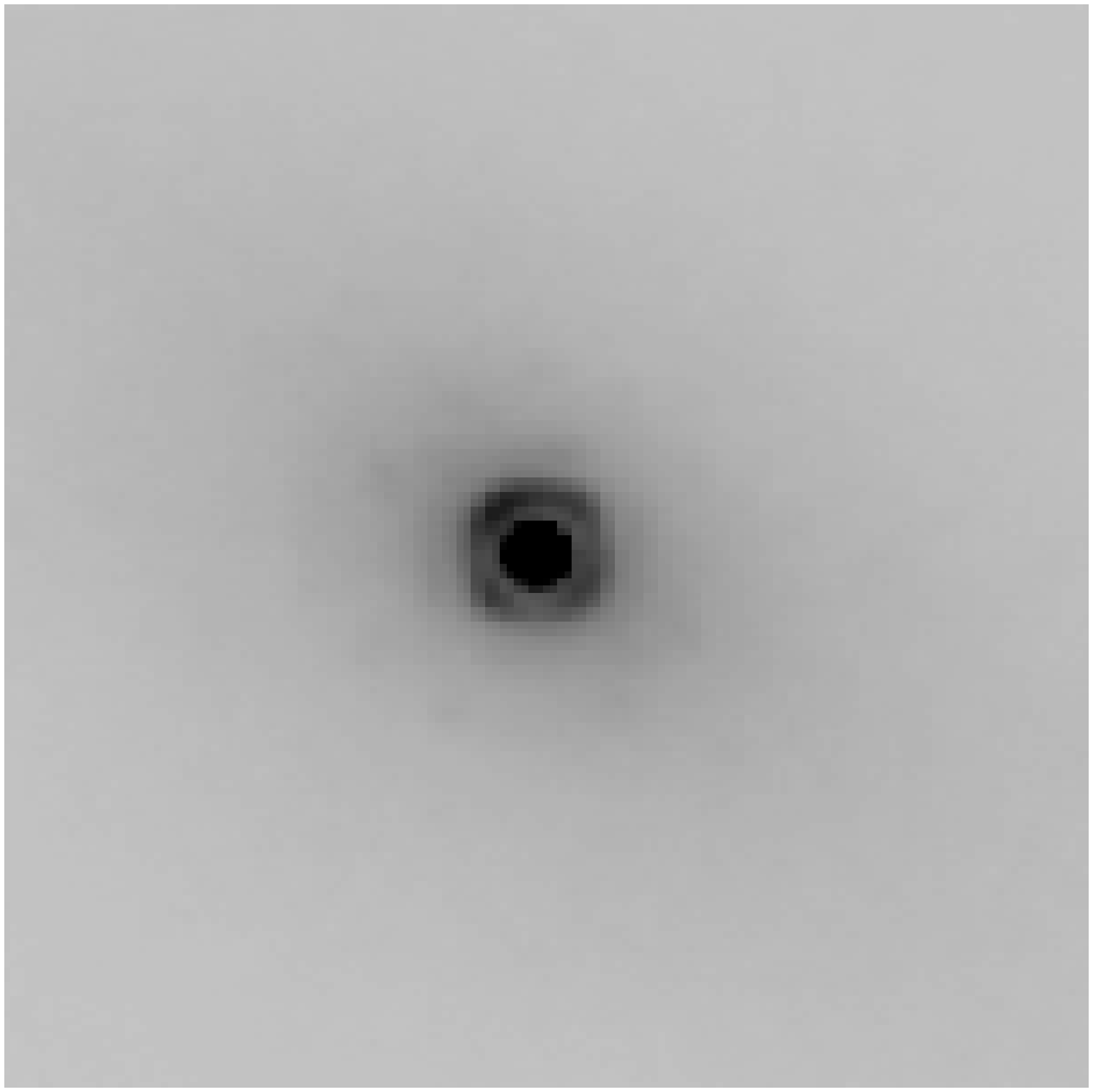}
\includegraphics[scale=0.2]{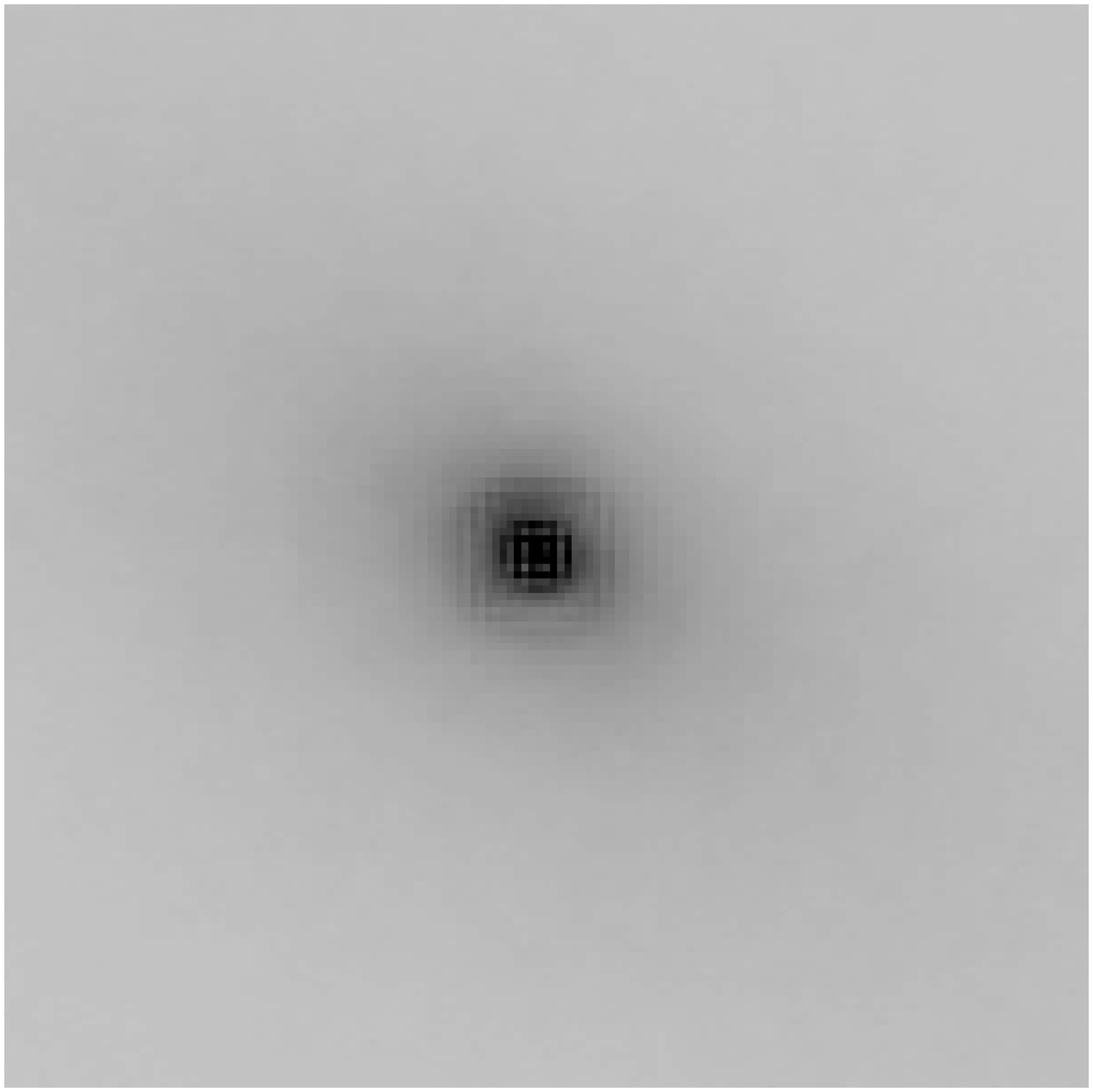}
\end{center}

\caption{{\slshape HST} NICMOS $K$-band images of the central
  $5\farcs7 \times 5\farcs7$ of NGC 4258, adapted from the images
  provided by \citet{ChaEtal00}. North is up, and East is to the
  left. {\slshape Top left:} The TinyTim model PSF shown at a
  nonlinear stretch (higher contrast at lower surface brightnesses) to
  emphasize the small-scale features. {\slshape Top right:} Same as
  before, but now scaled to the galaxy image and at the same linear
  stretch as in bottom row. {\slshape Bottom left:} The galaxy before
  PSF subtraction; the AGN is evident. {\slshape Bottom right:} The
  adopted PSF-subtracted galaxy image that was used to provide the
  central brightness profile in the $K$ band. In the end, the AGN
  could not be subtracted well enough in $K$, and the $J$-band profile
  was used for $r<0\farcs2$. \label{PSF}}

\end{figure}

%%%%%%%%%%%%%%%%%%%%%%%%%%%%%%%%%%%%%%%%%%%%%%%%%%%%%%%%%%%%%%%%%%%%%%%%%%%
% Fig CumPhotom

\begin{figure}

\begin{center}
\plotone{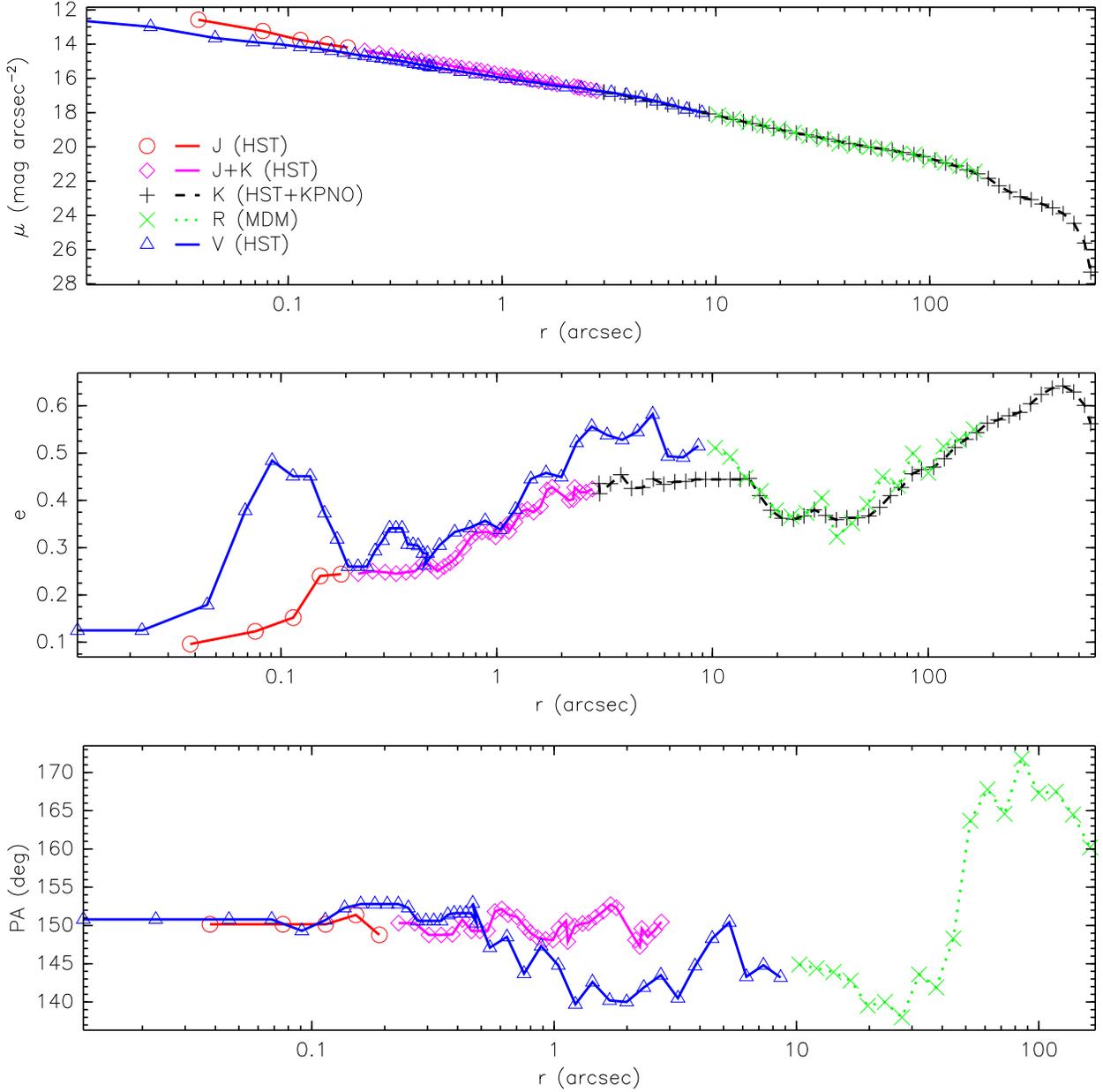}
\end{center}

\caption{NGC 4258 major-axis surface photometry {\em after}
  subtraction of prominent dust lanes, and correction for the AGN. The
  radius ($r$) is measured along the major axis. All {\slshape HST}
  profiles are deconvolved for $r<3\arcsec$. {\slshape Top:} NIR ($J$,
  $J$+$K$, $K$) and optical ($V$, $R$) surface brightness
  profiles. The $R$-band profile is calibrated using the {\slshape
    HST} $V$-band zeropoint between $r \simeq 5\arcsec$ and
  7\arcsec. The NIR profiles are scaled to match the optical profile
  at large radii to illustrate the absence of major color gradients in
  the disk. The almost power-law profile at $r \la 40\arcsec$ belongs
  to the bulge. Only data out to $r \approx 150\arcsec$ are used for
  the dynamical modeling. {\slshape Middle:} The ellipticity as a
  function of radius. {\slshape Bottom:} The position angle of the
  major axis as a function of radius. \label{CumPhotom}}

\end{figure}

%%%%%%%%%%%%%%%%%%%%%%%%%%%%%%%%%%%%%%%%%%%%%%%%%%%%%%%%%%%%%%%%%%%%%%%%%%%
% Fig GroundImages-R

\begin{figure}

\caption{Ground-based image of NGC 4258 in the $R$ band taken with the MDM 1.3-m McGraw Hill telescope. The square is centered on the nucleus, and is 8\arcsec\ on a side. The scale of the CCD was 0\farcs508 pixel$^{-1}$. A logarithmic stretch has been applied, and set to prevent ``overexposure'' of the bulge region. \label{GroundImages-R}}

\end{figure}

%%%%%%%%%%%%%%%%%%%%%%%%%%%%%%%%%%%%%%%%%%%%%%%%%%%%%%%%%%%%%%%%%%%%%%%%%%%
% Fig HSTImages-V

\begin{figure}

\begin{center}
\plotone{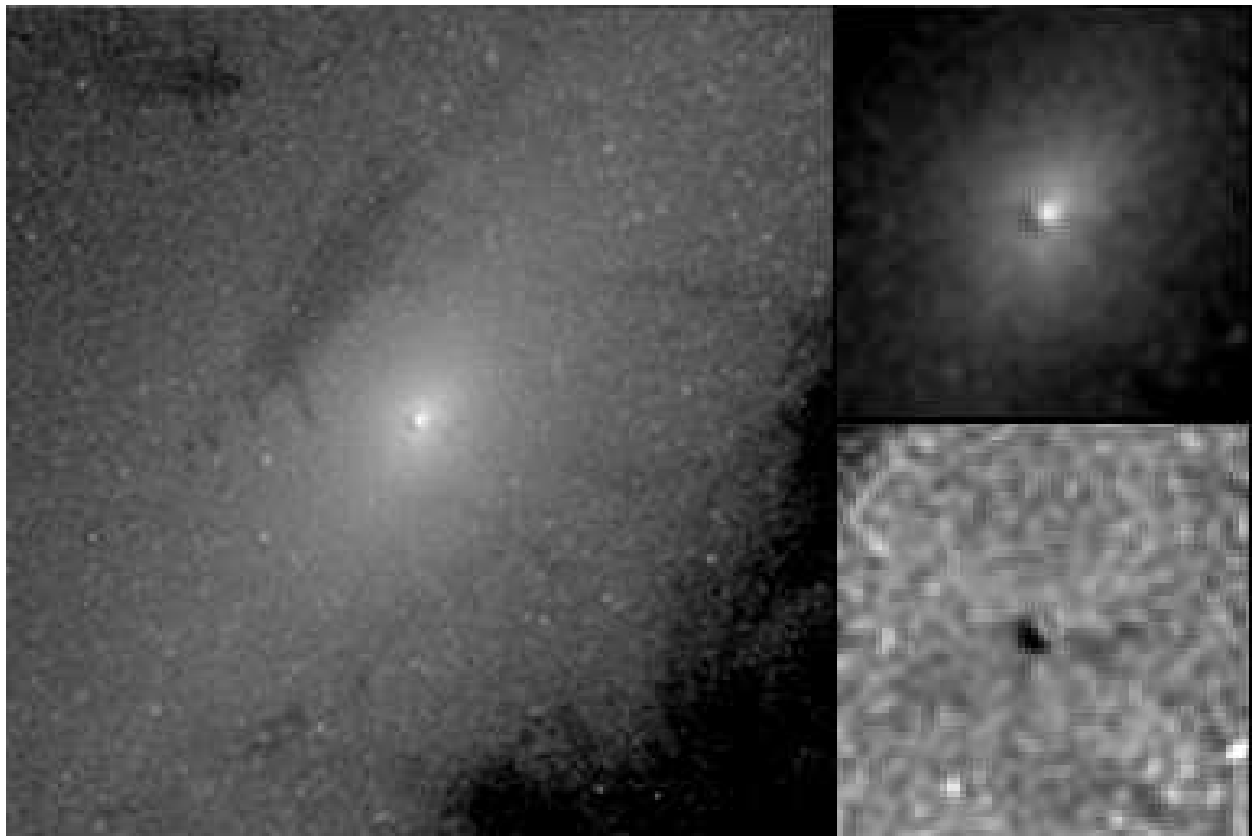}
\end{center}

\caption{{\slshape HST} WFPC2 image of NGC 4258. The large panel shows
  the central $256 \times 256$ pixels of the deconvolved F547M
  Nyquist-sampled image. It corresponds to $128 \times 128$ PC1
  pixels, or $5\farcs84 \times 5\farcs84$. An arbitrary logarithmic
  stretch has been applied. The small upper-right panel is the central
  $1\farcs48 \times 1\farcs48$ ($64 \times 64$ subpixels) of the same
  image, magnified by a factor of 2 compared to the larger panel. Again a
  logarithmic stretch has been applied, but with greater contrast over
  the nuclear region. The bottom-right panel shows the same region
  divided by a model reconstructed from the surface photometry. North
  is up and East to the left. \label{HSTImages-V}}

\end{figure}

%%%%%%%%%%%%%%%%%%%%%%%%%%%%%%%%%%%%%%%%%%%%%%%%%%%%%%%%%%%%%%%%%%%%%%%%%%%
% Fig ColorMap

\begin{figure}

\caption{Color map of the nuclear region of NGC 4258. It corresponds
  approximately to $V-I$, and was created from the ratio of the F547M
  image to the F791W image. Darker is redder. The map size is
  $8\farcs0 \times 7\farcs2$. The F791W image had each pixel resampled
  into 4 pixels to match the F547M image, and was then
  boxcar-smoothed. The central blue spike is not accurate because the
  center is saturated in the F791W data, and the F547M and F791W
  images are not perfectly aligned. \label{ColorMap}}

\end{figure}

%%%%%%%%%%%%%%%%%%%%%%%%%%%%%%%%%%%%%%%%%%%%%%%%%%%%%%%%%%%%%%%%%%%%%%%%%%%
% Fig HSTImages-color

\begin{figure}

\caption{Color composite image of NGC 4258. Red corresponds to F791W, green to F656W, and blue to F300W. The image has a size of $476 \times 526$ pixels, with a scale of 0.0455\arcsec~pix$^{-1}$. North is up, and East to the left. \label{HSTImages-color}}

\end{figure}

%%%%%%%%%%%%%%%%%%%%%%%%%%%%%%%%%%%%%%%%%%%%%%%%%%%%%%%%%%%%%%%%%%%%%%%%%%%
% Fig SpectraHST

\begin{figure}

\begin{center}
\includegraphics[scale=0.8]{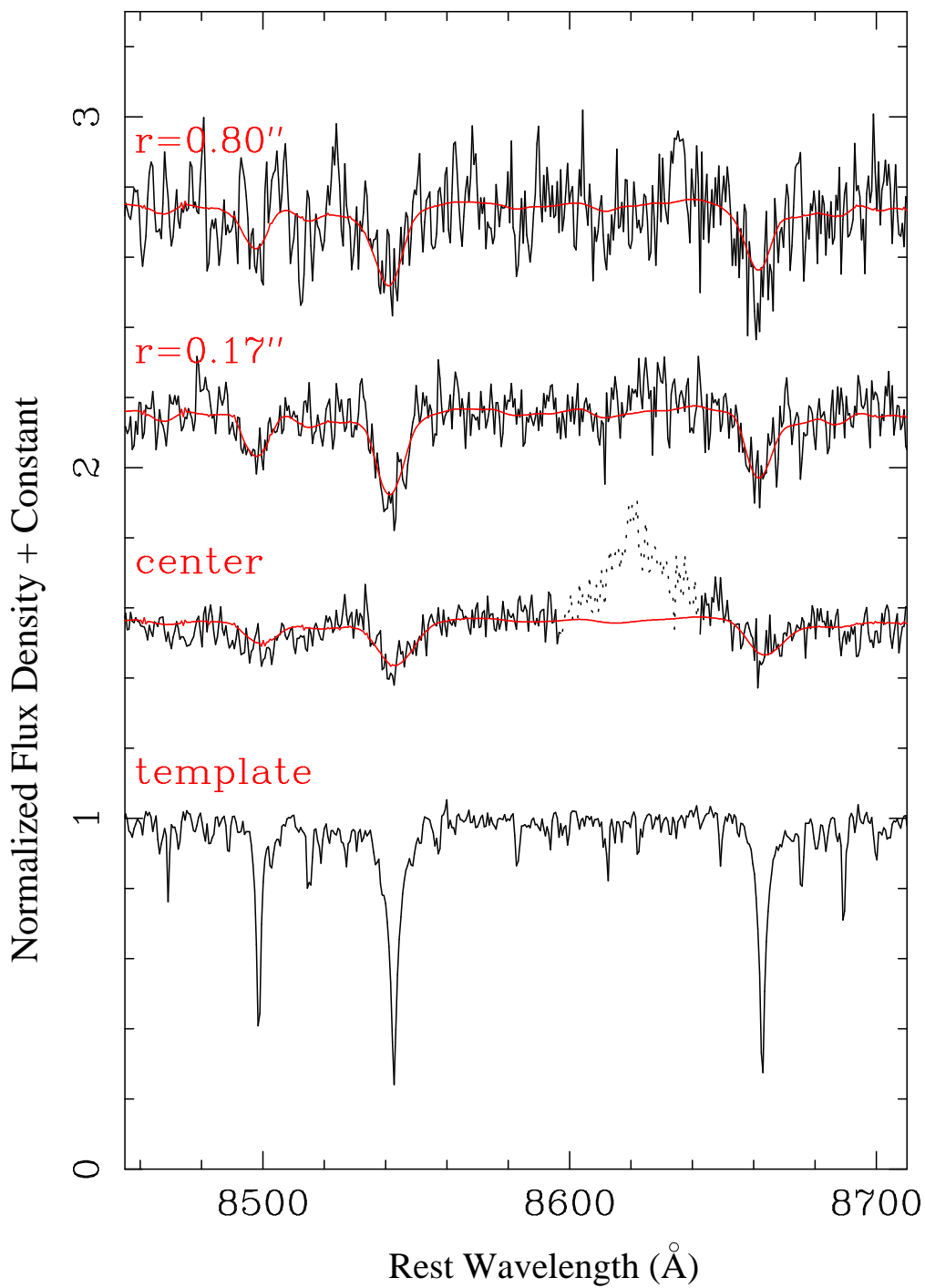}
\end{center}

\caption{Sample STIS spectra extracted from the central and two outer
  spatial bins on the approaching (SE) side of the galaxy, as well as
  of the template star HR 7615 which was used for the LOSVD
  deconvolution. The smooth solid lines superimposed on the galaxy
  spectra are the template stellar spectra convolved with the derived
  LOSVDs. The emission-line feature (probably Fe II) in
  the nuclear spectrum (dotted) was subtracted before the LOSVD
  analysis was performed. \label{SpectraHST}}

\end{figure}

%%%%%%%%%%%%%%%%%%%%%%%%%%%%%%%%%%%%%%%%%%%%%%%%%%%%%%%%%%%%%%%%%%%%%%%%%%%
% Fig SpectraGround

\begin{figure}

\begin{center}
\includegraphics[scale=0.8]{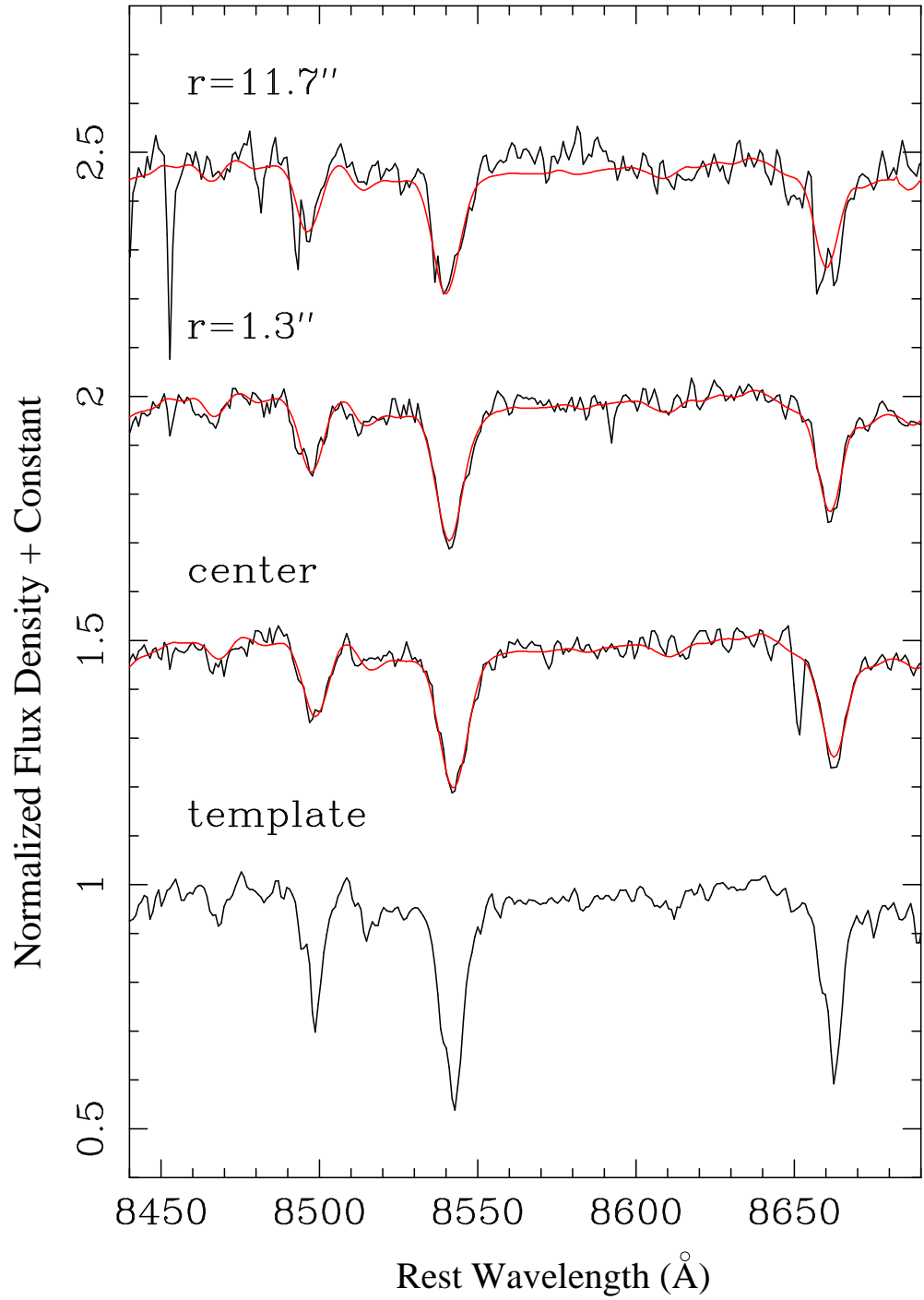}
\end{center}

\caption{As Figure \ref{SpectraHST} but for the ground-based spectra
obtained with Modspec at MDM. \label{SpectraGround}}

\end{figure}

%%%%%%%%%%%%%%%%%%%%%%%%%%%%%%%%%%%%%%%%%%%%%%%%%%%%%%%%%%%%%%%%%%%%%%%%%%%
% Fig VSigmaH3H4

\begin{figure}

\begin{center}
\includegraphics[scale=0.8]{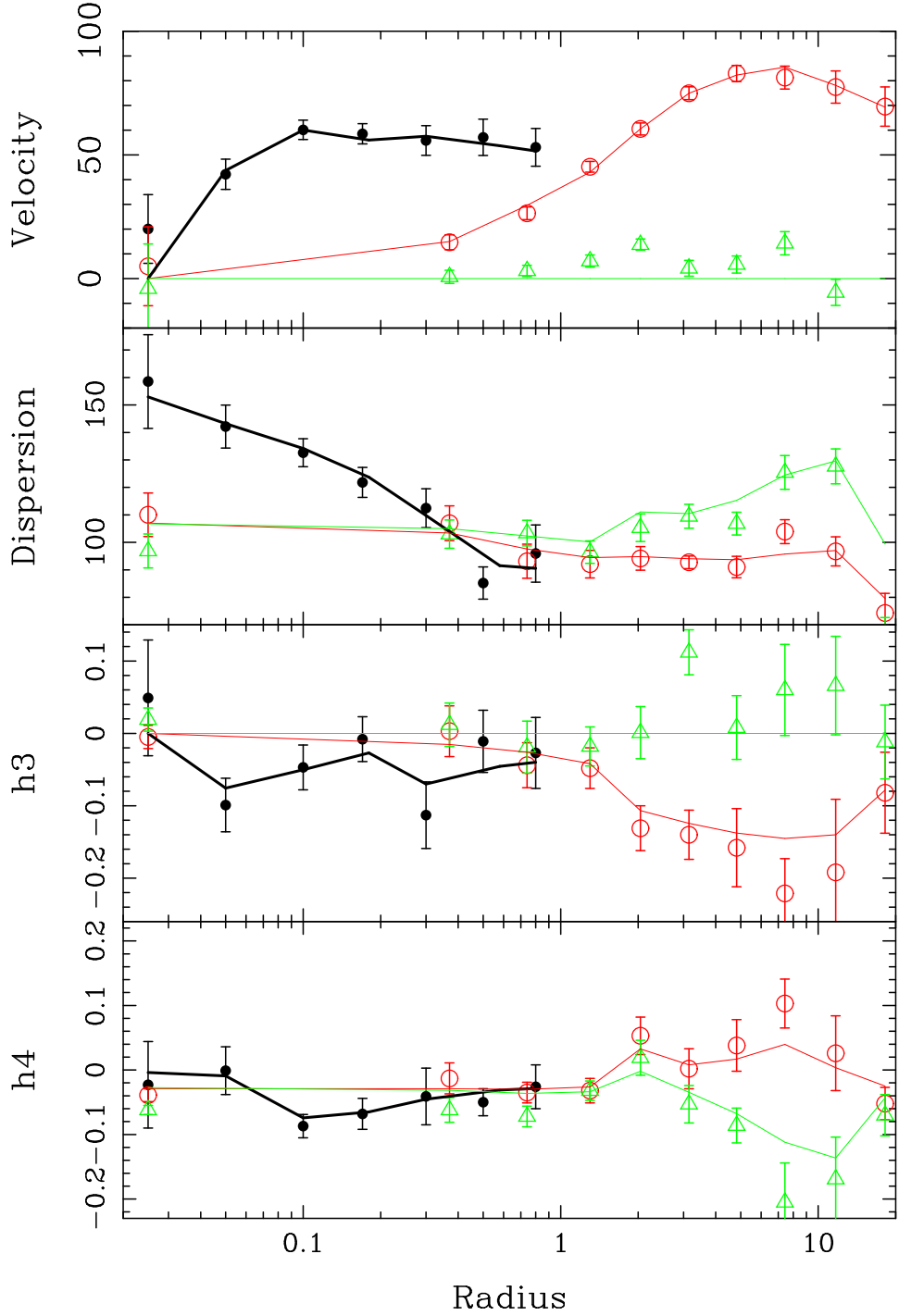}
\end{center}

\caption{Gauss-Hermite moments of the symmetrized LOSVDs as a function
  of position for the data and one model.  Black filled circles are
from the STIS data on the major axis.  Red open circles are ground-based 
major axis data.  Green triangles are ground-based 
minor axis data.  The solid lines represent the analogous quantities
derived from the LOSVDs of the base model described in \S 5.1.  
 \label{VSigmaH3H4}}

\end{figure}

%%%%%%%%%%%%%%%%%%%%%%%%%%%%%%%%%%%%%%%%%%%%%%%%%%%%%%%%%%%%%%%%%%%%%%%%%%%
% Fig CumPhotomCenter

\begin{figure}

\begin{center}
\plotone{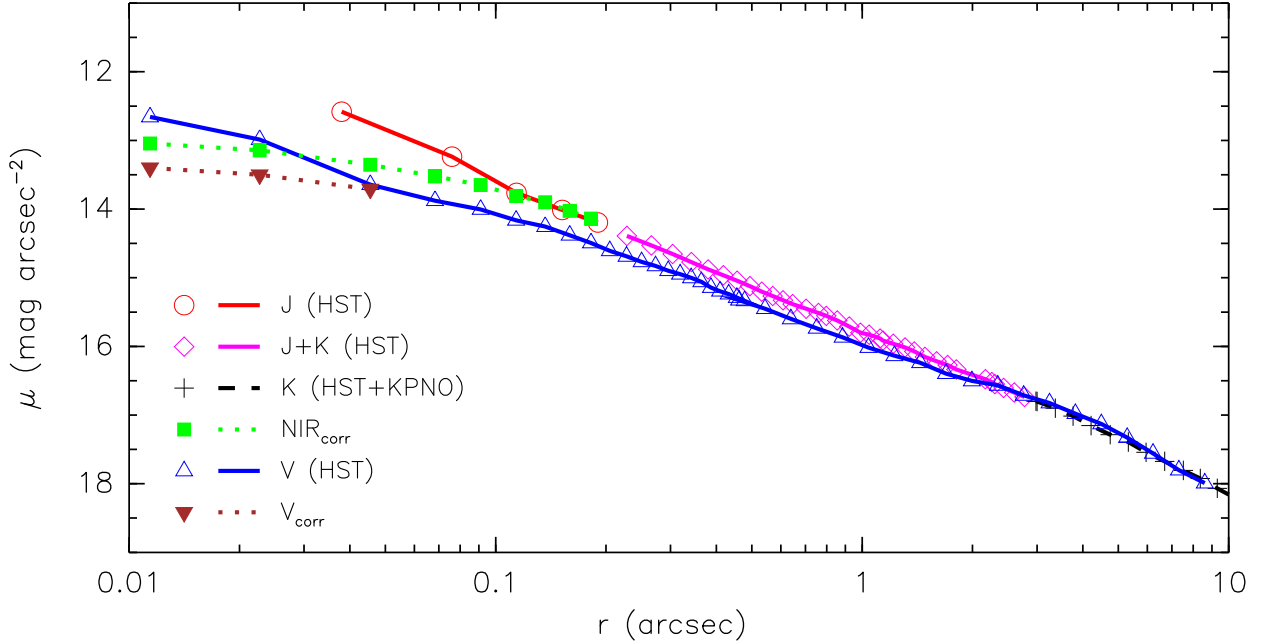}
\end{center}

\caption{NGC 4258 major-axis surface photometry: the central
10\arcsec. Photometric calibration is as in Figure
\ref{CumPhotom}. For $r \la 2\arcsec$ there is a color gradient
between the $V$ and the NIR ($J$, $J$+$K$, and $K$ spliced together)
profiles. The \Vcor\ profile is identical to $V$ except the inner
three points are replaced with an arbitrary ``core'' profile. The
\NIRcor\ profile is created from NIR by replacing the points inward of
$r=0\farcs2$ with points that follow the \Vcor\ profile, shifted by an
amount equal to the color difference between NIR and \Vcor\ at
$r=0\farcs2$. All profiles extend to $r>10\arcsec$ as shown in Figure
\ref{CumPhotom}. In the \NIRcor\ and \Vcor\ plots, only the points
that differ from the NIR and $V$ profiles, respectively, are
shown. \NIRcor\ and \Vcor\ are used to probe the effect of an
unresolved AGN on the black hole mass, as discussed in \S\ref{AmbigP};
they are meant to represent limiting cases, \emph{not} accurate
corrections to the AGN contamination. \label{CumPhotomCenter}}

\end{figure}

%%%%%%%%%%%%%%%%%%%%%%%%%%%%%%%%%%%%%%%%%%%%%%%%%%%%%%%%%%%%%%%%%%%%%%%%%%%
% Fig EW

\begin{figure}

\begin{center}
\plotone{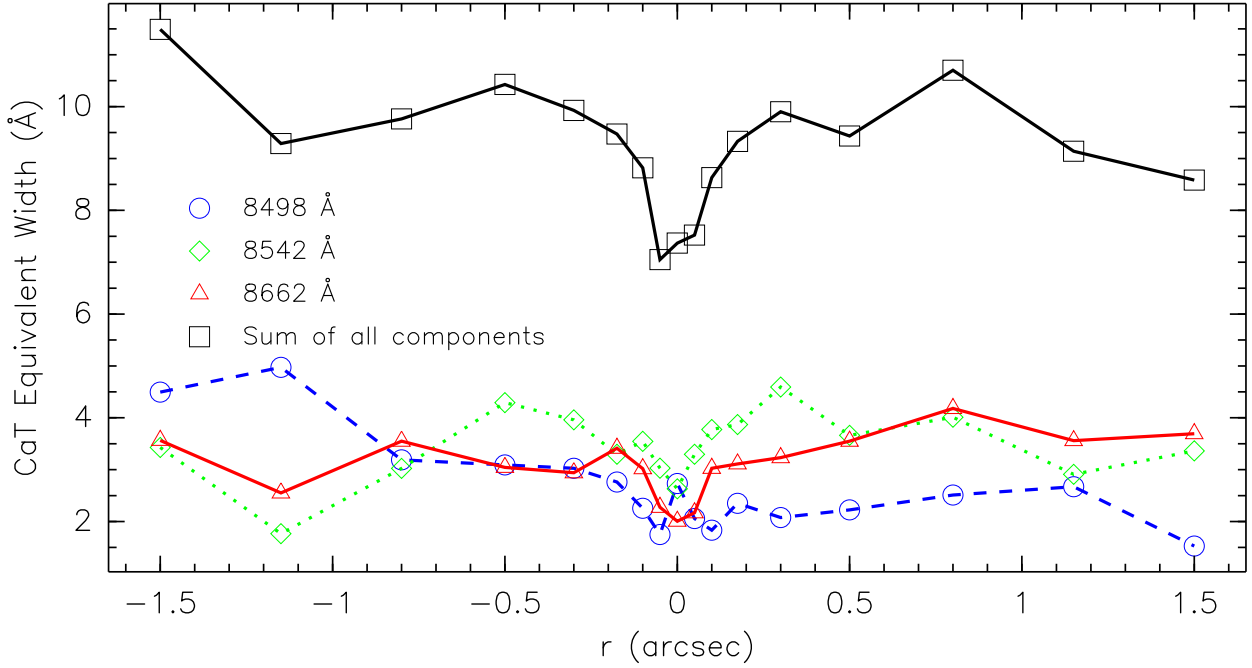}
\end{center}

\caption{Equivalent width (EW) profiles along the major axis for each component of the \ion{Ca}{2} triplet (CaT) line, and for the sum of their EWs. All data points are computed from the STIS spectrum. The subtraction of the emission line near the center (see Figure \ref{SpectraHST}) may have affected the EW estimate for the 8662 \AA\ component. The 8498 \AA\ component, although furthest away from the emission line, has a low S/N and hence its EW profile may be less reliable. Furthermore, the S/N of the spectra drops rapidly with increasing radius. Nevertheless, the summed profile shows clearly that the spectrum within $r < 0\farcs2$ is contaminated at the $\sim 30\%$ level by some source of smooth, lineless continuum light, such as an AGN or OB stars. The positive-$r$ direction corresponds to the approaching (SE) side of the galaxy. \label{EW}}

\end{figure}

\clearpage

%%%%%%%%%%%%%%%%%%%%%%%%%%%%%%%%%%%%%%%%%%%%%%%%%%%%%%%%%%%%%%%%%%%%%%%%%%%
% Fig Models

\begin{figure}

\begin{center}
\plotone{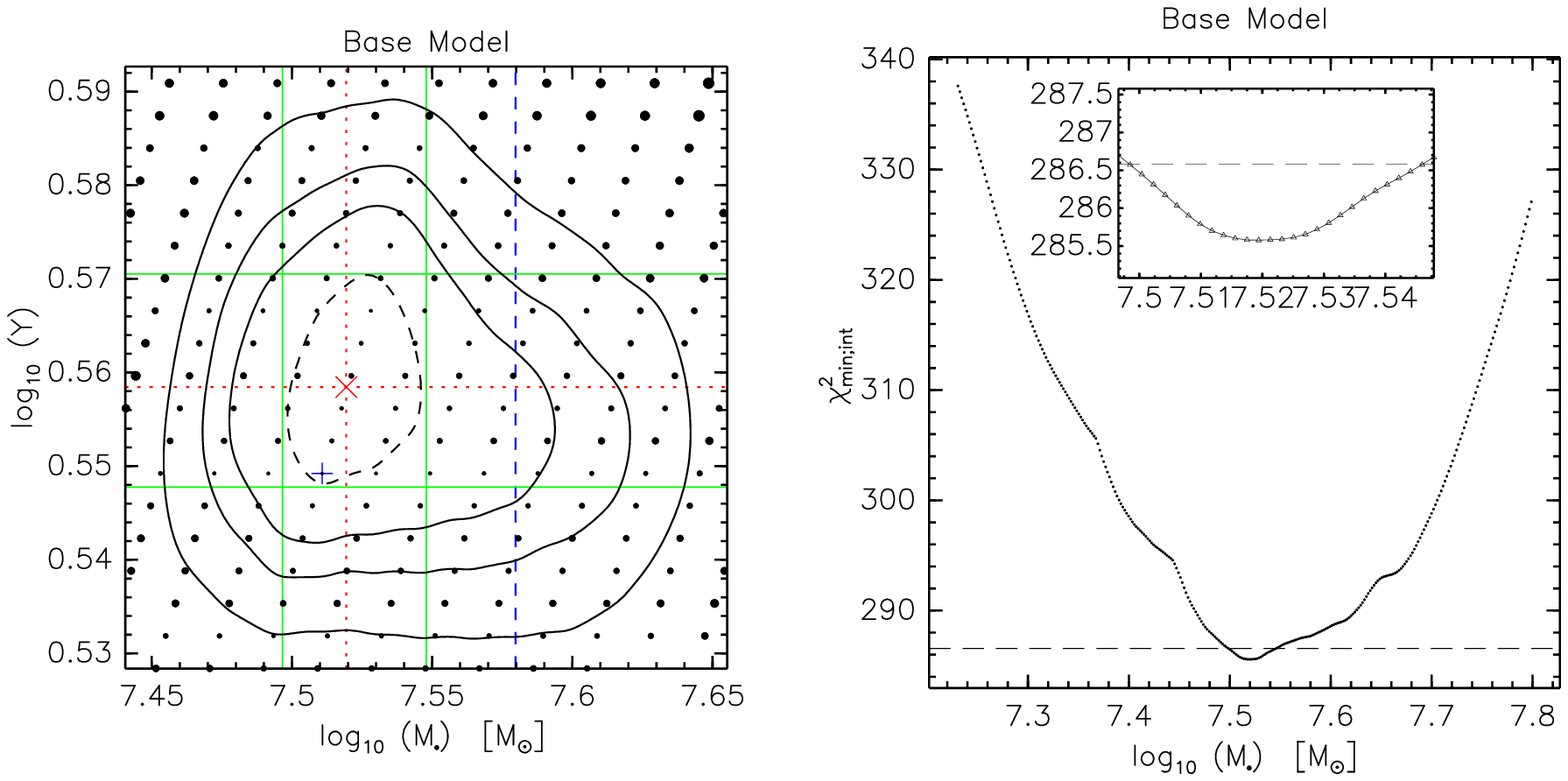}
\plotone{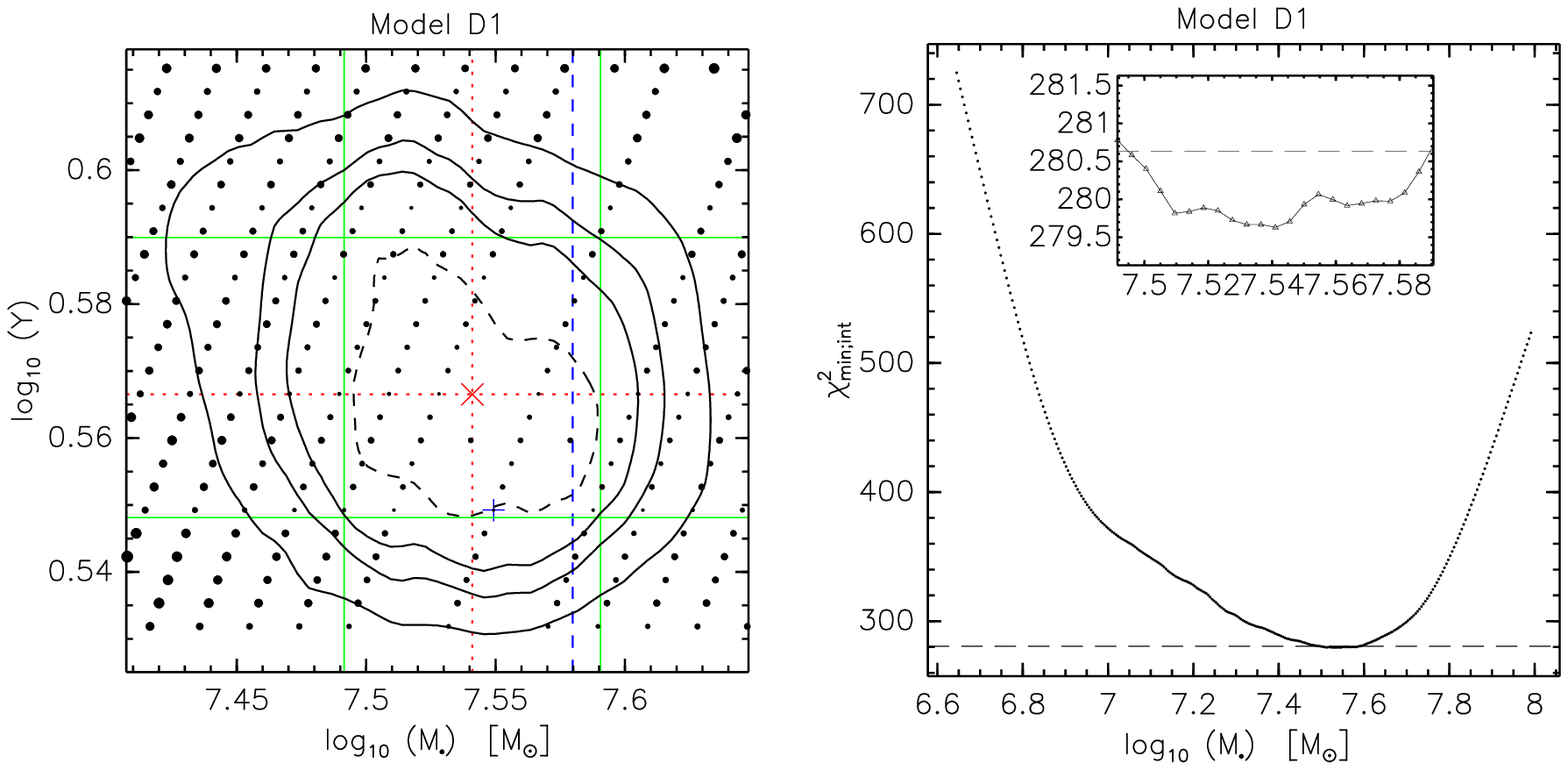}
\end{center}

\caption{{\slshape Left:} Contour maps of $\chi^2(\MBH,\ML)$ for the
  dynamical models bearing the parameters listed in Table
  \ref{ModelTable}. Model names are as in Table \ref{ModelTable}. The
  stellar $M/L$ ratio ($\ML$) refers to the $V$ band. Each dot
  represents a model, and dot size is proportional to the value of
  $\chi^2$ for that model. The symbol ``+'' corresponds to the
  ``best'' model ($\chi^2_{\rm min}$). The symbol ``X''
  corresponds to the minimum interpolated $\chi^2$.
  The contour levels are drawn for
  $\Delta\chi^2 \equiv \chi^2-\chi^2_{\rm min;int} =$ 1.0 (dashed),
  2.71, 4.0, and 6.63, corresponding to confidence levels of 68\%,
  90\%, 95\%, and 99\%, respectively, for one degree of freedom. The
  horizontal and vertical solid lines indicate the nominal
  ``1$\sigma$'' one-dimensional uncertainties. The vertical dashed line
  indicates the maser mass. {\slshape Right:} Values of $\chi^2_{\rm
    min;int}$ marginalized over the $\ML$ axis. The horizontal dashed
  line indicates the $\Delta\chi^2=1$ level. 
 \label{ModelPlots}}

\end{figure}

\begin{figure}
\figurenum{\ref{ModelPlots}}
\begin{center}
\plotone{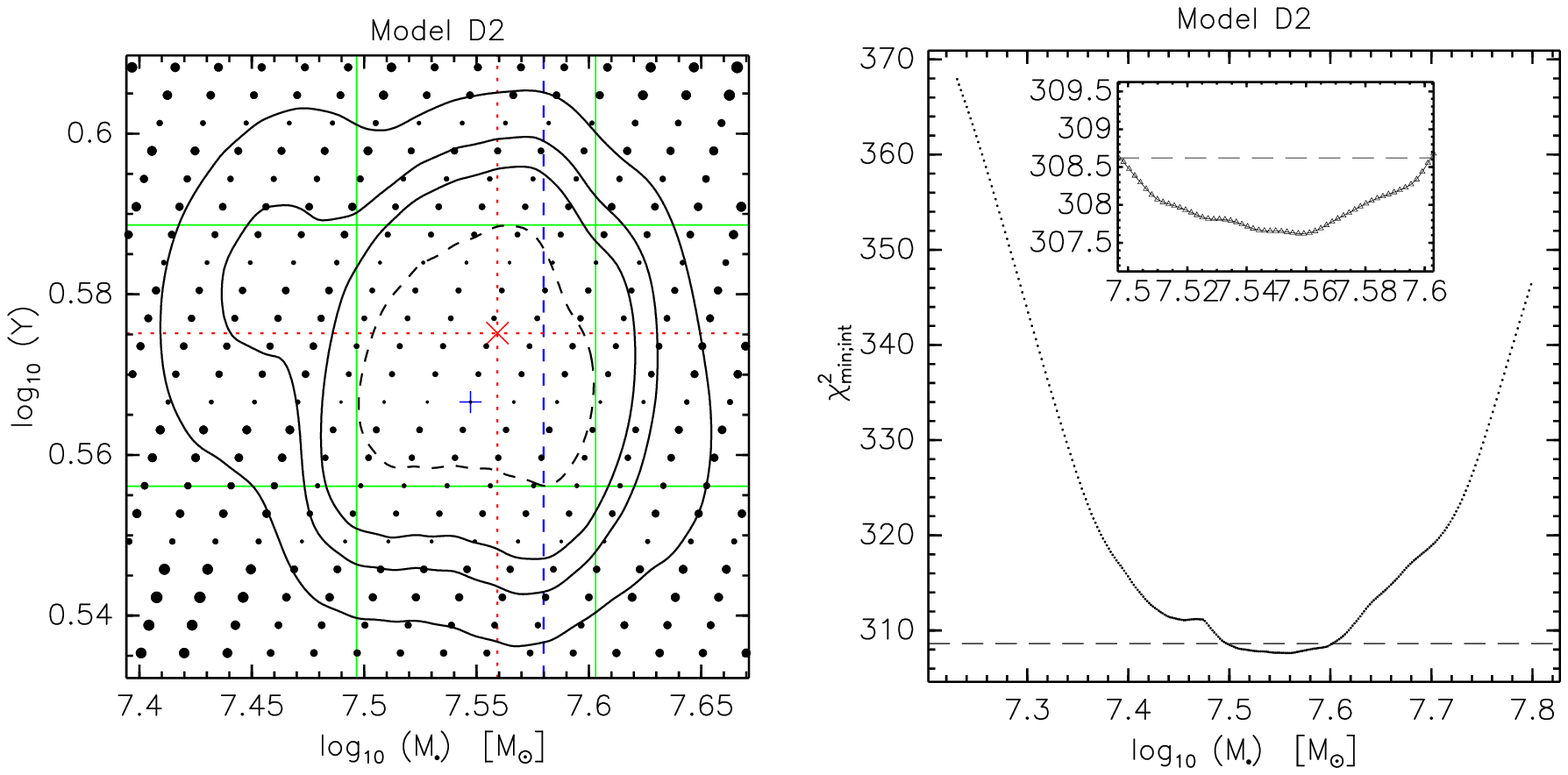}
\plotone{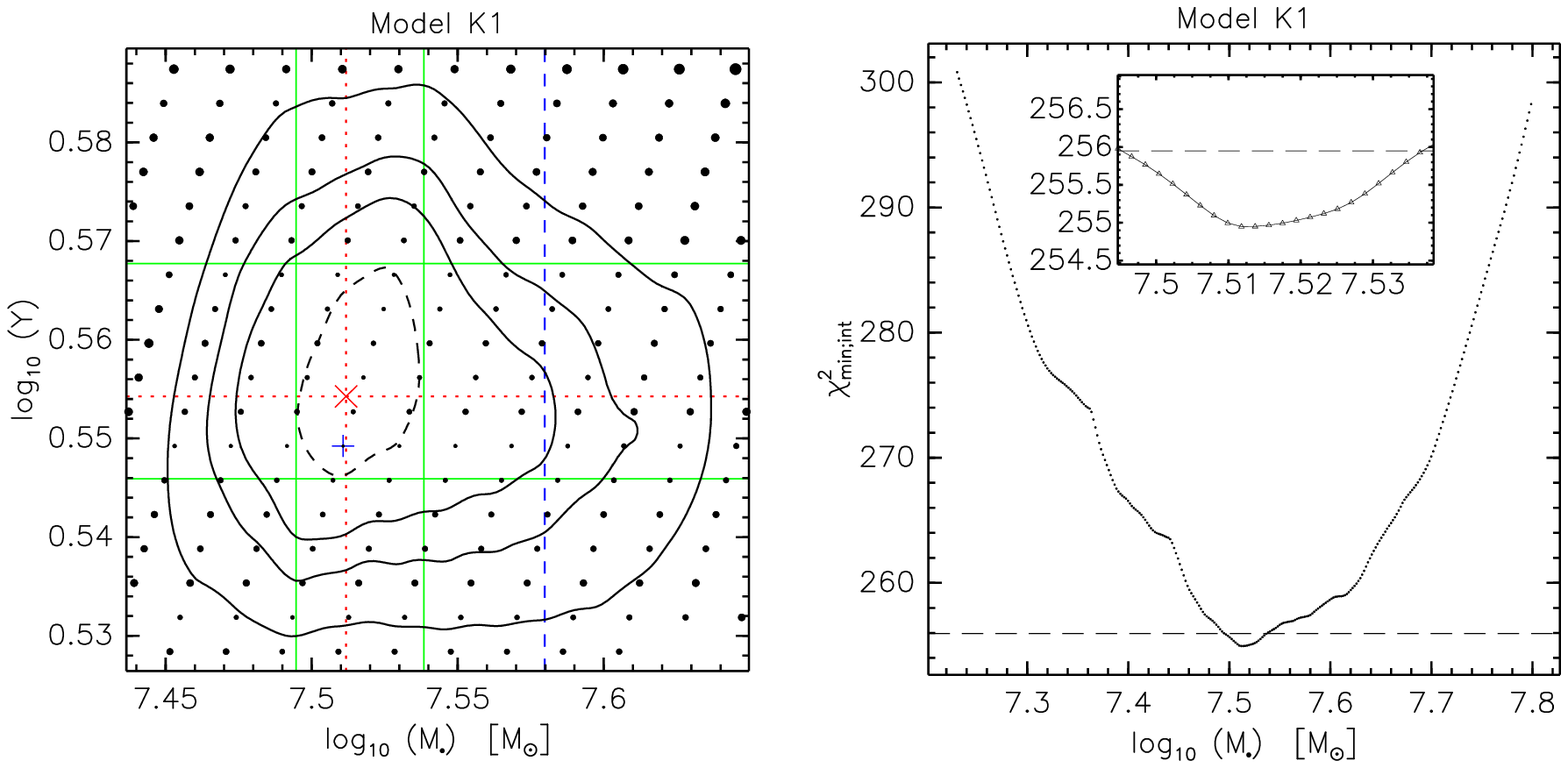}
\end{center}
\caption{Continued}
\end{figure}

\begin{figure}
\figurenum{\ref{ModelPlots}}
\begin{center}
\plotone{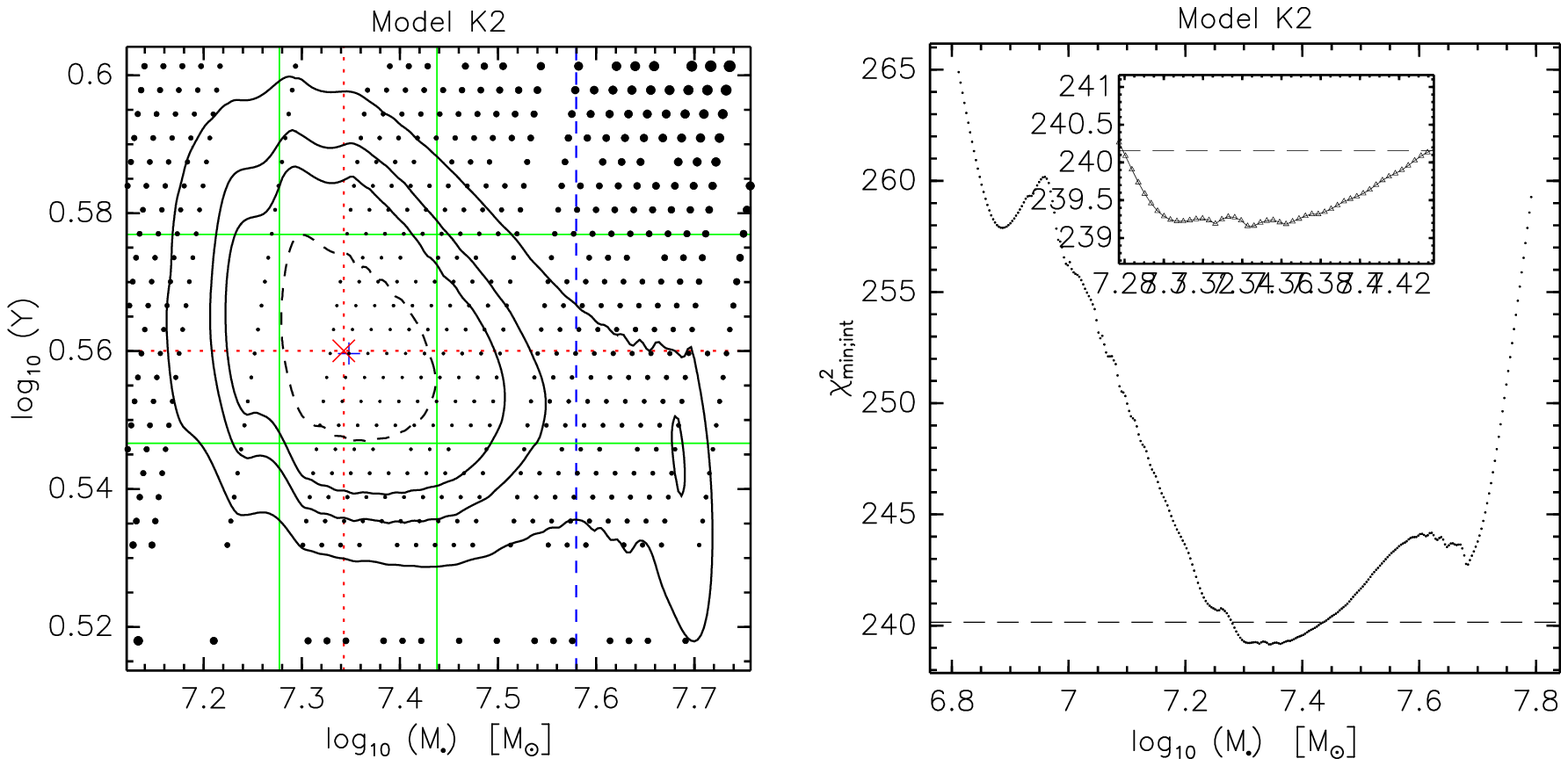}
\plotone{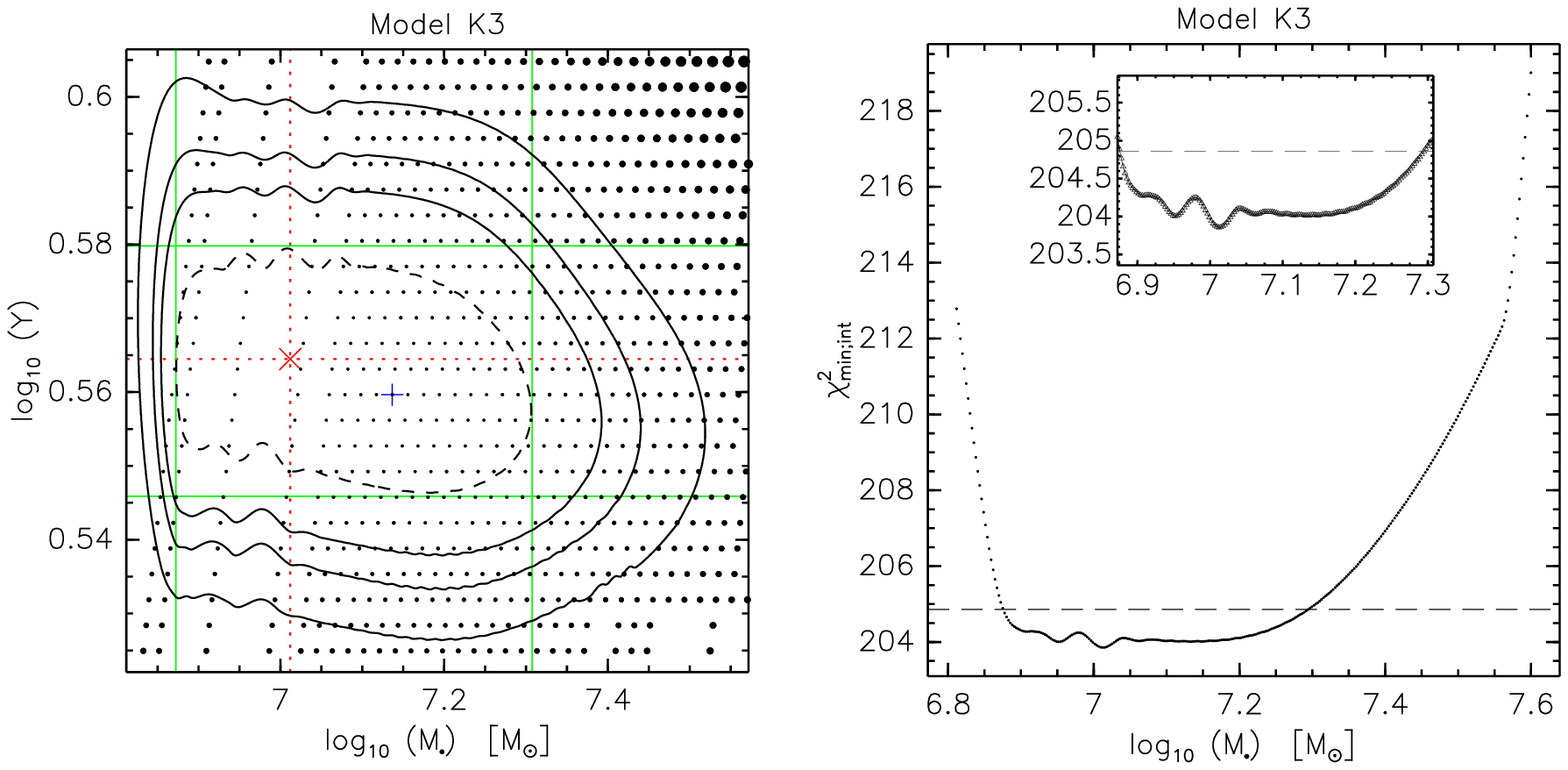}
\end{center}
\caption{Continued}
\end{figure}

\begin{figure}
\figurenum{\ref{ModelPlots}}
\begin{center}
\plotone{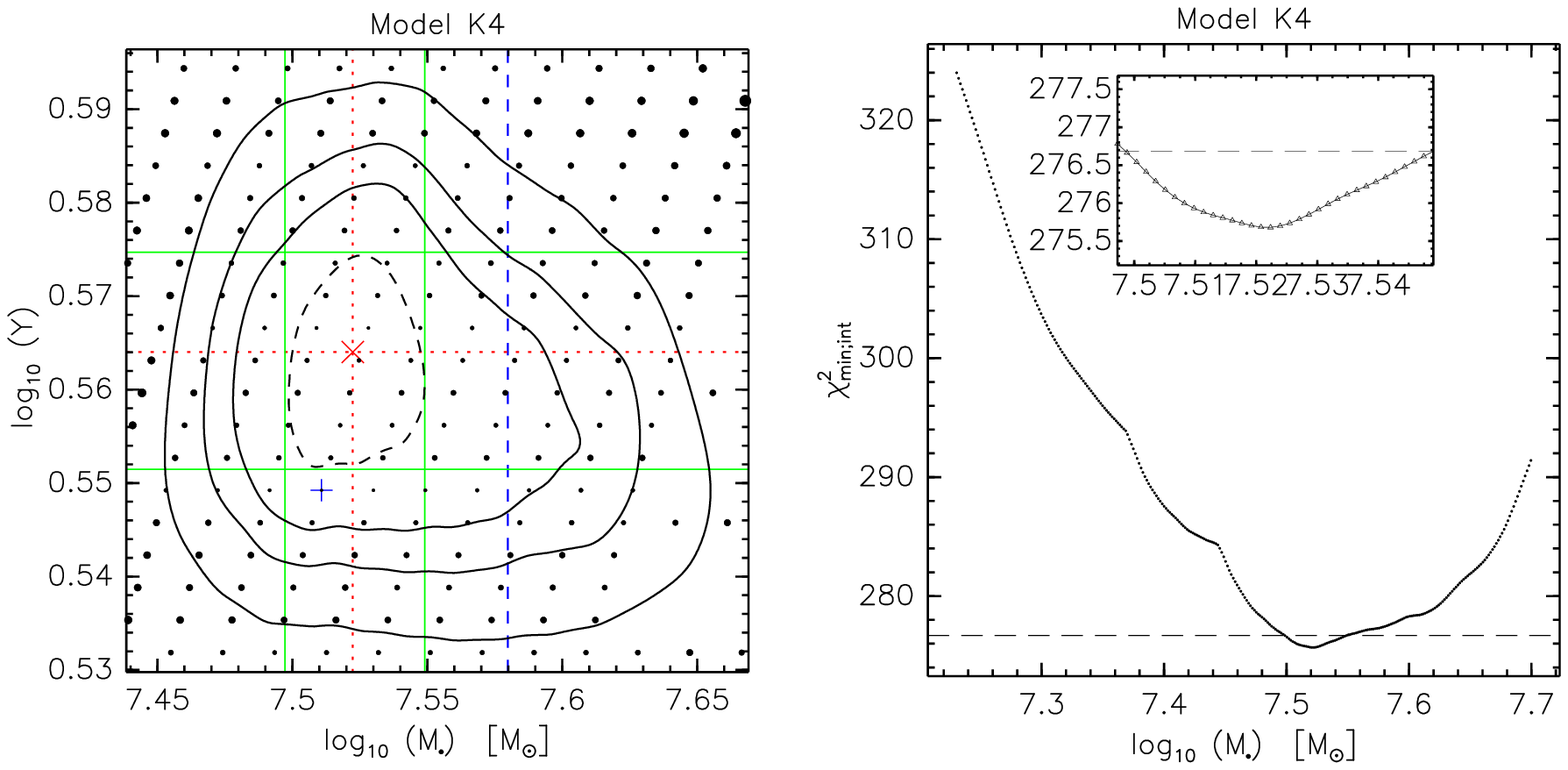}
\plotone{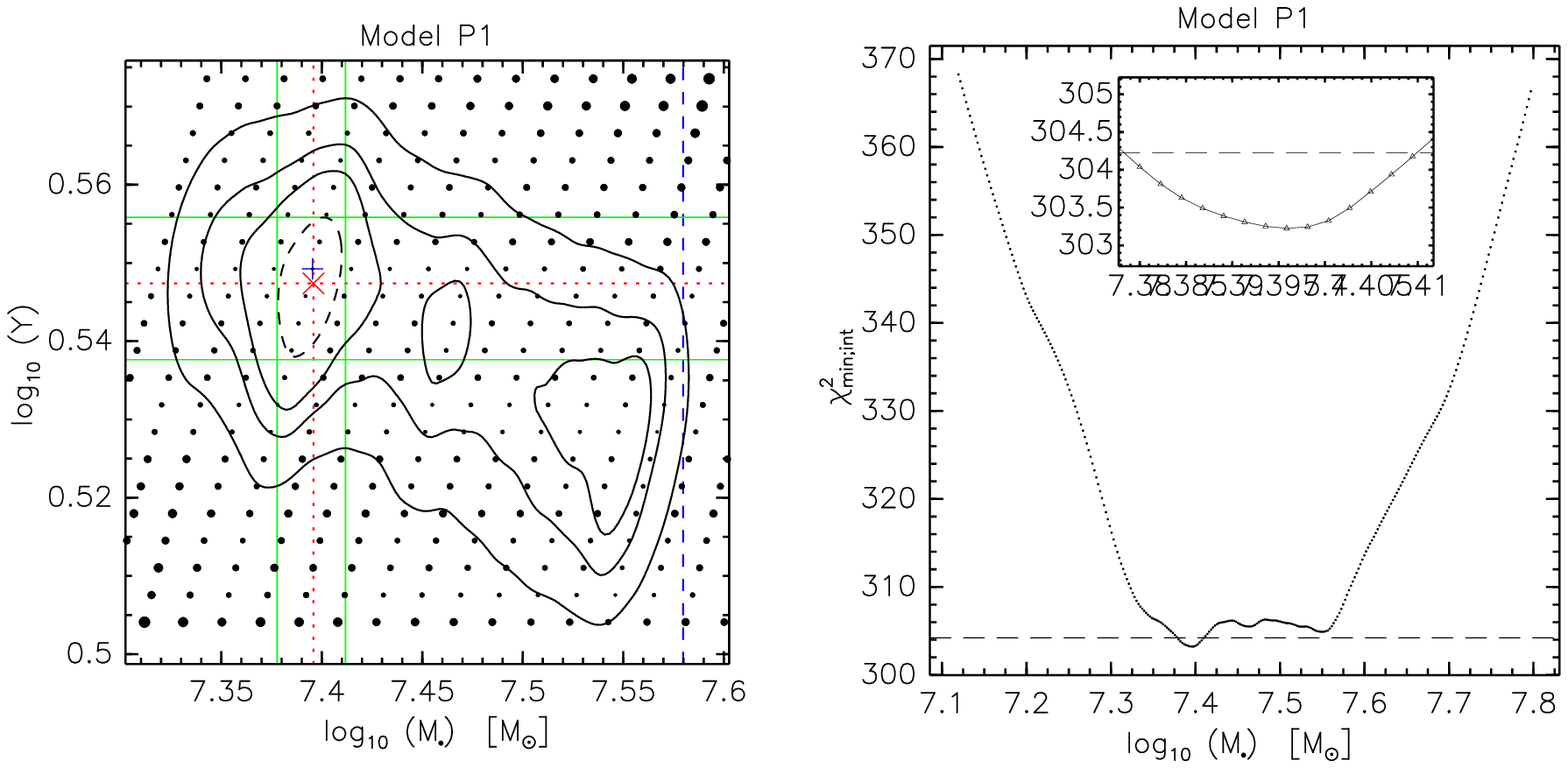}
\end{center}
\caption{Continued}
\end{figure}

\begin{figure}
\figurenum{\ref{ModelPlots}}
\begin{center}
\plotone{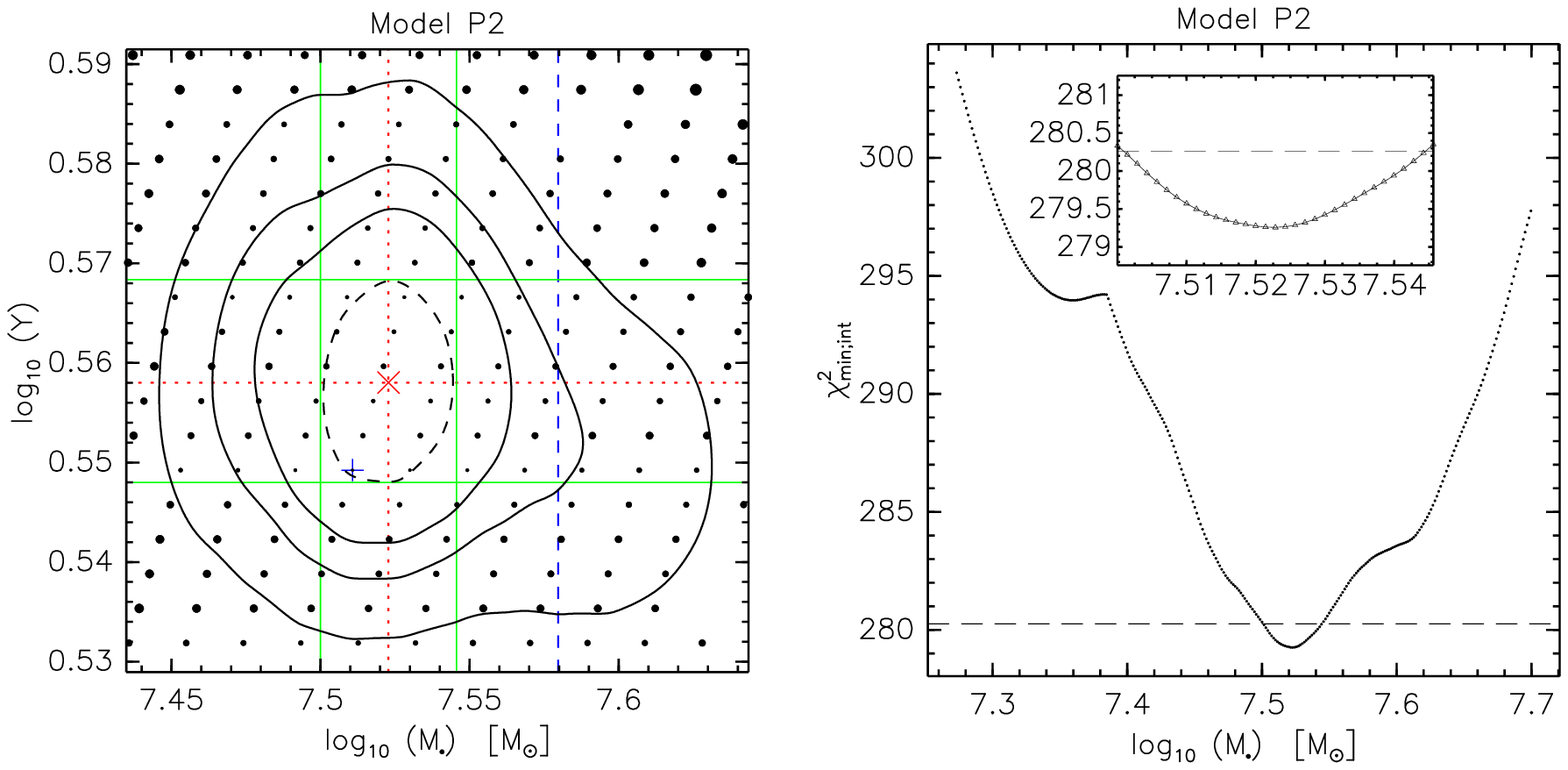}
\plotone{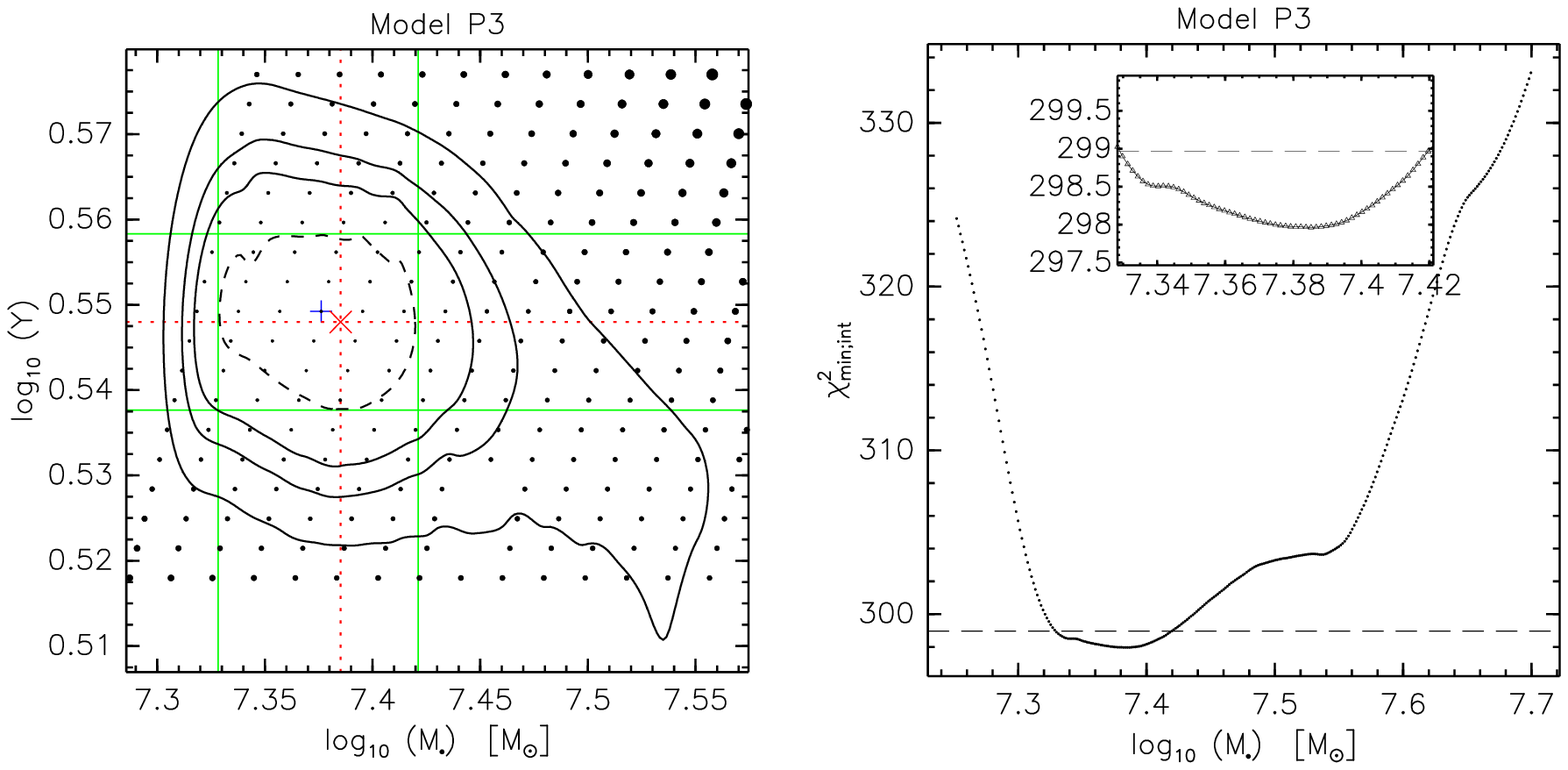}
\end{center}
\caption{Continued}
\end{figure}

\begin{figure}
\figurenum{\ref{ModelPlots}}
\begin{center}
\plotone{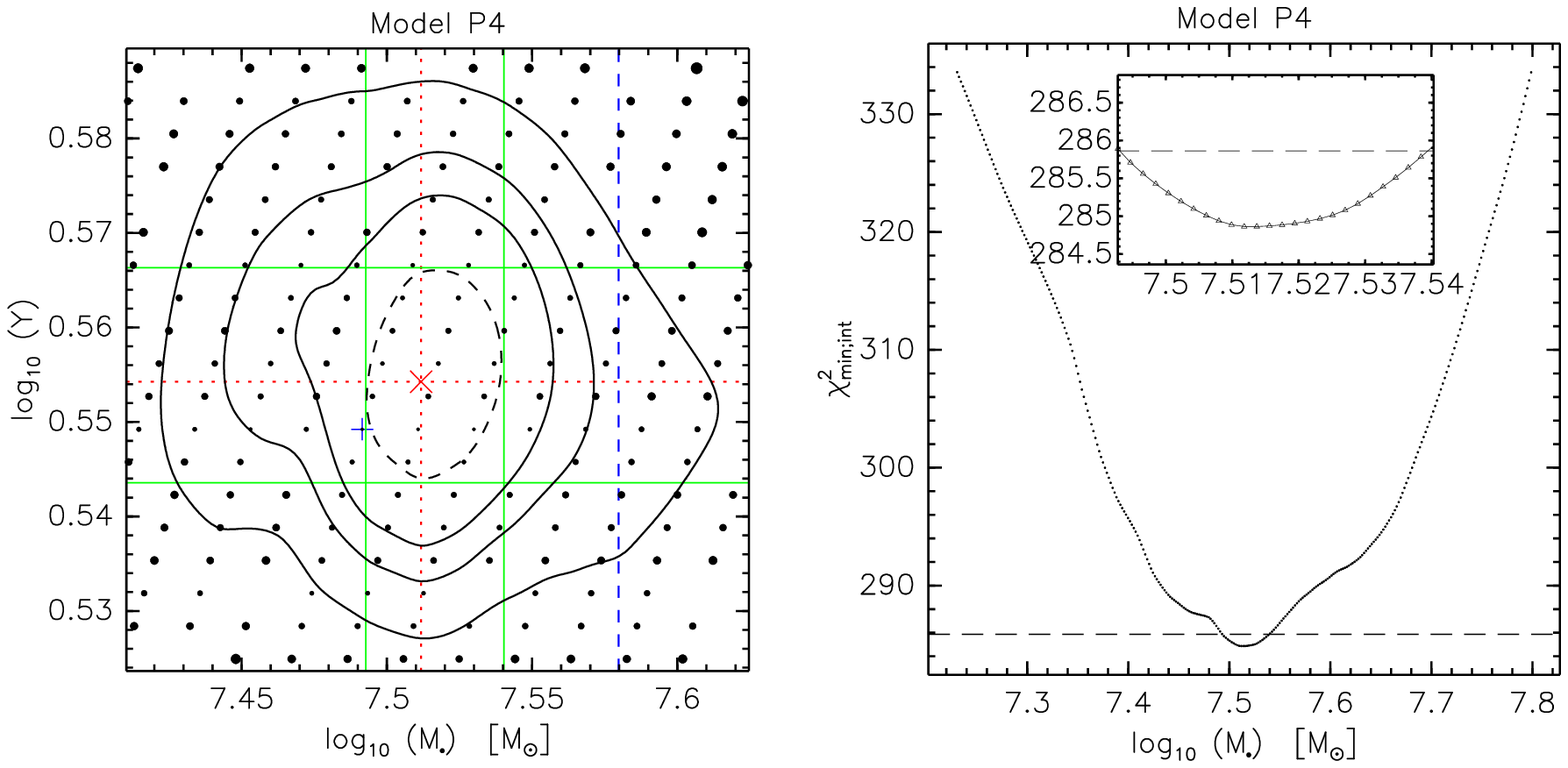}
\end{center}
\caption{Continued}
\end{figure}

\clearpage

%%%%%%%%%%%%%%%%%%%%%%%%%%%%%%%%%%%%%%%%%%%%%%%%%%%%%%%%%%%%%%%%%%%%%%%%%%%
% Fig moments 

\begin{figure}

\begin{center}
\plotone{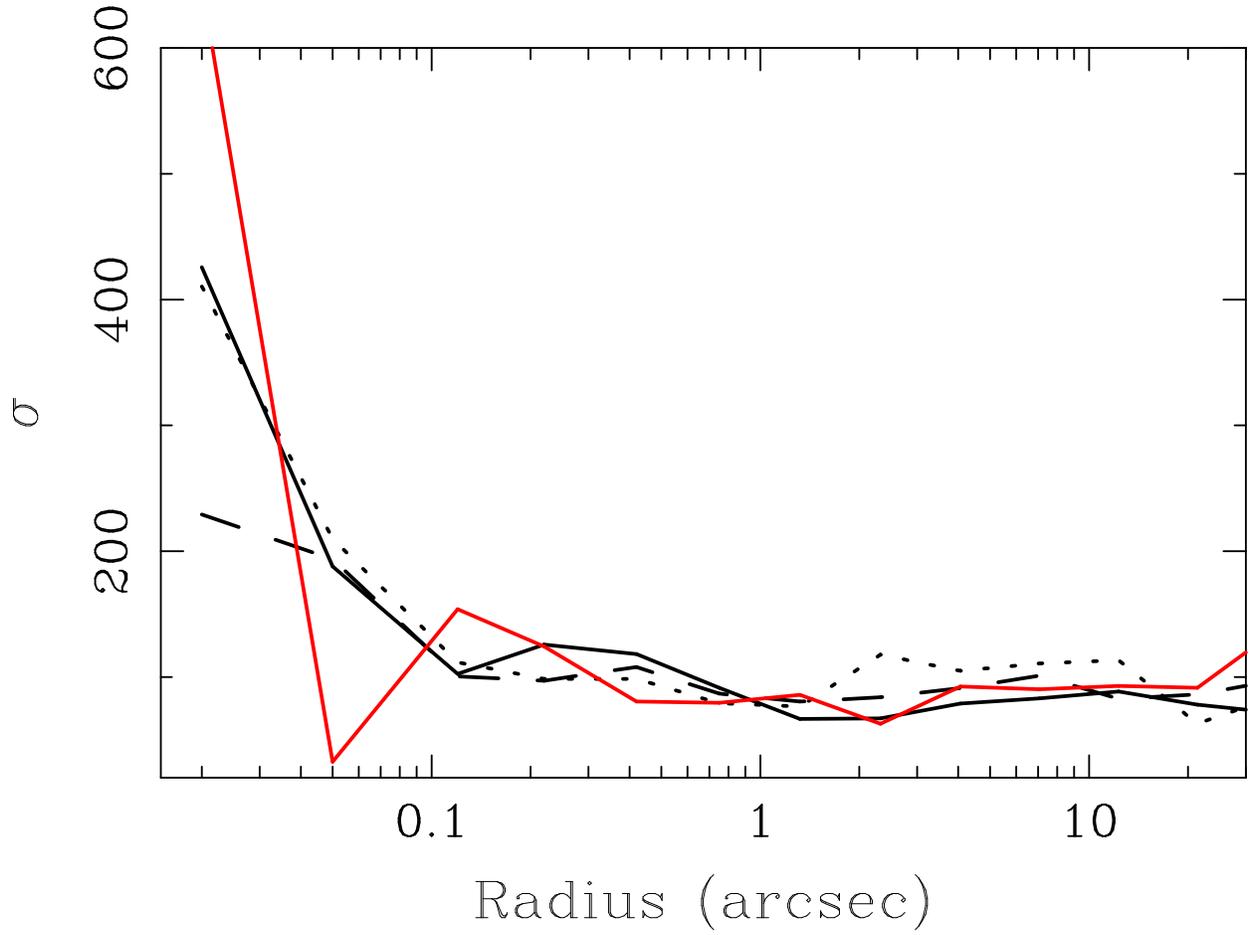}
\end{center}

\caption{
Internal moments of the baseline model in the equatorial plane.  
The red line illustrates the mean rotational velocity 
$\langle v_\phi \rangle$ and the dotted line is the 
velocity dispersion in the $\phi$ direction 
$\sigma_\phi = \sqrt{ \langle v_\phi^2 \rangle  - 
\langle v_\phi \rangle ^2}$.  
The solid line is the radial dispersion $\sigma_r$  
and the dashed line is $\sigma_\theta$.  
\label{Internalmoments}}

\end{figure}

%%%%%%%%%%%%%%%%%%%%%%%%%%%%%%%%%%%%%%%%%%%%%%%%%%%%%%%%%%%%%%%%%%%%%%%%%%%
% Fig mass profile 
\begin{figure}

\begin{center}
\plotone{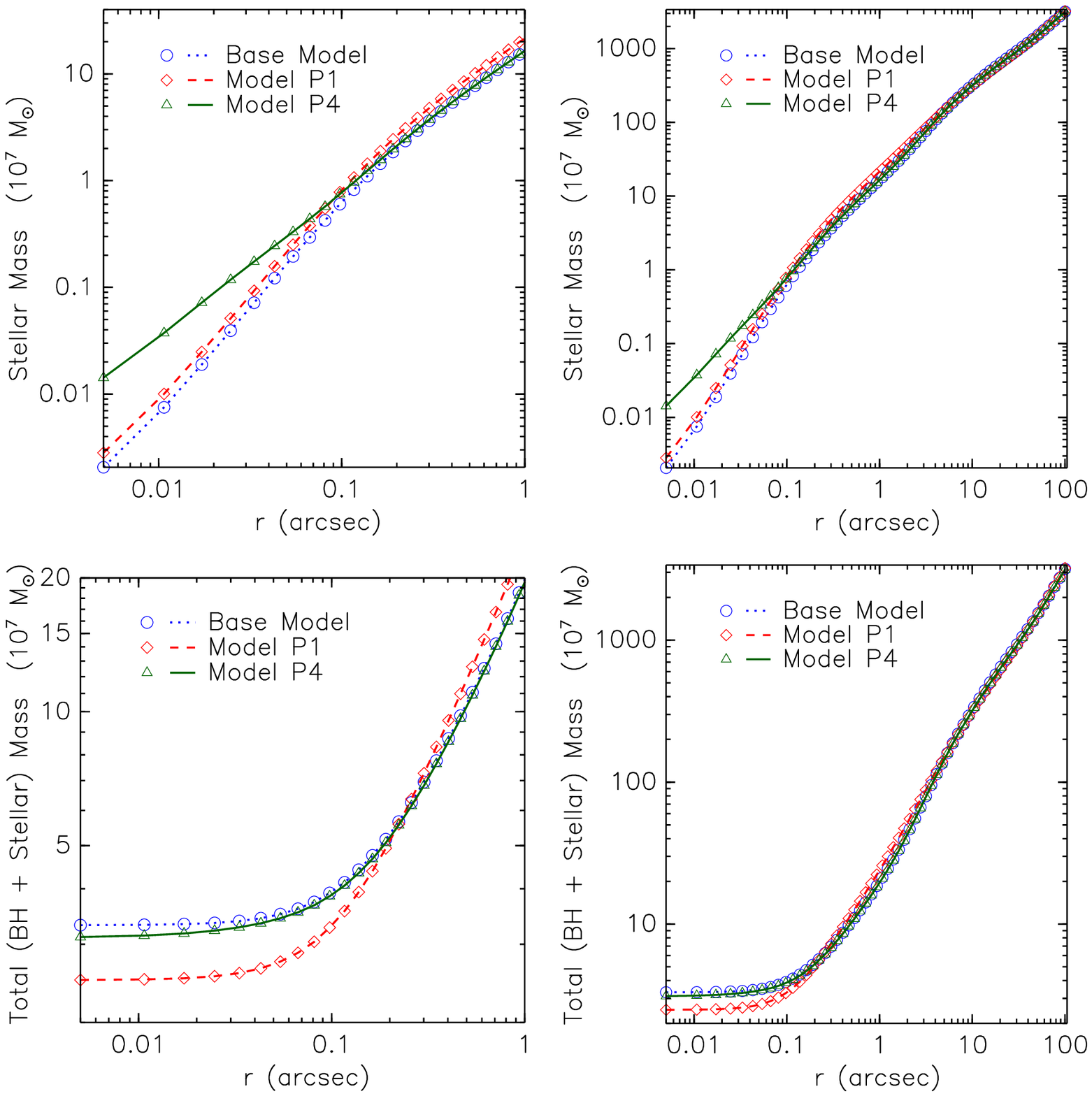}
\end{center}

\caption{The enclosed stellar mass $M_*$ (top row), and the sum of the
  black hole mass and the enclosed stellar mass $\MBH+M_*$ (bottom
  row), as a function of model radius. The base model and model P4 are
  constrained by the same kinematic dataset, and they share the same
  mass $V$-band profile for $r > 0\farcs04$ (see Table
  \ref{ModelTable}). Because of the degeneracy between $\MBH$ and
  $M_*$ within a small radius $r_0 \approx 0\farcs1$ from the center,
  comparable to the spatial resolution of the kinematic data, the
  modeling algorithm can only constrain the total mass inside that
  radius. As a result, model P4 is assigned a slightly lower $\MBH$ to
  compensate for the excess mass contained in the central ``bump'' of
  its mass profile. Model P1 is characterized by the \NIRcor\ mass
  profile, which differs from that of both P4 and the base model at
  $r >  r_0$. \label{MassProfile}}

\end{figure}

%%%%%%%%%%%%%%%%%%%%%%%%%%%%%%%%%%%%%%%%%%%%%%%%%%%%%%%%%%%%%%%%%%%%%%%%%%%

\begin{figure}

% Fig Appendix MCtest

\begin{center}
\includegraphics[angle=-90,scale=0.7]{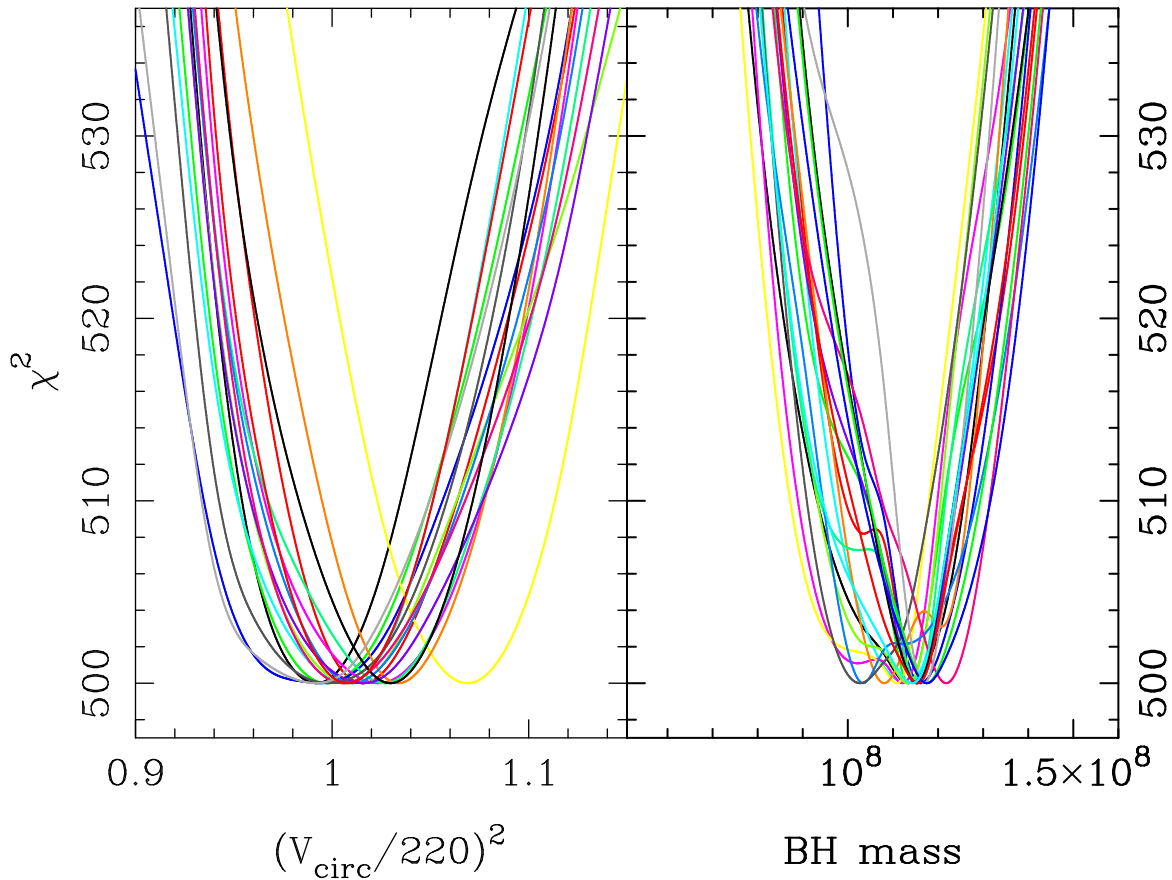}
\end{center}

\caption{$\chi^2$ profiles of orbit-based models of the Monte Carlo
  model described in the Appendix A.3. The left panel illustrates the
  recovery of the circular velocity of the potential for 20 different
  realizations of its LOSVDs.  The right panel illustrates the
  recovery of the black hole mass from the same set of
  realizations. The plots in each panel are marginalized over the
  variable in the other panel.  The mean of the derived black hole
  masses is $1.15 \times 10^8 \MSun$, consistent with the model
  mass. \label{MCtest}}

\end{figure}

%%%%%%%%%%%%%%%%%%%%%%%%%%%%%%%%%%%%%%%%%%%%%%%%%%%%%%%%%%%%%%%%%%%%%%%%%%%
% TABLES
%%%%%%%%%%%%%%%%%%%%%%%%%%%%%%%%%%%%%%%%%%%%%%%%%%%%%%%%%%%%%%%%%%%%%%%%%%%
% Table PhotoDataTable

\begin{deluxetable}{lccl}
\tablewidth{0pt}
\tablecaption{\sc Journal of Imaging Observations of NGC 4258 \label{PhotoDataTable}}
\tablehead{
\colhead{Instrument}  &
\colhead{Filter/Band}  &
\colhead{Radial Range}  &
\colhead{Source} 
}
\startdata
{\slshape HST} NICMOS\tablenotemark{a} & F110W, F160W, F222M\tablenotemark{b} & $r < 9\arcsec$ & \citet{ChaEtal00} \\
KPNO\tablenotemark{c} \, 2.1-m & $K$ & $2\arcsec \leq r < 32\arcsec$ & This paper \\
2MASS\tablenotemark{d} & $J$, $H$, $K$ & $r \ge 32\arcsec$ & \citet{JarEtal03} \\
{\slshape HST} WFPC2\tablenotemark{e} & F791W\tablenotemark{f} & $r < 9\arcsec$ & \cite{CecEtal00} \\
MDM 1.3-m & $R$ & $r \ge 9\arcsec$ & This paper \\
{\slshape HST} WFPC2 & F547M\tablenotemark{g} & $r < 9\arcsec$ & This paper \\
\enddata
\tablenotetext{a}{Near Infrared Camera and Multi-Object Spectrometer.}
\tablenotetext{b}{Corresponding, approximately, to the $J$, $H$, and $K$ bandpasses, respectively.}
\tablenotetext{c}{Kitt Peak National Observatory.}
\tablenotetext{d}{Two Micron All Sky Survey.}
\tablenotetext{e}{Wide Field and Planetary Camera 2.}
\tablenotetext{f}{Corresponding, approximately, to the $I$ bandpass.}
\tablenotetext{g}{Corresponding, approximately, to the $V$ bandpass.}
\end{deluxetable}

%%%%%%%%%%%%%%%%%%%%%%%%%%%%%%%%%%%%%%%%%%%%%%%%%%%%%%%%%%%%%%%%%%%%%%%%%%%
% Table SpectrConfigsTable

\begin{deluxetable}{llllccc}
\tablewidth{0pt}
\tablecaption{\sc Long-Slit Spectrograph Configurations \label{SpectrConfigsTable}}
\tablehead{
\colhead{Instrument \tablenotemark{a}} &
\colhead{Slit size} &
\colhead{$\lambda_{\mbox{cen}}$ \tablenotemark{b}} &
\colhead{$\lambda$-range \tablenotemark{b}} &
\colhead{Disp$^{-1}$ \tablenotemark{c}} &
\colhead{Comp.\ line $\sigma$ \tablenotemark{d}} &
\colhead{Scale \tablenotemark{e}}  \\
\colhead{+ grating} &
\colhead{$\arcsec \times \arcsec$} &
\colhead{\AA} &
\colhead{\AA} &
\colhead{\AA\ pix$^{-1}$} &
\colhead{\AA\ (\kms)} &
\colhead{\arcsec ~pix$^{-1}$}
}
\startdata
STIS + G750M     & \phn52$\times$0.1 & 8561 & 8275--8847 & 0.554 & 0.45 (12.4) & 0.051 \\
Wilbur + 831g/mm & 540$\times$0.9 & 8500 & 8100--9020 & 0.90\phn & 0.9 (32) & 0.37\phn \\
Echelle + 831g/mm & 540$\times$0.9 & 8500 & 8100--9020 & 1.46\phn & 0.9 (32) & 0.59\phn
\enddata
\tablenotetext{a}{Wilbur and Echelle are two of the CCDs used on the
  Modspec instrument at the MDM 2.4-m telescope.}
\tablenotetext{b}{Central wavelength and wavelength range. STIS values
  taken from the STIS Instrument Handbook \citep[pp.\ 231,
  234]{LeiEtal01}. For Modspec, we show the range extracted.}
\tablenotetext{c}{Reciprocal dispersion was measured using our own
  wavelength solutions. The distribution of dispersion solutions for a
  given dataset had a $\sigma \approx 0.00015$~\AA\ pix$^{-1}$. The
  average dispersion given in the Handbook for G750M is 0.56~\AA\
  pix$^{-1}$.}  \tablenotetext{d}{Instrumental line widths measured by
  fitting Gaussians to emission lines on comparison lamp
  exposures. This gives an estimate of the instrumental line width for
  {\em extended} sources.  We use $\sim$5 lines per exposure, and at
  least 5 measurements per line. All of the G750M observations used an
  unbinned 1024$\times$1024 pixel CCD. \citet{LeiEtal01} (p.\ 300)
  give an instrumental line width for {\em point} sources of
  $\sigma$=13.3 \kms\ for the STIS + G750M configuration.}
\tablenotetext{e}{The spatial scale along the slit is constant, but it
  varies across the slit from one grating to the next.  It is
  0\farcs05597 pix$^{-1}$ for G750M at 8561~\AA\ and 0\farcs05465
  pix$^{-1}$ for G750M at 6581~\AA\ (STIS ISR 98-23).}
\tablecomments{Some numbers for the CCDs Wilbur and Echelle on Modspec
  are calculated by the program ``modset'' by J.\ Thorstensen. The
  slit width, spectral resolution, and spatial resolution varied for
  the Modspec observations; typical measured values are shown here.}
\end{deluxetable}

%%%%%%%%%%%%%%%%%%%%%%%%%%%%%%%%%%%%%%%%%%%%%%%%%%%%%%%%%%%%%%%%%%%%%%%%%%%
% Table SpectrTable

\begin{deluxetable}{rrrrr}
%\tablewidth{0pt}
\tablecolumns{5}
\tablecaption{\sc Kinematic Parameter Profiles of NGC 4258
\label{SpectrTable}}
\tablehead{
\colhead{$R$ (\arcsec)} &
\colhead{$V$} &
\colhead{$\sigma$} &
\colhead{$H3$} &
\colhead{$H4$}
}
\startdata
\sidehead{{\slshape HST}\, STIS -- Major axis profile
(PA = 140\arcdeg) }
\tableline
0.00 & 20   $\pm$ 14  & 159  $\pm$ 17  & 0.05 $\pm$ 0.08 & -0.02 $\pm$    0.07 \\
0.05 &  42  $\pm$ \phn 6  & 142  $\pm$ \phn 8  & -0.10 $\pm$ 0.04 & 0.00 $\pm$ 0.04 \\
0.10 &  60  $\pm$ \phn 4  & 133  $\pm$ \phn 5  & -0.05 $\pm$ 0.03 & -0.09 $\pm$ 0.02 \\
0.17 &  59  $\pm$ \phn 4  & 122  $\pm$ \phn 5  & -0.01 $\pm$ 0.03 & -0.07 $\pm$ 0.02 \\
0.30 &  56  $\pm$ \phn 6  & 112  $\pm$ \phn 7  & -0.11 $\pm$ 0.05 & -0.04 $\pm$ 0.04 \\
0.50 &  57  $\pm$ \phn 7  & 85  $\pm$ \phn 6  & -0.01 $\pm$ 0.04 & -0.05 $\pm$ 0.02 \\
0.80 &  53  $\pm$ \phn 8  & 96  $\pm$ 10  & -0.03 $\pm$ 0.05 & -0.03 $\pm$ 0.03 \\
\sidehead{Ground-based (MDM 2.4-m with ModSpec) -- Major axis profile} 
%(PA = 150\arcdeg)}
\tableline
0.37 &  15  $\pm$ \phn 3  & 107  $\pm$ \phn 6  & 0.00 $\pm$ 0.04 & -0.01 $\pm$ 0.02 \\
0.74 &  26  $\pm$ \phn 3  & 93  $\pm$ \phn 6  & -0.04 $\pm$ 0.03 & -0.04 $\pm$ 0.02 \\
1.30 &  45  $\pm$ \phn 2  & 92  $\pm$ \phn 5  & -0.05 $\pm$ 0.03 & -0.03 $\pm$ 0.02 \\
2.04 &  60  $\pm$ \phn 2  & 94  $\pm$ \phn 4  & -0.13 $\pm$ 0.03 & 0.05 $\pm$ 0.03 \\
3.15 &  75  $\pm$ \phn 3  & 93  $\pm$ \phn 2  & -0.14 $\pm$ 0.03 & 0.00 $\pm$ 0.03 \\
4.82 &  83  $\pm$ \phn 3  & 91  $\pm$ \phn 4  & -0.16 $\pm$ 0.05 & 0.04 $\pm$ 0.04 \\
7.42 &  81  $\pm$ \phn 5  & 104  $\pm$ \phn 4  & -0.22 $\pm$ 0.05 & 0.10 $\pm$ 0.04 \\
11.69 &  77  $\pm$ \phn 7  & 97  $\pm$ \phn 5  & -0.19 $\pm$ 0.10 & 0.03 $\pm$ 0.06 \\
18.18 &  70  $\pm$ \phn 8  & 74  $\pm$ \phn 7  & -0.08 $\pm$ 0.06 & -0.05 $\pm$ 0.03 \\
\sidehead{Ground-based (MDM 2.4-m with ModSpec) -- Minor axis profile}
%(PA = 60\arcdeg)}
\tableline
0.00 & -4  $\pm$ 18  & 97  $\pm$ \phn 6  & 0.02 $\pm$ 0.02 & -0.06 $\pm$ 0.01 \\
0.37 &  1  $\pm$ \phn 3  & 103  $\pm$ \phn 5  & 0.01 $\pm$ 0.03 & -0.06 $\pm$ 0.02 \\
0.74 &  3  $\pm$ \phn 2  & 103  $\pm$ \phn 5  & -0.02 $\pm$ 0.04 & -0.07 $\pm$ 0.02 \\
1.30 &  7  $\pm$ \phn 2  & 96  $\pm$ \phn 4  & -0.02 $\pm$ 0.03 & -0.03 $\pm$ 0.02 \\
2.04 &  14  $\pm$ \phn 2  & 105  $\pm$ \phn 5  & 0.00 $\pm$ 0.04 & 0.02 $\pm$ 0.03 \\
3.15 &  4  $\pm$ \phn 3  & 109  $\pm$ \phn 4  & 0.11 $\pm$ 0.03 & -0.05 $\pm$ 0.03 \\
4.82 &  6  $\pm$ \phn 3  & 107  $\pm$ \phn 4  & 0.01 $\pm$ 0.04 & -0.09 $\pm$ 0.03 \\
7.42 &  14  $\pm$ \phn 5  & 125  $\pm$ \phn 6  & 0.06 $\pm$ 0.06 & -0.21 $\pm$ 0.06 \\
11.69 &  -6  $\pm$ \phn 5  & 128  $\pm$ \phn 6  & 0.07 $\pm$ 0.07 & -0.17 $\pm$ 0.07 \\
18.18 &  -58  $\pm$ 11  & 58  $\pm$ 14  & -0.01 $\pm$ 0.05 & -0.07 $\pm$ 0.03 \\
\enddata
\end{deluxetable}

%%%%%%%%%%%%%%%%%%%%%%%%%%%%%%%%%%%%%%%%%%%%%%%%%%%%%%%%%%%%%%%%%%%%%%%%%%%
% Table ModelTable

\begin{deluxetable}{cllcccccccc}
\tablewidth{0pt}
\tablecaption{\sc Dynamical Models \label{ModelTable}}
\tablehead{
\colhead{} &
\colhead{Mass\tablenotemark{a}} &
\colhead{Tracer\tablenotemark{a}} &
\colhead{Kinem.\tablenotemark{b}} &
\colhead{$i$\tablenotemark{c}} &
\colhead{} &
\colhead{$\MBH$\tablenotemark{e}} &
\colhead{} &
\colhead{}                           \\
\colhead{Model} &
\colhead{Profile} &
\colhead{Profile} &
\colhead{Dataset} &
\colhead{(\arcdeg)} &
\colhead{$\epsilon$ \tablenotemark{d}} &
\colhead{($10^7 \MSun$)} &
\colhead{$\ML$\tablenotemark{e}} &
\colhead{\S}            
}
\startdata
Base & \Vcor & \Vcor & S+G & 72 & 0.35 & $3.31^{+0.22}_{-0.17}$ & $3.6^{+0.1}_{-0.1}$ & \ref{DynMethod} \\
D1 & \Vcor & \Vcor & S+G & 72 & 0.45 & $3.48^{+0.42}_{-0.38}$ & $3.7^{+0.2}_{-0.2}$ & \ref{DynModEll} \\
D2 & \Vcor & \Vcor & S+G & 62 & 0.35 & $3.62^{+0.38}_{-0.49}$ & $3.8^{+0.1}_{-0.2}$ & \ref{DynModIncl} \\ \hline
K1 & \Vcor & \Vcor & S$_1$+G & 72 & 0.35 & $3.25^{+0.21}_{-0.13}$ & $3.6^{+0.1}_{-0.1}$ & \ref{AmbigK} \\
K2 & \Vcor & \Vcor & S$_3$+G & 72 & 0.35 & $2.20^{+0.54}_{-0.31}$ & $3.6^{+0.1}_{-0.1}$ & \ref{AmbigK} \\
K3 & \Vcor & \Vcor & G & 72 & 0.35 & $1.03^{+1.00}_{-0.28}$ & $3.7^{+0.1}_{-0.2}$ & \ref{AmbigK} \\
K4 & \Vcor & \Vcor & S+G$_1$ & 72 & 0.35 & $3.33^{+0.21}_{-0.19}$ & $3.7^{+0.1}_{-0.1}$ & \ref{AmbigK} \\ \hline
P1 & \NIRcor & \Vcor & S+G & 72 & 0.35 & $2.49^{+0.09}_{-0.10}$ & $3.5^{+0.1}_{-0.1}$ & \ref{AmbigP} \\
P2 & \Vcor & \NIRcor & S+G & 72 & 0.35 & $3.33^{+0.18}_{-0.17}$ & $3.6^{+0.1}_{-0.1}$ & \ref{AmbigP} \\
P3 & \NIRcor & \NIRcor & S+G & 72 & 0.35 & $2.43^{+0.21}_{-0.30}$ & $3.5^{+0.1}_{-0.1}$ & \ref{AmbigP} \\
P4 & $V$ & $V$ & S+G & 72 & 0.35 & $3.25^{+0.22}_{-0.14}$ & $3.6^{+0.1}_{-0.1}$ & \ref{AmbigP} \\ \hline
\enddata
\tablenotetext{a}{Mass and tracer profiles named as in Figure \ref{CumPhotomCenter} (cf.\ \S\ref{AmbigP}).}
\tablenotetext{b}{S = All STIS kinematic data listed in Table \ref{SpectrTable}. \\
\hspace{4mm} S$_1$ = As in S but without the central kinematic datapoint. \\
\hspace{4mm} S$_3$ = As in S but without the 3 central kinematic datapoints.\\
\hspace{4mm} G = All ground-based (Modspec) kinematic data listed in Table \ref{SpectrTable}. \\
\hspace{4mm} G$_1$ = As in G but without the central minor-axis datapoint.}
\tablenotetext{c}{Disk inclination (cf.\ \S\ref{DynModIncl}). Edge-on corresponds to $i=90\arcdeg$.}
\tablenotetext{d}{Isophote ellipticity, assumed constant throughout the galaxy (cf.\ \S\ref{DynModEll}).}
\tablenotetext{e}{Black hole mass and stellar $M/L$ ratio ($\ML$), assumed constant throughout the galaxy (cf.\ \S\ref{DynMethod}). Uncertainty values correspond to the nominal ``1$\sigma$'' one-dimensional uncertainties from the contour maps (Figure \ref{ModelPlots}). The $M/L$ ratio calibration is in the $V$ band.}
\end{deluxetable}

\end{document}